\documentclass[aps, pre, onecolumn, 10pt, showkeys]{revtex4-1} % 10pt font size

\usepackage{graphicx}% Include figure files
\usepackage{dcolumn}% Align table columns on decimal point
\usepackage{bm}% bold math
\usepackage{amsmath} % Required for align, tfrac

\begin{document}

\preprint{APS/123-QED}

%%%%%%%%%%%%%%%%%%%%%%%%%%%%%%%%%%%%%%%%%%%%%%%%%%%%%%%%%%%%%
%%%%%%%%%%%%%%%%%%%%%%%%%%%%%%%%%%%%%%%%%%%%%%%%%%%%%%%%%%%%%
%%%%%%%%%%%%%%%%%%%%%%%%%%%%%%%%%%%%%%%%%%%%%%%%%%%%%%%%%%%%%
%%%%%%%%%%%%%%%%%%%%%%%%%%%%%%%%%%%%%%%%%%%%%%%%%%%%%%%%%%%%%
%%%%%  FINAL PAPER PhysRevE: EB11720:
%%%%%  Resonance-driven Ion Transport and Selectivity 
%%%%%  in Prokaryotic  Ion Channels
%%%%%  r.l.westra, 
%%%%% accepted 5 November 2019
%%%% after acceptance updated : november 22, 2019
%%%% after acceptance updated : november 26, 2019
%%%% alignment with proofread T/m 6. In line 701 : december 10, 2019
%%%% this final version : december 11, 2019
%%%%   We 11/12/2019 om 11:58: klaar!
%%%%%%%%%%%%%%%%%%%%%%%%%%%%%%%%%%%%%%%%%%%%%%%%%%%%%%%%%%%%%
%%%%%%%%%%%%%%%%%%%%%%%%%%%%%%%%%%%%%%%%%%%%%%%%%%%%%%%%%%%%%
%%%%%%%%%%%%%%%%%%%%%%%%%%%%%%%%%%%%%%%%%%%%%%%%%%%%%%%%%%%%%%
\title{Resonance-Driven Mechanisms of Ion Transport and Selectivity }

\author{Ronald L. Westra }

\affiliation{
 Department of Gravitational Waves and Fundamental Physics,\\
 Maastricht University, Maastricht, the Netherlands. \\
 \href{https://orcid.org/0000-0003-1547-9117}{ORCID: 0000-0003-1547-9117} }

\date{\today}
\email{westra@maastrichtuniversity.nl}

%\date{November 5, 2019} % date may be explicitly specified
\date{\today} % date may be explicitly specified

\begin{abstract}

Ion channels exhibit an extraordinary ability to selectively transport specific ions with remarkable accuracy. However, the underlying mechanisms of ion selectivity remain poorly understood, with current explanations relying primarily on molecular biology and bioinformatics.  

Here, we propose a physical model for ion selectivity based on the \textit{Driven Damped Harmonic Oscillator} (DDHO). In this framework, the driving force originates from self-organizing turbulent structures within the dehydrating ionic flow through the ion channel—manifesting as oscillating pressure waves in one dimension and toroidal vortices in two and three dimensions. These density fluctuations efficiently transfer energy to aqua-ions that resonate with the driving frequency, causing them to shed their hydration shells and pass through the ion channel as free ions.  

Existing modeling approaches fail to capture the necessary spatiotemporal complexity of this process. To address this, we introduce a novel macroscopic continuum model for ionic dehydration and transport, incorporating the hydrodynamics of a dissipative ionic flow subject to electrostatic and amphiphilic interactions. This model integrates three classical physical frameworks: the Navier-Stokes equations from hydrodynamics, Gauss’s law from Maxwell’s theory, and the convection-diffusion equation from continuum physics.  

Numerical simulations of mixed ionic species with varying hydration states reveal the emergence of strong oscillations in the ionic flow. These oscillations drive efficient dehydration and generate a directed ionic jet into the cell, providing a natural engine for the DDHO mechanism. Theoretical predictions of our model align well with empirical patch-clamp data.  

The DDHO framework predicts a standard response curve with a unique resonance frequency that depends on the ion’s mass and charge. As a result, the driving oscillations act as a selective filter, distinguishing between different ion species. Applying the model to experimental ion transport data reveals a clear separation between chemical species and between hydrated and bare ions, quantified by a large Mahalanobis distance and high oscillator quality.  

Furthermore, the DDHO model offers insight into how single nucleotide polymorphism (SNP) mutations can disrupt ion selectivity, leading to severe genetic disorders. Mutations that alter the ion channel’s geometry shift the resonance peaks, impairing the proper selection of the required ion type.  

\end{abstract}

\maketitle

%%% \tableofcontents
%%% ToC:
\tableofcontents

\clearpage

\section{Introduction}
\label{sec:introduction}

%%% subsection ION CHANNELS - A
\subsection{The biological relevance of ion channels}
\label{sec:SOC-introduction}

Ion channels are nanoscale biological mechanisms that regulate the transport of hydrated ions (aqua-ions) across the membranes of animal cells, either from the extracellular environment into the cell or vice versa. These channels play a fundamental role in electro-physiological processes, including signal transmission in the nervous system and the propagation of electrical impulses in the heart \cite{Heijman-2013}. The study of ion transport mechanisms has revealed complex interactions governing their function, as detailed in previous work \cite{WestraPRE2019}.  

Aqua-ions consist of positively or negatively charged bare ions surrounded by one or more layers of dipolar water molecules \cite{Thompson-2009, Picollo-2009}. As part of their function, ion channels selectively dehydrate the incoming ions before allowing them to pass through. Each type of ion channel is highly specialized, selecting for a specific ion species, such as potassium or sodium (positive ions) or chloride or iodide (negative ions) \cite{Catterall-2015}. Additionally, ion channels operate on different time scales—slow, rapid, and ultra-rapid—depending on their physiological role.  

Cardiac cells (myocytes) and nerve cells (neurons) typically contain thousands of ion channels of varying types, enabling intricate electrical signaling. Evolutionary selection has shaped the genetic variations underlying ion channel function, leading to a diversity of exon structures. These genetic variations, depending on their expression levels, give rise to distinct ion channel types across different organisms, reflecting their ecological roles as predators or prey \cite{Kaufman-1993}.

%%% subsection ION CHANNELS - B
\subsection{Ion channels: function and control }
The main role of an ion channel is to generate an inward or outward flux of specific target ions through the membrane. In doing so, they change the electric charge inside the cell - and therefore the potential difference along the cell membrane. This potential itself can trigger changes in the cell and in the ion channel state, as in voltage-gated ion channels. This feedback mechanism is at the basis of macroscopic conductance of electrical signals \cite{Heijman-2011}. 
Numerous sources, both experimental and theoretical, report the inherently stochastic nature of ion transport through nanochannels. Siwy and Fulinski \cite{REF1SUG11} show that the flicker noise observed in synthetic nanopores and biological channels originates from its opening-closing process, and that it is related to the underlying motions of channel wall constituents. This provides clear experimental evidence that the underlying dynamics is entirely stochastic. \\

%%% subsection ION CHANNELS - C
\subsection{Architecture and geometry of ion channels}
\begin{figure}
\begin{center}
\includegraphics[width=0.7\columnwidth]{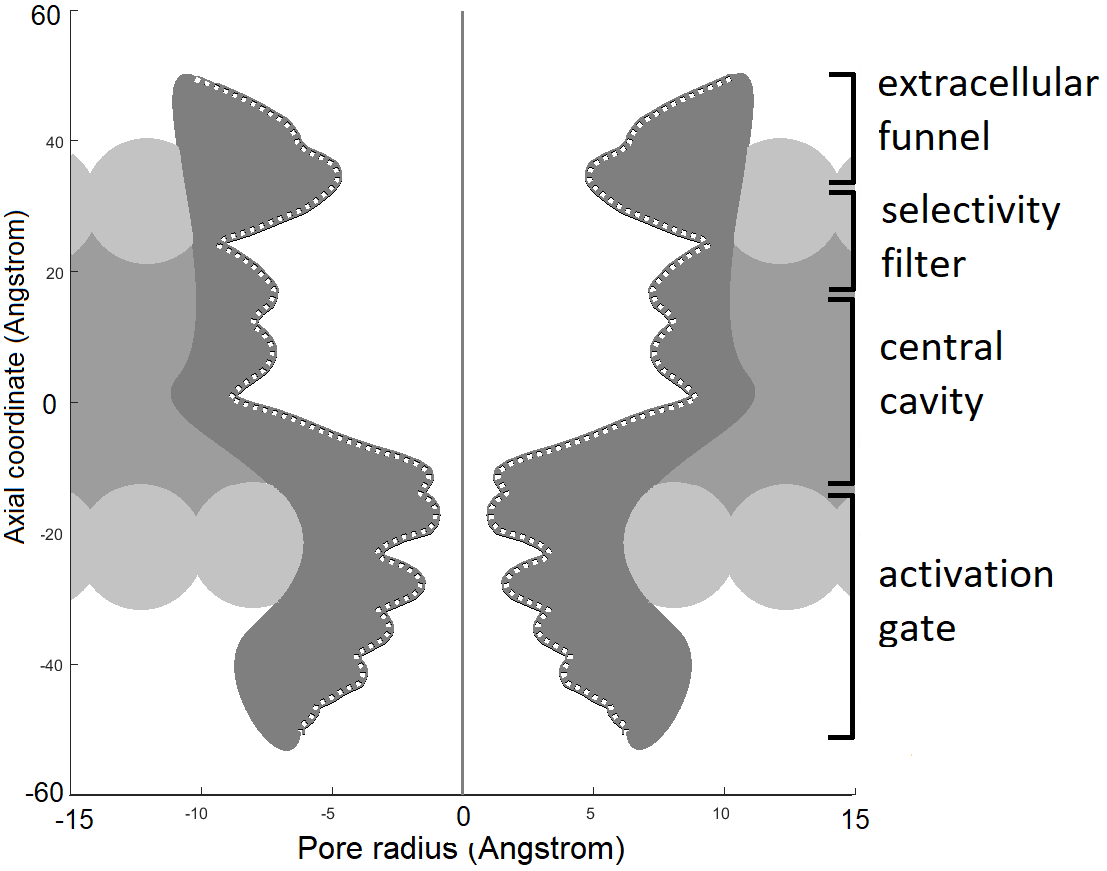}
\caption{Schematized lateral view of the ligand-gated ion channel "ELIC" from {\em Erwinia chrysanthemi} \cite{Song-2010} with channel axis vertical. Functional parts indicated right. Bottom side is intracellular, upper side extracellular. Dotted: interior walls, dark gray: $\alpha$-subunit protein envelope, light gray:  lipid bilayer of cell membrane. Protein subunits $\beta, \gamma$, ... not depicted }
\label{fig: ELIClateralview}
\end{center}
\end{figure}

\begin{figure}
\begin{center}
\includegraphics[width=0.47\columnwidth]{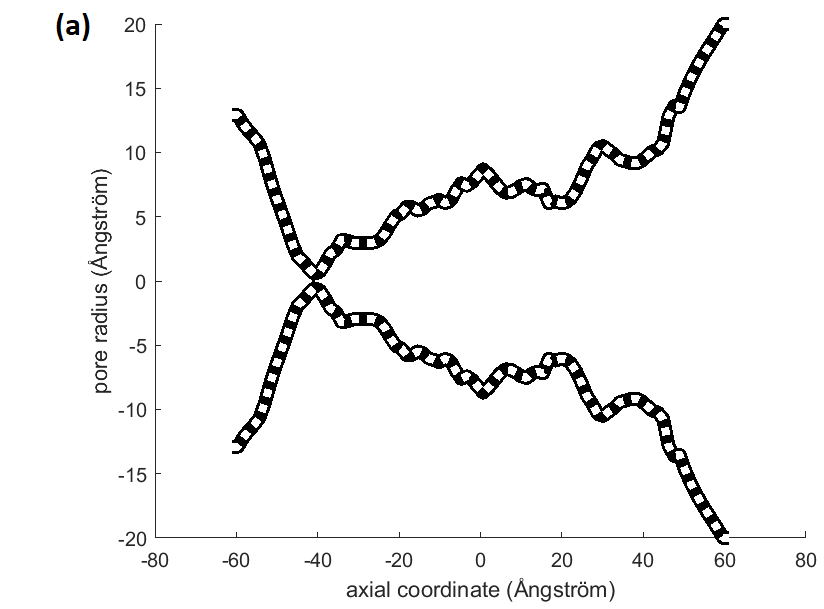}
\includegraphics[width=0.47\columnwidth]{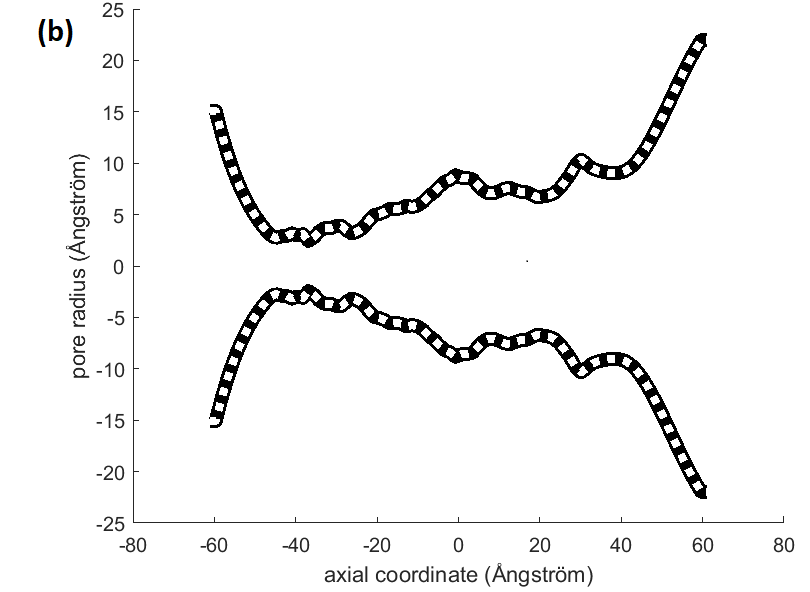}
\caption{Schematic representation of the lateral view of the GLIC proton-activated channel from {\em Gloebacter violaceus} \cite{Song-2010} used in EAH modeling. (a): closed state (GLIC1), (b): open state (GLIC2). Right side: extracellular, left side: intracellular environment. Dotted: interior walls. Cell membrane and subunits not depicted. 
%Note the different scales on horizontal and vertical axes.
}
\label{fig: GLIC12}
\end{center}
\end{figure}

Ion channels exist in a wide variety of geometries and architectures \cite{IONstandardlit}. Here, we focus on the bacterial ligand-gated ion channels ELIC from {\em Erwinia chrysanthemi} (Fig \ref{fig: ELIClateralview}) and GLIC from {\em Gloeobacter} (Fig \ref{fig: GLIC12}), and the prokaryotic potassium channel KcsA from {\em Streptomyces lividans} (Fig \ref{fig:KcsAporeradius}).  
An ion channel is composed of four functional units: (1) an extracellular funnel, (2) a selectivity filter, (3) a central cavity, and (4) an activation gate \cite{MART2008}. 
Note that ELIC and GLIC have particularly wide selectivity filters (SF) \cite{SPUR2013}\cite{FRITS2011} compared to KcsA that exhibits a narrow tubular SF \cite{Kuang2015}. 

%%%% figure ELIClateralview: NavAB ion channel.jpg
Ion channels are situated in the membrane of the cell. As integral membrane proteins, they typically consist as assemblies of several individual homologous proteins, in a more-or-less rotational-symmetric assembly \cite{Catterall-2017}, closely packed around a water-filled pore through the cell membrane. 
The core geometry of an ion channel consists of $\alpha$ subunits that are transcribed and translated from a single gene \cite{IONstandardlit}\cite{Catterall-2010}\footnote{For instance, the KCNQ1 gene directly codes for the KCNQ1 potassium voltage-gated ion channel \cite{Heijman-2011B}}. 
These subunits combine to form a slightly skewed helically winding but otherwise mostly cylindrical symmetric structure \cite{Shaw-2013}. 
Besides the pore-forming $\alpha$ subunits, they contain various auxiliary subunits, denoted $\beta$, $\gamma$, and so on.
The internal geometry is defined in terms of the equipotential surface of the tertiary structure based on the individual 3D positions of the thousands of ligands of the proteins. 
The geometrical configuration of the ion channel exists in a number of conformal states, varying between the extremes 'open' (full conduction of ions) and 'closed' (no ion flux) \cite{Kapetis-2017}, see Fig \ref{fig: GLIC12}. Crystal structure data is available for an ever increasing number of ion channels \cite{Song-2010}\cite{Song-2013}.

%%%% Figure GLIC12: GLIC-ionchannels GLIC1-CLIC2

%%% subsection IOC selectivity
\subsection{What is ion channel selectivity }
\label{sec:whatIOSis}
The mechanism behind ion channel selectivity \footnote{We define the {\em selectivity} of an ion channel as the fraction $CI/(CI + FI)$. Here, $CI$ is the amount of the correct ion species (true positives) it selects from an ambient mixture of various aqua-ions in the extracellular suspension  into the intracellular space. $FI$ is the amount of incorrect ions the channel erroneously selects (false positives). The associated {\em flux} is defined as the rate in which the correct ion species enters the cell.} is not well understood \cite{Padhi_2015}. There is a vast literature on theories explaining ion selectivity, predominantly in the biomolecular and genetic domain.  
While values for ion selectivity for an ion channel in the open state attain values exceeding 99\%  \cite{Corry2018}\cite{ROUX2017}, the simultaneous rate of transport easily surpasses $10^7$ ions per second by magnitudes \cite{Shaw-2013}. Measurements on the bacterial ELIC channel exhibit transport rates peaking at $2.5.10^9$ ions/sec \cite{ZIM2011}, see Fig \ref{fig:ZIMM2011FIGS4B}. Some scholars report ionic speeds up to 39 m/s \cite{CHUNG1998}.
The high ion species selectivity combined with a large conductance rate is called the 'central paradox of ion channels' \cite{Kuyucak2003}\cite{MacKinnon2003}. \\
There is overwhelming experimental evidence and general consensus amongst scholars that in most ion channels ions pass in a single file through the narrow selection filter (SF), see \cite{Corry2018}.
For such channels the challenge is to find a 1D physical mechanism that simultaneously explains fast throughput and high selectivity. 
Pawe\l{}ek et al. \cite {PAW2010} modeled asynchronous motion of ions and water molecules in the SF of KcsA. Their model correctly estimates the experimentally determined electrical current through the channel.
Kopec et al. \cite{Kopec2018} recently presented numerical simulations of direct knock-on of desolvated ions that favours K$^+$ over Na$^+$ ions with high accuracy and efficiency. Water molecules play an essential role in their model, as the energetic cost for dissolving a K$^+$-ion is considerably lower than for a Na$^+$ ion. Their work thus exposes the deep-rooted role of dehydration and the thermodynamic potential in the selectivity process. \\
Yet, there are also numerous other types of ion channels, for instance the bacterial ligand-gated ion channels ELIC and GLIC, that do not exhibit narrow SFs but a more spacious cavity that actually over-arches the SF. \\
%%% PPLG = Prokaryotic Pentameric Ligand-Gated
The question thus is whether there is a more fundamental physical principle that explains selectivity in both cases, viz. narrow and broad SFs. 
Central in this principle must be the ionic transport, de- and re-hydration, and thermodynamics. 
This demands some additions to current physical explanations, notably the Poisson-Nernst-Planck (PNP) model, that cover these concepts.

%%% subsection previous work IO-SEL	
\subsection{Survey of physical models for ion channel selectivity} 
\label{sec:Survey}
The conductance and selectivity of ion channels is a wide and intensively studied multidisciplinary field, accompanied by a substantial and comprehensive literature. In the perspective of the biomedical field, notably structural and molecular biology and bioinformatics, emphasis is on the effects of binding sites, protein structures and genome sequences on selectivity and conductance \cite{Thompson-2009}\cite{Picollo-2009}\cite{MAFF2012}\cite{RUDY2018}. See \cite{Catterall-2017} for a recent survey. 
However, most models in the bio-molecular domain have problems explaining the small nanoscale time-scales involved in ion selection.
\\
We focus on the perspective of the physical-chemical domain, where fundamental physical principles are applied to obtain mathematical representations and computational solutions. In this perspective, the three major theoretical frameworks are: deterministic continuum models; many-particle system molecular dynamics; and the stochastic Brownian dynamics (BD). Continuum theories describe a.o. ionic permeation \cite{Koepfer-2011} or diffusion of charged chemical species in a fluid medium \cite{Song-2011}, a.o. using free energy barriers.  \\
The Poisson-Nernst-Planck (PNP) model is based on a mean-field approximation of ion interactions and continuum descriptions of concentration and electrostatic potential \cite{Zheng2011}. It provides qualitative explanation and increasingly quantitative predictions of experimental measurements for the ion transport problems.
The system of PNP equations provide the most appropriate and widely accepted continuous description framework for ion transport problems, despite its notorious technical difficulties, both in simulations and numerical solution.
Kuyucak and Bastug \cite{Kuyucak2003} and Sahu et al. \cite{Sahu-2017} employ PNP and the Poisson-Gauss equation to determine the electrostatic potential to describe electrodiffusion and electrophoresis. 
Kosinska et al. \cite{REF1SUG10} provide an analytic expression for ion current rectification using model reduction on 3D PNP and obtain a 1D model of synthetic nanopores that adequately describes experimental data. \\
Molecular dynamics is a fundamental {\em ab initio} approach, and essentially a Classical Mechanics many-particle system description \cite{Cheng-2009} \cite{Cory-2013}. However, like in continuum models, it suffers from numerous formal and practical setbacks, such as the inability to derive various key properties, including ion channel conductance, and severe computational complications \cite{Kuyucak2003}. 
BD \cite{Cooper1985}\cite{Corry2002} and random walk models \cite{Berneche-2003} \cite{Song-2009} provide a semi-microscopic alternative to continuum theories \cite{Kuyucak2003}. In this approach, the microscopic motion of ions is averaged over a large ensemble in phase space, starting from the master equation \cite{vanKampen-2007}, using the Langevin equation, while the rest of the system is treated in the continuum approximation. This models the average motion near the interior surface, taking various constraints in account, including finite size. 
However, BD simulations are computationally much more demanding than solving the Poisson-Boltzmann (PB) and PNP differential equations. \\
In current modeling, most researchers focus entirely on the effect of the electric field on ion conduction and totally neglect the impact of geometric constraints on channel gating. In our work, the ionic flow characteristics depends strongly on the channel geometry, and thus it belongs to a family of recently published models \cite{RUBI2017}-\cite{Wawr2018}, where the influence of channel protein’s geometry on its activity are discussed.

\begin{figure}
\begin{center}
\includegraphics[width=0.48\textwidth]{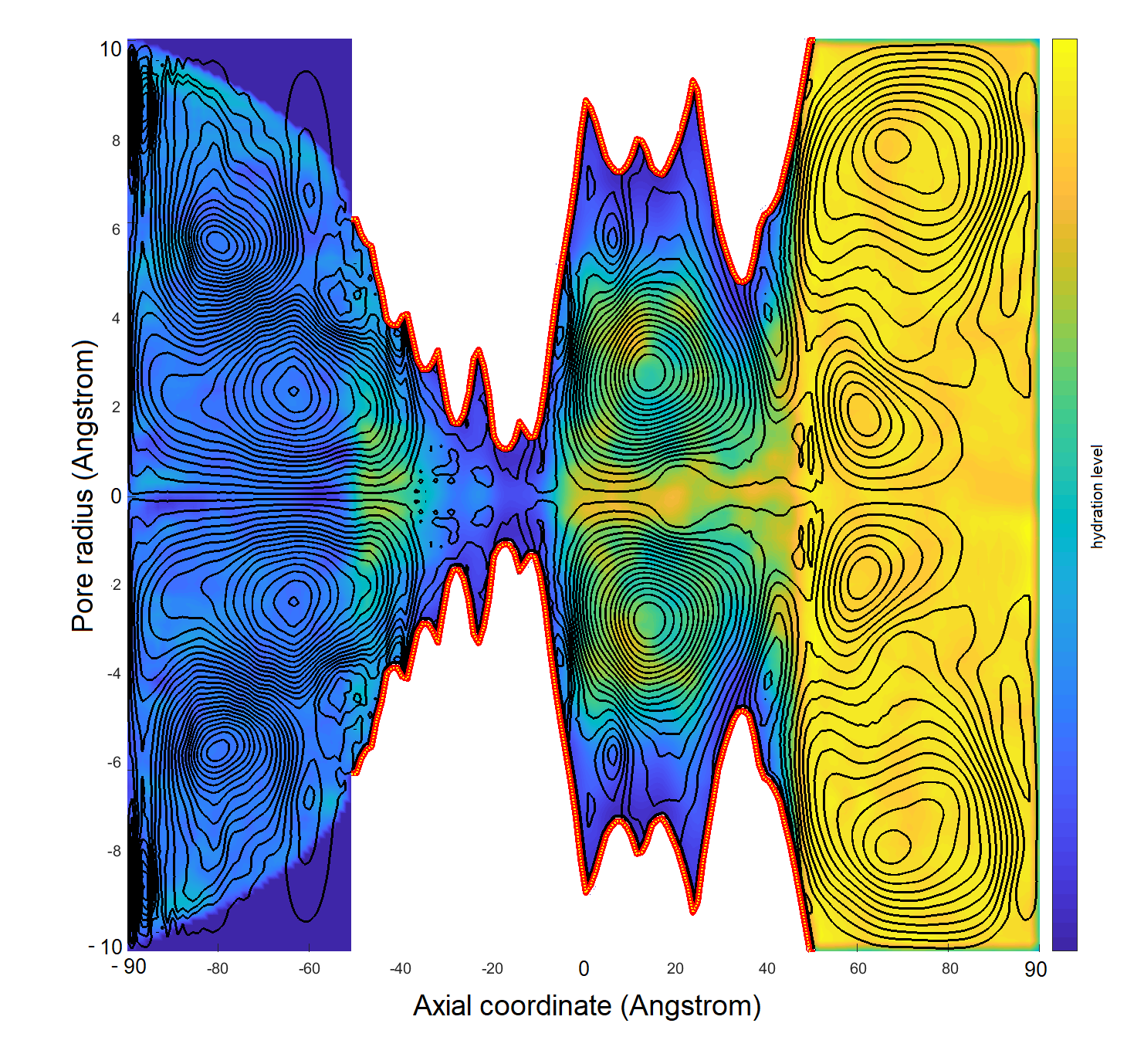}
\caption{Flow-lines of the fluid velocity field $\textbf{u}(\textbf{r},t)$ (solid black) and hydration $\eta(\textbf{r},t)$ in a cross-section of the ELIC ion channel at $t$=20 ms, showing a strong central toroidal vortex. With Re = 600, after 250K iterations. Gray scale (online color) depicts hydration. Left: intracellular, right: extracellular space, middle (white): cell membrane. Figure made with the EAH model.}
\label{fig:ELICflow}
\end{center}
\end{figure} 

\subsection{The Physical Validity of Continuous Models at the Nanoscale}
\label{subsecContModels}

Macroscopic continuum models are appealing as they bypass the mathematical and computational challenges posed by atomic-scale discreteness and stochastic fluctuations. By averaging microscopic quantities over the entire accessible phase space, such models offer a tractable framework for describing complex systems \cite{Colin2014}.  

However, the validity of the continuum assumption relies on the premise that the phase space is both ergodic and fully explored. This assumption fails in various dynamical systems, including those governed by the Navier-Stokes equations. Over time, such systems become constrained to only a minute subset of phase space due to \textit{ergodicity breaking} (EB) \cite{BOUC1992}. This raises fundamental concerns regarding the applicability of continuum models in our context.  

Furthermore, a key question arises: what is the smallest physical scale at which a continuum description remains valid? Identifying this threshold is crucial for assessing the predictive power of macroscopic models at the nanoscale.  

In this section, we examine these two issues in detail, addressing both the impact of EB on continuum assumptions and the fundamental physical limits of continuum modeling.  

%%%%%%%%%%%%%% SubSec F1 :
\subsubsection{Ergodicity Breaking and the Validity of Macroscopic Models}
\label{ErgodBreaking}

The validity of the proposed macroscopic \textit{Electro-Amphiphilic Hydrodynamic} (EAH) model, derived from underlying stochastic microscopic dynamics, rests on the assumption that it constitutes a truly ergodic dynamical system. This implies that the system's dynamics explore the accessible phase space uniformly, both temporally and spatially.  

However, understanding ergodic properties in dynamical systems remains an open challenge in mathematical physics \cite{FLAND2005}. One of the most notorious unresolved questions in this context is whether turbulence—fully developed from either microscopic or macroscopic models—exhibits ergodicity \cite{ADZH2003}. This is central to our explanation of ion channel selectivity.  

The most extensively studied system in this regard is the Navier-Stokes equation, which underlies the EAH model. Although individual particle trajectories generated by the EAH equations exhibit highly complex, and in some cases fractal, structures, this fine-grained detail is largely irrelevant for understanding the full system. Instead, statistical properties of the collective motion—particularly the system’s attractor set in phase and configuration space—are of greater significance \cite{FLAND2005}.  

The phase and configuration spaces of such systems are extraordinarily high-dimensional (e.g., $10^{80}$). However, ergodicity breaking (EB) fragments them into numerous smaller subdomains, which may be either fully disconnected (\textit{hard} EB) or weakly connected (\textit{weak} EB) \cite{BEL2006}. This is a well-documented feature of Navier-Stokes systems \cite{BARK2006}, raising concerns about whether a mean-field derivation remains valid, given that only a restricted subspace of the phase space is occupied at any time.  

Despite these concerns, evidence from turbulent flows suggests that ergodicity may still hold at a statistical level. B. Galanti et al. \cite{GALANTI2004}, studying the Navier-Stokes equations, provided arguments supporting the ergodic hypothesis in turbulence. Moreover, observed statistical properties of turbulent flows are highly reproducible and ergodic, despite deterministic forces acting to break ergodicity at larger scales \cite{Tsinober2003}.  

Importantly, even if EB influences the trajectory of a specific system realization, it does not necessarily impact the \textit{derivation} of the EAH equations. The system’s macroscopic behavior is not dictated by an accidental, transient subspace of phase space occupied after EB but rather by the ensemble of all possible such subspaces. Averaging over this complete set restores the validity of the macroscopic EAH equations, making them robust against both hard and weak EB.  

\subsubsection{From Microscopic Discrete Dynamics to Macroscopic Continuum Models}

The applicability of continuous models at the nanoscale remains a topic of ongoing debate \cite{REF1SUG3}. Corry, Kuyucak, and Chung \cite{REF1SUG1,REF1SUG2} investigated mean-field approximations in Poisson-Boltzmann (PB) and Poisson-Nernst-Planck (PNP) theories, comparing them with Brownian Dynamics (BD) simulations in cylindrical nanopores. Their findings indicate that continuum models significantly overestimate shielding effects when the pore radius is smaller than two Debye lengths, suggesting that such models fail to accurately describe electrolyte physics in nanopores.  

Furthermore, continuum models exhibit theoretical limitations. They neglect dielectric self-energy effects, which play a crucial role in conductance saturation via self-energy barriers \cite{Sahu-2017, Cherstvy2006, Kuyucak2001}. Additionally, they face practical computational challenges, further questioning their suitability for modeling nanoscale ion channel dynamics.  

This issue is critical for our explanation of ion selectivity, which relies on the emergence of nanoscopic vortices and eddies. In Section \ref{ContParams}, we address this concern by identifying physical parameters that define the smallest length scales at which turbulence, and consequently micro-vortices, can exist.  

Ultimately, at even smaller scales dominated by quantum phenomena, continuum modeling transitions to a different formalism—namely, the Schrödinger equation—where it remains fully applicable.  

%%%% SS motivation
\subsection{Motivation and structure of the paper}
In this paper we study the physical foundations of ion channel selectivity and to this ends we propose the classical physical mechanism of the  the \textit{Driven Damped Harmonic Oscillator} (DDHO). 
Current models for ionic transport and selectivity in ion channels fail to provide the required harmonic driving force for generating the resonances. 
Therefore, in Section \ref{ModelingIOC} we introduce a new physical continuum model for the hydration electrohydrodynamics of a dissipative ionic flow through an ion channel in the amphiphilic environment of its interior walls.
This model is called the \textit{Electro-Amphiphilic Hydrodynamics} (EAH) model. 
This approach yields a macroscopic continuum model, that is new in that it allows for the internal motion in the channel. As it is expressed in terms of the fluid vector field, it can represent streamlines, so oscillations in 1D, and vortices and eddies in 2D and 3D. This can not be achieved in BD and MD.
In Section \ref{NumExp} we study the EAH model numerically in 1D, 2D, and 3D, and compare its predictions with experimental patch clamp data. 
In Section \ref{sec:DDHO-model}, we venture to explain ion channel selectivity as result of frequency entrainment by the classical DDHO mechanism. The driving frequency is provided by persistent fluid oscillations, such as standing waves and toroidal vortices in the dissipative ionic flow through the ion channel. 
In Section \ref{SEC-Discusion}, we discuss how the observed oscillatory fluid dynamics actually acts as the core of the ion selectivity control mechanism.
Thus, we come to the conclusion that hydrated ions (aqua-ions) and dehydrated ions (bare ions), as they differ in mass and effective electrical charge, respond differently to the driving frequency of the fluid oscillations when trapped inside. In case of resonance this response is maximal, which defines a resonance frequency that depends on the mass and charge of the ion. In this way, the oscillator, vortex or oscillating pressure wave, acts as a selective mechanism that uniquely filters out one specific ion.

%%%%%%%%%%%%%%%%%%%%%%%%% %%%%%%%%%%%%%%%%%%%
%%%%%%%%%%%%%%%%%%% SECTION II: Electro-Amphiphilic Hydrodynamics (EAH): A Physical Model for the spatiotemporal dynamics of a dehydrating ionic fluid
%%%%%%%%%%%%%%%%%%%%%%%%%%%%%% %%%%%%%%%%%%%%%%%%%%%%%%%%%%%
\section{Electro-Amphiphilic Hydrodynamics (EAH): A Physical Model for the spatiotemporal dynamics of a dehydrating ionic fluid}
\label{ModelingIOC}

We introduce a macroscopic continuum model for ion transport through an ion channel, and show that it exhibits strong and persistent oscillations in the form of oscillating pressure waves in 1D and vortices in 2D and 3D. These oscillations act as the driving force for self-organized ion species selectivity. 

\subsection{Model assumptions and generalizations}
\label{SectionModelAssumptions}

\subsubsection{Ions as fluid: from discrete events to a continuum model}
Rather than the microscopic dynamics of individual ions, we here focus on a macroscopic continuum model in the ergodic subspace  of a microcanonical ensemble of a large number of ions. It is well known from Computational Physics that even for relatively small numbers of molecules, viz. a few dozen, a macroscopic model adequately describes the overall systems dynamics \cite{Allen1993}\cite{HOOV2008}. \\
Our approach is justified by: (i) the aforementioned high rate of ion transport, peaking to 2.5E9 ions/sec for ELIC; (ii) the fast selection ($>$ 99\%) and small length and time scales ({\AA} and nanosec) involved in the selection process that are more characteristic for a physical than for a molecular-biological phenomenon \cite{MILO215}-\cite{SHAM2016}; and (iii) the huge number of ion channels per individual cell - typically many tens of thousands \cite{Heijman-2011}\cite{BUCH2002}. The latter statistically  averages out the ion channel dynamics per cell. \\
The thus obtained average ionic motion exhibits highly detailed fluid streamlines as in Fig \ref{fig:ELICflow}, despite the comparable size of an ion and the diameter of a channel - the pore radius - of only a few \AA ngstr\"{o}ms \cite{Naranjo-2016}.  
Macroscopically, our model constitutes an electrically charged incompressible fluid of several ion species, spatially contained within the confinement of the ion channel. The interactions between the ions are governed by electrostatic interactions and collisions, while the interactions of ions with the surface of the ion channel are dominated by amphiphilic interactions. 
Gravitational and magnetic forces are considered too weak and are not included. Another assumption is that the process is considered to be isothermal.

%%%%%%%%% SUBSEC phys Foundations
\subsubsection{Physical foundations of the EAH model}
Starting point for constructing a macroscopic model is the microscopic Boltzmann master equation for a mixture of multiple interacting chemical species in a fluid. Following the derivation of M. Bisi and others
% \cite{Bisi-2014}\cite{Bisietal2008}\cite{Bisietal2014}\cite{Desvillettes-2009}\cite{Desv17}\cite{Breuer-1993} 
\cite{Bisi-2014}-\cite{Breuer-1993} 
in an electrohydrodynamic setting following A.I. Zhakin \cite{Zhakin-2013}, we obtain a modified Navier-Stokes equation (NSE) for the fluid velocity vector field $\textbf{u}(\textbf{r},t)$ of an incompressible fluid in equation \ref{eq:EHANS1}. Incompressible, because of the strong repelling forces between the ions and the ambient water. 
In our case, however, we allow for a continuum of species, namely the dehydration series of the ions, measured by $0 \leq \eta \leq 1$. Internal collisions between the molecules are quantified by the diffusion term in the NSE and gauged by the dynamic viscosity $\nu$ \cite{Lavalette-1999}.  \\
Starting point for the hydration equation \ref{eq:EHANS2} is the chemical reaction: 
\begin{equation}\label{eq:EHANS0}
X^{K}.(H_{2}O)_{n_H} \rightleftharpoons X^{K}+n_{H} H_{2}O
\end{equation}
for the (de/re)hydration of ion $X$ with chemical valence $K$ ($+$ or $-$) and its hydration number $n_H$ \footnote{For example $X^{K}=Ca^{2+}$ has $n_H = 7.2$}. This leads to a convection–-diffusion equation for $\eta$  with terms for convective transport (including vorticity), orthotropic diffusion, and sources and sinks. \\
Furthermore, the geometry of the ion channel imposes strict boundary conditions. First, a Neumann boundary condition on $\textbf{u}$, as we assume that the flow is entirely parallel to the surface on the set of all ion channel boundaries ${\mathcal{B}}$. So, $\textbf{u}$ is perpendicular to the normal vector of ion channel interior surface at the boundary. Moreover, we assume that at the entrance of the ion channel, with cylindrical coordinate $z=L$, the velocity faces horizontally inward with a velocity $U_0$. Second, Dirichlet conditions on: (i) $\eta$ and $\textbf{u}$ because of a constant influx of fresh fully hydrated ions at the entrance of the ion channel, i.e. $z = L$; and: (ii) the electric potential $\phi_E(\textbf{r},t)=0$ on ${\mathcal{B}}$.

%% SS: Mathematical equations of EAH
\subsubsection{Mathematical equations of Electro-Amphiphilic Hydrodynamics }
Combining both macroscopic processes, ion transport and ion (de)hydration, we obtain the set of coupled PDEs in equations \ref{eq:EHANS1}-\ref{eq:PNPEquation}. They model the electro and amphiphilic hydrodynamics of an incompressible mixture of aqua and bare ions, immersed in ambient H$_2$O molecules.
We refer to the obtained continuum model as the {\em Electro-Amphiphilic Hydrodynamics}: EAH. 
The dynamic EAH model is expressed in terms of: 
\begin{enumerate}
  \item The joint spatiotemporal fluid velocity vector field $\textbf{u}(\textbf{r},t)$ pertaining to {\em all} types and species of ions and all other molecules - including H$_2$O.
  \item The scalar field $\eta_s(\textbf{r},t)$: the local degree to which ion species $s$ is hydrated, so a number between 0 and 1 \footnote{Note that individual ions contain an integer number of surrounding $\text{H}_2\text{O}$-molecules, but that in the statistical average of our approach this number is real, called the {\em hydration number} $n_H$.}.
  \item The spatiotemporal scalar electric potential field $\phi_E(\textbf{r},t)$. 
  \item The ion concentration, i.e. the particle density, $n_s$ of ion species $s$.
\end{enumerate}
%%% \noindent 
The EAH-model is expressed as a system of four PDEs and three associated conditions and constraints:

%%%%%%%%%% EHANS EQUATIONS:
\vspace{4mm} 
\noindent SOLVE: $\{\bf{u},\eta_s,n_s,\phi_E\}$ for all $(\textbf{r},t) \in {\mathcal{S}}\times{\mathcal{T}}$ from: 
\begin{eqnarray} 
\tfrac{D\textbf{u}}{Dt} = \overline{\nu}{\nabla^2}\textbf{u}-\nabla P/\rho-\overline{\chi}\nabla \phi_E- \overline{a}\nabla \phi_A  \label{eq:EHANS1} \\
\tfrac{D\eta_s}{Dt} = (1-\eta_s)\ast\sigma^+_s  - \eta_s\ast\sigma^-_s + \nabla\cdot(D_s\nabla\eta_s)    \label{eq:EHANS2}  \\
\varepsilon_0\nabla\cdot({\varepsilon}_r\nabla\phi_E) = -(n\overline{\chi} + \lambda_E) \label{eq:EHANS3}\\
\tfrac{Dn_s}{Dt} = -\nabla\cdot\Phi_s 
\label{eq:PNPEquation}
\end{eqnarray}
\noindent
SUBJECT TO:
\begin{itemize}
  \item incompressible fluid:  $\nabla\cdot\textbf{u} = 0$,   $\tfrac{D\rho}{Dt} = 0$.
  \item boundary conditions:   $\eta_s|_{z=L} = 1$, ${\bm u}|_{z=L} = - U_0\hat{\textbf{e}}_z$, $\tfrac{\partial\textbf{u}}{\partial n}|_{\mathcal{B}} = \bm{0}, \hspace{3 mm} \phi_E|_{\mathcal{B}} = 0$.
  \item initial conditions:   $\eta_s(\bm{r},0) = 1$,  $\textbf{u}(\bm{r},0) = \bm{0}$ 
\end{itemize}

\noindent
With convective derivative: $\tfrac{D }{Dt} = \frac{\partial }{\partial t} + (\textbf{u}\cdot\nabla) $, and the Hadamard-Schur i.e. component-wise vector product: $(\textbf{a}\ast \textbf{b})_i = a_ib_i$. $\mathcal{S}$ represents the set of spatial coordinates, inside the interior of the ionchannel and left and right where necessary. $\mathcal{T}$ represents the time domain in our modeling, starting from $t$=0. 
The plane $z=L$ (in cylindrical coordinates with $z$-axis aligned with the ion channel symmetry axis) marks the extracellular entrance of the ion channel. 
A mixture of $S$ species of ions is represented by the hydration array $[\eta_1,\eta_2,..,\eta_S]$. Overline characters $\overline{\eta}, \overline{\nu}, \overline{\chi}, \overline{a}$ describe weighted ensemble averages over $\eta$, see equations \ref{eq:etamean}-\ref{eq:amean}. 
All other elements in these equations are discussed in detail in the Sections below.

%%%%%%%%%%%%%%%%%% SubSec: Explanation of the terms in EAH
\subsection{Parameters and terms in the Electro-Amphiphilic Hydrodynamics Equations}
\label{SubSecEHANSterms}
Equation \ref{eq:EHANS1} gives the spatiotemporal equation for the vector field $\bf{u}$ that applies for all species of the mixture. The LHS of the equation represents the convective derivative of the flow. The right-hand side (RHS) contains various terms. The first term is the isotropic diffusion measured by the viscosity $\nu$. The second term represents the force, exerted on the fluid, due to pressure gradients. Note that this is the only term that explicitly contains the mass density $\rho$. The next two terms are the body forces exerted by respectively the electrical field and the amphiphilic interaction, and will be discussed in more detail below.  
The hydrodynamic pressure $P(\textbf{r},t)$ in equation \ref{eq:EHANS1} is a scalar field used for regularization to match the LHS and RHS.\\
\noindent The system of PDEs is solved under constraints that the flow $\textbf{u}$ is divergence-free, and the net mass density $\rho$ is conserved - as, following equation \ref{eq:EHANS0}, only the frequencies of the aqua versus bare ions interchange. Consequently, the particle density $n$ is {\em not} conserved, as aqua ions break up in bare ions and water molecules.
For a mixture of bare and aqua ions in a substrate of inert (organic and inorganic) molecules (the 'rest') in ambient water, the integral densities $\rho$ and $n$ are given by: $\rho = \rho_{\text{H2O}} + \rho_{\text{rest}} + \sum_k \rho_k + \sum_k n_k h_k m_{\text{H2O}}$, and: 
$ n = n_{\text{H2O}} + n_{\text{rest}} + \sum_k n_k + \sum_k (1-\eta_k)h_k n_k $. \\
Here, $\rho_*$ and $n_*$ are resp. the mass and particle density of component *, $m_*$ the (free, i.e. unhydrated) mass of an individual molecule of the component. Note that $\rho_k = n_k m_k$. For ion species $k$, $h_k$ denotes its hydration number (a.k.a. $n_H$: 'shell number'). 
Note that several variables in the equations explicitly depend on the degree of hydration. Besides the particle density $n$, these include the charge density $\chi$ and the viscosity $\nu$. \\
The boundary conditions, including Neumann or Dirichlet, model the ion interactions and the electric field with the ion channel wall. As initial conditions we start from a fully hydrated field at rest. \\
An important auxiliary local variable in the model is the orthogonal or minimal distance $d$ to the ion channel wall. This metric is a function of position $\textbf{r}$ only, and gives the minimum distance from $\textbf{r}$ to any point on the ion channel walls. The gradient $\nabla d(\textbf{r})$  gives the direction towards the wall, and after normalization it becomes the local orthogonal unit vector $\hat{\textbf{e}}_{\perp}(\textbf{r})$ directed towards the wall, and by implication also defines two local unit vectors 
$(\hat{\textbf{e}}_{\parallel1}(\textbf{r}),\hat{\textbf{e}}_{\parallel2}(\textbf{r}))$ parallel to the wall \footnote{We chose unit vector $\hat{\textbf{e}}_{\parallel1}(\textbf{r})$ to be co-incident in the $(z,r)$ plane in cylindrical coordinates.}.

%%%% SUBSEC Hydration DYna
\subsubsection{Hydration dynamics}

Equation \ref{eq:EHANS2} represents the dynamics of de/re-hydration and consists of three terms. 
The LHS expresses the convective motion of ions. 
The first term of the RHS denotes the sources and sinks for aqua-ions near the walls. 
We introduce the amphiphilic source/sink function $\sigma$ acting on the hydrated fraction $\eta$ of the ions. It depends only on the minimum distance $d \ge 0$ to the wall and the ion radius $R_s$ of ion species $s$ \cite{Dove1997}. 
For a mixture of $S$ species of ions, this becomes an array $\sigma_s(d)$, $s=1 \dots S$. For a particular species of ion $s$ we define this function as: $\sigma_s(d) = K_s^+(1-\eta_s)\sigma^+_s(d) -K_s^-\eta_s \sigma^-_s(d)$, with positive functions $\sigma^+$ (source: hydrophobic) and $\sigma^-$ (sink: hydrophilic), both ranging between 0 and 1. The positive constants $K_s^{+/-}$ represent the strength of source/sink $s$. 
In our approach, we model a sink $\sigma^-(d)$ with the spherical cap function: $\sigma^-_s(d) = (R_s-d)^2(2R_s+d)/2R_s^3$ for $d \le R_s$, and zero elsewhere. This is the fraction of a sphere of radius $R_s$ behind a plane at distance $d$ from its center, and specifies the penetration of an ion of type $s$ into the ion channel wall. Rehydration of dehydrated bare ions occurs ubiquitous and constant with kinematic reaction constant $K^+_s$. It is thus represented by the (implicit) source function $\sigma^+_s(d) = 1 - \sigma^-_s(d)$. Without the convective term it would lead to a dehydrated zone alongside the walls. \\
The last term denotes the diffusion of aqua-ions. Diffusion of aqua-ions is considered to be generally isotropic with diffusivity $\delta_0$. However, near the ion channel walls it is considered to be orthropic, with preferred diffusion $\delta_1$ alongside the surface. This is modelled with a mixed diffusion matrix $\text{D}(d) = \beta(\textit{d})\text{D}_0 + (1-\beta(\textit{d}))\text{D}_1$. The scalar function $\beta$ is a function of the orthogonal distance $d$ to the ion channel wall, and monotonically increases from $\beta$=0 for $d$=0 to $\beta$=1 in the deep interior of the channel. Here we define $\beta$ as the above introduced spherical cap function: $\beta(d)\equiv\sigma^-(d)$. Matrix $\text{D}_0=\delta_0\text{I}$ is the constant global isotropic diffusion matrix \footnote{Here I stands for the identity matrix}. The local orthotropic diffusion matrix $\text{D}_1$ is defined as: $\text{D}_1 = 
\delta_1\hat{\textbf{e}}_{\parallel1}\otimes\hat{\textbf{e}}_{\parallel1}^\text{T}
+ \delta_1\hat{\textbf{e}}_{\parallel2}\otimes\hat{\textbf{e}}_{\parallel2}^\text{T}
+ \delta_0\hat{\textbf{e}}_{\perp}\otimes\hat{\textbf{e}}_{\perp}^\text{T}$, with diffusivity $\delta_1$ in directions parallel to the wall, and $\delta_0$ orthogonal to the wall; we assume that $\delta_1 \gg \delta_0$. The orthogonal unit vectors $\hat{\textbf{e}}_{\parallel1}(\textbf{r})$, $\hat{\textbf{e}}_{\parallel2}(\textbf{r})$, and $\hat{\textbf{e}}_{\perp}(\textbf{r})$ are the two local tangent and one normal unit vectors to the surface at position $\textbf{r}$ as defined above.
Note that diffusion works in two opposite ways: it transports freshly dehydrated ions into hydrated regions, but it also dilutes dehydrated regions with inflowing hydrated ions. 

%%%%%%%%%%%%%%%  SubSec AMPHI
\subsubsection{Amphiphilic Interactions and the entropic force}
\label{sec:AmphInteractions}
Amphiphilic - i.e. hydrophilic and hydrophobic - interactions constitute a central role in the EAH paradigm and are represented by various terms in equations \ref{eq:EHANS1} and \ref{eq:EHANS2}. Though well understood qualitatively in a chemical and biological context, in a strict physical sense they are poorly described quantitatively \cite{Sun-2017}-\cite{Yoo-2014}. The hydrophobic force is of particular physical interest as it does not result from one of nature's four fundamental interactions, but an 'entropic force', i.e. a statistical-physical effect resulting from the thermodynamic tendency to maximize the system entropy \cite{Sokolov-2010}\cite{NRoos-2014}. 
In the case of the ion channels, amphiphilic interactions take place near the walls, i.e. inner ‘surface’ and in the selectivity filter. 
The strength of unwrapping decreases as one moves away from the surface \cite{Donaldson-2014}. Consequently, there are net hydrophilic gradients directed towards the surface of the channel and axially towards the inlet of the channel.
Eisenberg et al. \cite{Eisenberg-1982} introduced the mean helical hydrophobic moment to quantify amphiphilicity, and Silverman \cite{Silverman-2001} showed the importance of first- and second-order hydrophobic moments \cite{Lin-2007}. \\
The hydrophilic interaction is modeled with a scalar function $\phi_A$ \cite{Lin-2007}. The 'amphiphilic potential' $\phi_A(\textbf{r})$ denotes the ability to de- or re-hydrate an ion at position $\textbf{r}$ \cite{Eisenberg-1982}-\cite{Phoenix-2002}. It relates to the gain or loss in entropy and internal energy per water molecule in ion de/re-hydration.
The energy $\Delta E$ released or required in de- or re-hydration is ion-species dependent \cite{Kopec2018}, such that for species $s$:  $\Delta E = a_s\Delta \phi_A=a_s\nabla\phi_A\cdot\textbf{u}$, where $a_s$ can best be described as the 'amphiphilic charge' of species $s$. $\Delta \phi_A$ is the amphiphilic potential step involved in the de/re-hydration, and relates to the changes in entropy $\Delta S$ and internal energy $\Delta U$. \\
For an isolated amphiphilic source or sink, $\phi_A$ depends on its orthogonal distance $d$ to position $\textbf{r}$. Following Donaldson \cite{Donaldson-2014}, the amphiphilic potential decays exponentially with $d$ as: $\phi_A(d) = C_0.\text{exp}(-d/\lambda_A)$, where $\lambda_A$ is the range of the amphiphilic interaction, and $C_0$ is respectively negative for hydrophilic, and positive for hydrophobic sources. 
In our case there is not one single amphiphilic source or sink, but the entire hydrophilic wall of the ion channel. Therefore, we parameterize the amphiphilic sources/sinks with parameter(s) $\xi$, such that the amphiphilic strength at ligand position $\textbf{x}(\xi)$ is $C_A(\xi)$, and we obtain: $\phi_A(\textbf{r}) = \int C_A(\xi)\text{exp}(-\|\textbf{r}-\textbf{x}(\xi)\|/\lambda_A)d\xi$. Here, parameter $\xi$ describes the spatial parameterization of the 1D-ligand position on the $\alpha$-subunits that constitute the wall of the ion channel. 
Note that we assume that the decay-length $\lambda_A$ is ion-species independent.\\
The amphiphilic potential $\phi_A$ acts as the cause for the amphiphilic force $\textbf{F}_A$, namely as the negative spatial gradient of $\phi_A$. It acts \textit{solely} on \textit{hydrated} ions, so the fraction $\eta$ of the local mixture of molecules, as a \textit{body force} $\textbf{f}_A = \textbf{F}_A/\rho = -\eta\nabla\phi_A$ that features in equation \ref{eq:EHANS1}. Note that the amphiphilic potential $\phi_A(\textbf{r})$ is entirely different from the dehydration source and sink functions $\sigma^{\pm}(\textbf{r})$ used in equation \ref{eq:EHANS2}. We do not include higher-order tensorial terms in the EAH model, such as the amphiphilic moment.   \\

%%%% SS Electric potential
\subsubsection{Electrostatic interactions}
Equation \ref{eq:EHANS3} describes how the electric potential $\phi_E(\textbf{r},t)$ derives from Poisson's formulation of Gauss's law, with the vacuum permittivity  constant $\varepsilon_0$ and local relative permittivity $\varepsilon_r(\bm{r},t)$, and 
mean local electric charge density $\overline{\chi}(\bm{r},t)$. $\lambda_E$ is the time-independent surface charge density of the interior ion channel wall, consisting of the $\alpha$-subunits. As we focus on charged ion interactions, we do not include the relative weak Lennard-Jones potential.\\
The multi-ionic species mixture composes an electrolyte which, due to Debye screening, effectively suppresses the Coulomb potential of an individual ion by an exponential factor. This collective effect results from the combined attraction of local counter-ions and repulsion of neighboring co-ions, leading to a net counter charge density. In the EAH formalism, Debye screening of the electrostatic potential is expressed in equations \ref{eq:EHANS3} and \ref{eq:ensemblechi} by the spatiotemporal fields $\varepsilon_r$  and $\chi$. 
Thus, the mixture of ions constitutes a dielectric medium, with ambient induced charges  screening the electric field. The bound charge density consists of the hydrated ions, polarising the medium. This bound charge acts as a source for the electric field, so the total charge density (per ion species) in Gauss's law reads:  $\chi = \chi_{\text{bound}} + \chi_{\text{free}} = n\eta q_A + n(1-\eta)q_B$. 
Here $q_A$ and $q_B$ are the effective electrical charges of an aqua-ion and a bare ion respectively, and $n$ is the local particle density. This difference quantifies the effect of electrostatic screening on hydrated ions by the envelope dipole water molecules on short range electrostatic interactions \cite{WOHLERT2004} \cite{YANG2013}. The electrostatic potential is calculated concurrently with equations \ref{eq:EHANS1} and \ref{eq:EHANS2} by solving Poisson’s equation in the interior of the channel with Dirichlet boundary conditions.
The electric force follows from $\phi_E$ as: $\textbf{F}_{E}(\textbf{r}) = -\chi\rho\nabla\phi_E$. This gives the body force $\textbf{f}_{E} = \textbf{F}_{E}/\rho = -\chi\nabla\phi_E$ in equation \ref{eq:EHANS1}. 
Magnetic forces are not included as they are magnitudes smaller than the electric and amphiphilic forces in this setting.  
 
%%%%%%%% MIXTURES
\subsubsection{Mixture of Different Species of Ions}
\label{Mixtures}
Let us now consider a mixture of $S$ different species of ions of different mass and charge, viz. $\text{Ca}^{2+}, \text{Na}^+, \text{Cl}^-$, each with a particle density $n_s$. Let $f_s=n_s/n$ with $n=\sum_sn_s$, ($s = 0\dots S$) be the (numerical) fraction of the species of the total mixture. $n_0$ denotes the non-ionic rest of the mixture, notably H$_2$O. Now, the hydration levels of the aqua-ions can be represented by an array $[\eta_1, \eta_2, \dots, \eta_S]$, where $\eta_s$ is the level of hydration of species $s$.
All species in EAH equation \ref{eq:EHANS1} share the same velocity vector field $\textbf{u}$, but each species has its own particular dehydration dynamics as expressed in EAH equation \ref{eq:EHANS2}. It may happen that one species becomes predominantly dehydrated and others not, based on the setting of the EAH-parameters. 
The ensemble averages are given by:
%%%% ENSEMBLE averages EQNS
\begin{align}
&\overline{\eta} = \sum_{s=1}^{S} f_s\eta_s   \label{eq:etamean} \\
&\overline{\nu}(\eta) = \sum_{s=1}^{S} f_s\nu_s(\eta) \label{eq:ensembleviscosity}\\ 
&\overline{\chi}(\eta) = \sum_{s=1}^{S} (f_s\eta_sq_{A,s} + f_s(1-\eta_s)q_{B,s})  \label{eq:ensemblechi}  \\
&\overline{a}(\eta) = \sum_{s=1}^{S} f_sa_s\eta_s   \label{eq:amean} 
\end{align}
% \vspace{5mm}
Note that the density $\rho_s$ per partially hydrated species $s$ depends on $\eta_s$ but that for the {\em entire} ensemble a loss of one aqua-ion is the gain of one bare ion plus its number of released shell water molecules, so the total density $\rho = \sum_sf_s\rho_s$ remains constant. 
The de/re-hydration depends on the different physical properties of the species; its mass $m_s$, charges $q_{A,s}$, $q_{B,s}$ (aqua and bare), amphility $a_s$, and their interaction with the membrane: dehydration speed, and the natural timescale of re-hydration away from the ion-channel, defined by the source and sink functions $\sigma^+(d)$ and $\sigma^-(d)$. 
The species interact indirectly through the dehydration dynamics at the ion chamber walls. The only direct interaction between the different species is via their electrical charge: the ions are subject to the electrical forces of the charged mixture: $n_s(q_{As}\eta_s + q_{Bs}(1-\eta_s))$. Hydrated ions are attracted to the wall, due to the entropic force $-\overline{a}\nabla\phi_A$.
\\
Important for the ion channel dynamics is the effective net ion flux through the channel. This flux follows from the continuity equation: $dn_s/dt = \Phi^{in}_sA^{in}_s-\Phi^{out}_sA^{out}_s + B_s$, with $A^{in/out}$ the surface of the cross section of in/out-let of the ion channel, $\Phi^{in/out}_s$ the associated flux, and $B_s$ the creation/annihilation of species $s$ inside. $B_s$ follows from equation \ref{eq:EHANS2}. In the EAH model, we assume that $\Phi^{out}_s$ is proportional to $n_s$.
Species $s$ that are very efficient in dehydration and with a high outflux jet $\Phi^{out}_s$  will become rare inside the ion channel, depending on their production $B_s$, and their influx $\Phi^{in}_sn_s$, which we assume to be constant for all species. See Fig \ref{fig:ZIMM2011FIGS4B}.

%%%% SUBSEC BD resonances
\subsubsection{Brownian Dynamics Framework }

It seems attractive to augment this approach using the framework of BD, as the stochastic fluctuations on the microscopic scale are immense. Indeed, it is fully justified to insert Langevin terms in the EAH equations \ref{eq:EHANS1}-\ref{eq:PNPEquation}, as we know the physical cause of the fluctuations \cite{Mountain_1977}.
However, from standard theory \cite{Landau_Lifschitz}\cite{Landau_1971} it is known, and confirmed by our simulations, that this impedes and ultimately destroys the formation and evolution of stable oscillations and vortices, even for low levels of noise, due to the non-linearity of the hydrodynamic Navier-Stokes equation \ref{eq:EHANS1}, caused by the kinematical term $(\bm{u}\cdot\nabla)\bm{u}$ \cite{vanKampen-2007}. For this reason, we do not include stochastic fluctuations.

%%%% SubSec C : Particle densities and ionic transport in the EAH framework}
\subsection{Particle densities and ionic transport in the EAH framework}
The EAH equations describe the evolution of the flow velocity field and spatiotemporal hydration. 
In order to analyze the particle density flux and electric current, we augment the classical PNP equations with additional velocity and hydration related terms.

\subsubsection{EAH-augmented particle flux}
In order to find the evolution of the local particle densities $n_s$, we start from the conventional PNP equation \ref{eq:PNPEquation} in terms of the particle flux $\Phi_s$ of ion species $s$, and augment it with terms caused by the fields $\{\textbf{u},\eta\}$. 
This adds an amphiphilic term to $\Phi_s$. Moreover, the charge (and valence $K$) now becomes hydration-dependent through electrostatic screening: $q_s(\eta_s)=\eta_sq_{A,s}+(1-\eta_s)q_{B,s}$, and $K_s(\eta_s)=q_s(\eta_s)/e$.
The augmented flux $ \Phi_s$ of species $s$ thus becomes \footnote{we assume that the contribution of the magnetic vector potential; $\partial\mathbf{A}/\partial t$, is magnitudes smaller (namely 1/{\em c}$^2$) and can be ignored.}:
\begin{equation}
   \Phi_s =  n_s\textbf{u}-D_s(\nabla n_s - \alpha K_sn_s\nabla\phi_E - \beta a_s\eta_sn_s\nabla\phi_A)
\label{eq:AugFluxEqation}
 \end{equation}
where $D_s$ denotes the ion species-dependent local 3$\times$3 anisotropic diffusion tensor.
Here we postulate the same diffusion tensor $D_s$ for particle and hydration transport, i.e. equations \ref{eq:EHANS2}  and \ref{eq:AugFluxEqation}.
The constants $\alpha$=$e/k_BT$ and $\beta$ = $1/k_BT$ gauge the electric and amphiphilic potentials to the density flux. 
The amphiphilic charge $a_s$ was introduced in Section \ref{sec:AmphInteractions}.
Note that the convective derivative term $\textbf{u}\cdot\nabla n_s$ in the left-hand side (LHS) of equation \ref{eq:PNPEquation} cancels, due to the assumption of an incompressible fluid: $\nabla\cdot\textbf{u}=0$. 
The modified PNP-flux $\Phi_s$ now explicitly contains the fields $\textbf{u}$ and $\eta$, and thus relates to the geometry of the ion channel. 
This augmented PNP equation is processed simultaneously with the EAH equations \ref{eq:EHANS1}-\ref{eq:PNPEquation} and gives the concurrent evolution of the spatiotemporal concentrations $n_s(\textbf{r},t)$ and fields $\textbf{u}(\textbf{r},t)$ and $\eta_s(\textbf{r},t)$.
This can be employed, a.o., to determine electrical currents across the channel, and express the particle flux of dehydrated ions  $\textbf{j}_s = (1-\eta_s)n_s\textbf{u}$.

%%%% SubSec augmented GHK equation and ionic permeability
\subsubsection{The augmented GHK equation and ionic permeability }
The (absolute) permeability $p_s$ of an ion species $s$ is a measure for the degree to which ion $s$ is permeant through a given ion channel.
The  permeability is experimentally determined by determining the reversal potential, a.k.a. Nernst potential, $V_{N}$ across the cell membrane. $V_{N}$ is described by the Goldman-Hodgkin-Katz (GHK) voltage equation that defines the cross-channel potential $V$ that makes the total outward current $I_{tot}$ zero.
The {\em relative} permeability of a mixture of different ionic species is the ratio of their absolute permeabilities \cite{NAYL2016}. \\
It must be noted that the question has been raised as to the reliability of reversal potential measurements for determining permeability ratios, particularly given the use of an equation such as the GHK equation, which is often used to calculate such ratios \cite{BARRY2006}.
\\
In order to find $V_N$ for our model, we consider a mixture of $S$ ionic species where $n_s$ is the particle density and $K_s$ the valence of species $s$. Also, assume that the diffusion constant differs per species and has a value $D_s$ for species $s$. 
For a given rotational-symmetric ion channel, the particle flux $\Phi_s$ of species $s$  along the central $z$ axis is given by combining the Stokes-Einstein equation with Fick's law of diffusion:
$$
\Phi_s(z) =\alpha K_sp_sl\frac{dV_E}{dz}n_s(z) - p_sl\frac{dn_s(z)}{dz}
$$
with constant $\alpha = F/RT$ and the ionic permeability $p_s \equiv D_s/l$. $V_E(z)$ is the electric potential inside the channel, $l$=2$L$ its length, $T$ is the absolute temperature, $R$ the molar gas constant, and $F$ Faraday's constant. \\
This can be solved as a differential equation in $n_s(z)$ for assumed constant $\Phi_s$, with boundary conditions at $z$=$-L$ (inside the cell): $n_s$(-$L$)=$n_s^{i}$, and $z$=$L$ (outside): $n_s(L)$=$n_s^{o}$, yielding the 'classical' Goldman equation:
\begin{equation}
\Phi_s = \alpha K_sp_sV_E \frac{n_s^{o}-n_s^{i}e^{\alpha K_sV_E}}{1-e^{\alpha K_sV_E}}
\label{ClassGoldmanEq}
\end{equation}
where $V_E$ now denotes the entire cross-channel electric potential. 
In the mixture of $S$ ionic species, each with specific electrical charge $q_s$=$K_s$e, the total outward electric flux $I_{\text{tot}}$ as function of the electric potential difference $V$, a.k.a. the Goldman curve, becomes:
$$
I_{\text{tot}}(V) = \sum_{s=1}^Sq_s\Phi_s
$$
The Nernst potential $V_N$ can then be found by solving: 
\begin{equation}
   I_{\text{tot}}(V_N) = 0
\label{eq:GHKEqation}
 \end{equation}
in terms of the electric potential $V$. This defines the GHK equation. \\
In the framework of EAH we also involve the degree of hydration $\eta$, the amphiphilic potential $V_A$, and the (axial) velocity field $u$, so that equation \ref{eq:AugFluxEqation} for the 1D-axial particle flux becomes (per ion species):
$$
\Phi(z) = - pl\frac{dn}{dz} + \alpha Kpn\frac{dV_E}{dz} + \beta ap\eta n\frac{dV_A}{dz} + nu
$$
Where $V_A$ indicates the average amphiphilic potential \footnote{$V_A(Z)$ is the amphiphilic potential averaged over the vertical plane $z=Z$ in cylindrical coordinates.} at axial position $z$ and $nu$ is the convective particle flux. $a$ is the amphiphilic charge of the ion.
In a similar derivation as above, this leads to an augmented Goldman equation:
\begin{equation}
 \Phi = \frac{f^o-f^iQ}{1-Q}
\label{eq:AugGoldmanEq}
\end{equation}
with: $f^o$=$pn^o(\alpha K^oV_E + \beta a\eta^oV_A + u^o)$, $f^i$= $pn^i(\alpha K^iV_E + \beta a\eta^iV_A + u^i)$, $Q$=exp$(\alpha K^iV_E + \beta a\eta^iV_A + u^i)$, and '$o$' refers to extracellular entry of the channel (\textbf{o}utlet), and '$i$' to its intracellular exit (\textbf{i}nlet), so $K^i = K(\eta^i)$, etc. Here, $V_E$ and $V_A$ are, respectively, the full electric and amphiphilic potential across the entire channel: $V_{\{E,A\}} = V_{\{E,A\}}(L)-V_{\{E,A\}}(-L)$. In our approach, we assume full hydration at the entry, so: $\eta^o=1$. In the ideal case, the hydration is zero at the exit, $\eta^i=0$. In practical cases, we can define the efficiency of hydration of the channel as: $\epsilon\equiv 1-\eta^i/\eta^o$.  
The total current $I_{tot}$ for the mixture of $S$ ion species entering the cell becomes:
$$
I_{tot}(V_E,V_A) = \sum_{s=1}^Sq_s^{in}\Phi_s
$$
now with {\em screened} charges $q_s^{in} = \eta_s^{in}q_{A,s}+(1-\eta_s^{in})q_{B,s}$. 
The augmented Nernst potential $V_N$ again follows from solving the GHK-equation \ref{eq:GHKEqation} in terms of $V_E$.  \\
Note that including the hydration shifts the Nernst-potential (and the isolated singularity for resp. $V_E$=0 or $Q$=1) relative to the 'classical' case.

%%%%%%%%%%%% subsec:  EAH framework in different Spatial Dimensions %%%%%%%%%%%
\subsection{The EAH framework in different Spatial Dimensions}
Though the EAH-equations \ref{eq:EHANS1}-\ref{eq:PNPEquation} are expressed dimension-free, their physical realization and effects vary in different spatial dimensions. Here, we separately describe 1D-, 2D-, and 3D-systems.

\subsubsection{1D-EAH Systems}
\label{EAH1Dmodeling}
The narrow non-curved SFs observed in numerous ion channel architectures, viz. KcsA, are essentially thin tubes of only one spatial dimension; the axial coordinate $z$ \cite{PDB-KcsA}\cite{REF1SUG7}. 
In case of a pure 1D-system, the incomprehensibility of the fluid, $du_z/dz = 0$, dictates that the axial velocity $u_z$ must be constant along the axis $z$. 
Fluctuations in $u_z$ cause density waves that propagate with the velocity of sound $c = \sqrt{\partial P/\partial \rho}$, which is higher than the flow speed $u_z$. 
A more realistic approach is to model narrow nanoscale tubes as cylinders with small diameter, and only study the axial dynamics. This means that we only take the $z$-component of the Navier-Stokes equation \ref{eq:EHANS1}.
Note that the total dimension of the EAH-space is $D_{EAH}=2S+2$, namely describing $S$ different ion species, the flow velocity $u(z)$, and the electric field $\phi_E(z)$. 
As  $D_{EAH}>3$, the system may still exhibit complex dynamics, despite its D=1 spatial dimension \footnote{In continuous dynamical systems, the Poincar\'{e}-Bendixson theorem shows that chaotic behaviour can only arise in three or more dimensions. See: EA Coddington, N Levinson, {\em The Poincar\'{e}-Bendixson Theory of Two-Dimensional Autonomous Systems}, McGraw-Hill, pp. 389-403 (1955)}. 

\subsubsection{2D-EAH Systems}
The dynamics in two spatial dimensions exhibits the emergence of vortices and eddies in the flow in the fluid, as is well-known experimentally observed and mathematically described, e.g. in the Navier-Stokes equation.
The complexity of such flows is adequately studied and analyzed, e.g. in the seminal work of Landau and Lifschitz \cite{Landau_Lifschitz}\cite{Landau_1971}. 
It must be noted that in rotation-symmetric systems, the vortices can form 3D co-axial toroids, which, however, are azimuthally segmented \footnote{I.e. in direction of the cylindrical $\theta$-coordinate.} and strictly not-connected.
As such, they do not allow the different 2D-structures to interact in the azimuthal dimension or exchange energy and information (viz. entropy). 
In a real physical system, embedded in ambient noise, symmetry breaking will swiftly connect these cycles into left- or right-handed rotating toroidal vortices, subject to the physical conservation laws, notably angular momentum and energy. 
The rotational energy in these structures can be converted into internal energy, viz. heat and chemical energy, and thus provides a powerful source for dehydration.

%%%%% SS Rotating ion channel
\subsubsection{3D-EAH Systems and rotations in the flow}
\label{sec:Rot3DEAH}
In 3D-systems the ensuing dynamics becomes more complicated if rotation is involved. Rotation can be caused by residual angular momentum in the incoming fluid, internal helically skewed walls, or by rotation of the ion channel itself. 
If the ion channel is fully rotation-symmetric, the mathematical description can be reduced to 2D, assuming that the angular momentum is conserved. In this case, rather than the isolated not-connected structures in 2D, rotational transport is around the symmetry axis and the vortices and eddies are full 3D-objects. \\
Rotation introduces new terms in the EAH-equations, 
namely by including bulk centrifugal and Coriolis forces exerted on the flow. 
Thus, with angular velocity vector $\bm{\Omega}$ this augments the EAH main equation \ref{eq:EHANS1} with two rotational kinematic terms:
$$ 
\tfrac{D\textbf{u}}{Dt} = \overline{\nu}{\nabla^2}\textbf{u}-\nabla P/\rho-\overline{\chi}\nabla \phi_E- \overline{\eta}\nabla \phi_A 
-  \bf{\Omega}\times(\bf{\Omega}\times\bf{r})-\text{2}\bf{\Omega}\times\textbf{u} 
$$
Introduction of these terms result in strong vortices that spin fast around the central symmetry $z$-axis.
The rotational forces only apply {\em inside} the ionchannel. They are proportional to the mass density. An individual aqua-ion has a higher mass than a bare ion, due to its shell of $n_H $water molecules. 
This difference does, however, not cause centrifugal separation of aqua and bare ions, as the contribution of the rotational forces is mass-dependent, but as mass acts on both LHS and RHS of Newton’s second law - the microscopic basis of the NSE - it cancels exactly.  
However, the combination of hydration-dependent viscosity and hydration-independent rotational forces will cause centrifugal separation of ions causing more hydrated ions farther away from the rotational axis, pushing hydrated ions to the amphiphilic walls of the ion channel \footnote{The kinematic viscosity $\nu(\overline{\eta})$ relates to the molecular mass. The Stokes-Einstein equation \cite{Einstein-1905} relates kinematic viscosity to the microscopic diffusion coefficient $D$ as: $\nu$ $\propto$ $1/\text{D}$, while $D$ itself is mass dependent: $D$ $\propto$ $({\sum}_s{M_s^{-1}})^{1/2}$ \cite{Fuller-1966}, with $M_s$ the molecular mass of species $s$ in the mixture ($s=1 \dots N_S$). This mass depends on the fractions $n_s$ and $\eta_s$ of hydrated molecules for species $s$: $M_s(\eta_s) = M_{s,0} + \eta_s n_{H,s}M_{H_2O}$. Here $M_{s,0}$ is the mass of the bare ion of species $s$, $n_{H,s}$ its maximum aqua-shell number, and $M_{H_2O}$ is the mass of a water molecule.}. \\
An interesting new hypothesis for understanding ion transport in ion channels is introduced by R. Shaw with the concept of spinning ion channels \cite{Shaw-2013}. According to this idea, ion channels in the open state spin with ultrahigh frequencies of 10 -– 100 kHz.  
The EAH equations, augmented with above rotational terms, can adequately describe this paradigm, and they result in powerful turbulent phenomena. However, even without the rotational terms, i.e. $\bf{\Omega}$=$\bf{0}$, turbulence appears for high enough Reynolds number $Re$, and in this way the EAH model does not depend on external rotation and the spinning ion channel paradigm.

%%%  \subsubsection{Justification of the EAH framework}
\subsubsection{Physical markers for the existence of micro-turbulence in EAH dynamics}
\label{ContParams}

In the EAH framework, the fields $\textbf{u}$ and $\eta$ define self-organized oscillations that for high Reynolds numbers can be very small \cite{Schlichting2017}.
Here we quantify the lowest admissible length-scales for our case, the {\em Debye length} for the range of the electrostatic field, and the {\em Kolmogorov length} for the smallest hydrodynamic length scale in a  turbulent flow.
More detailed calculations in Appendix \ref{Appendix1}. 
\begin{itemize}
    \item     %(i) The Debye length 
    The Debye length $\lambda_D$ measures how far the electrostatic effect of a charge in an electrolytic solution or plasma persists. In electrohydrodynamics, it measures the lower bound where the theory applies. For our context inside an ion channel, we find a Debye length of $\lambda_D \approx 0.34$\AA. This is less than the grid-size of the EAH model of approximately 1\AA, so allows for the level of detail exhibited. 
    Note that this is substantial smaller than the values of Corry et al. \cite{REF1SUG1} of 5.6 - 7.9 {\AA}.
    \item  %%(ii) The Kolmogorov length
    The Kolmogorov length $\kappa$ is the smallest hydrodynamic length scale in turbulent flows.  The Kolmogorov length implicitly defines Re, namely as the scale at which the Re becomes equal to 1, so the onset of turbulence. For our context of EAH inside an ion channel, we find $\kappa \approx 0.025$\AA. This is substantially less to the EAH grid-size of 1\AA.
\end{itemize}
Both markers indicate that the ionchannel environment is in the appropriate turbulent regime, and thus support the validity  of the macroscopic mean-field EAH framework for exhibiting meaningful micro-vortices in the streamlines at scales of approximately 1{\AA}.

%%%%%%%%%%%%%%%%%%%%%%%%%%%%%%%%%%%%%%%%%%%%%%%%%%%%%%%%%%%%%%%%
%%%%%%%%%%%%%%%%%%%%%%%%%%%%%%%%%%%%%%%%%%%%%%%%%%%%%%%%%%%%%%%%
%%%%%%%%%% SECTION NUMERICAL EXPERIMENTS and Observed Phenomena 
%%%%%%%%%%%%%%%%%%%%%%%%%%%%%%%%%%%%%%%%%%%%%%%%%%%%%%%%%%%%%%%%%%%%
%%%%%%%%%%%%%%%%%%%%%%%%%%%%%%%%%%%%%%%%%%%%%%%%%%%%%%%%%%%%%%%%
\section{Numerical experiments and Comparison with Empirical Results } 
\label{NumExp}

\subsection{Computational model of the ion channel}
Our interest is in the underlying mechanism of self-organized selectivity of the ion channel, and therefore we  focus mainly on the relation between turbulence formation and ion dehydration, especially near the interior walls \cite{Stephens-2015}. In our simulations we represent the ionchannel in cylindrical coordinates $(r,z,\theta)$, as we assume that it is axially symmetric. The axial coordinate $z$ is aligned with the symmetry axis of the ion channel. The extracellular entrance of the ion channel is positioned at $z = L$, and the intracellular exit at the origin $z=-L$, where $l$=2$L$ is the length of the ion channel. Consequently, the flow through the channel has a {\em negative} sign.
\\
For our physical model we used published data on the crystal structure of ion channels, notably represented as the {\em pore radius}. The pore radius also indicates whether the structure represents an open or a closed state of the ion channel. This provides data for the EAH model for $z$ as function of $r$ in cylindrical coordinates. 
As prototypes we selected published data on x-ray crystal structures of prokaryotic homologs of ligand-gated ion channels from {\em Erwinia chrysanthemi} (ELIC) \cite{PDB-ELIC}, and three structures of a proton-activated channel from {\em Gloebacter violaceus} (GLIC1, GLIC2, and GLIC1M) \cite{PDB–GLIC} from Song and Corry \cite{Song-2010}. 
ELIC, depicted in Fig \ref{fig: ELIClateralview}, exhibits a minimum pore radius of 1.2 \AA, located near the extracellular side of the membrane, which implies that it is a closed state structure.  
A schematic representation of the GLIC ion chanel in the open state (GLIC1) and closed state (GLIC2) is depicted in Fig \ref{fig: GLIC12}, excluding the subunit appendices and the protein $\alpha$ subunits envelope. GLIC1M is the E221A mutation of GLIC1, and it mirrors the open state GLIC1.
For studying the essentially 1D transport of a single file of ions in a narrow SF, we selected the x-ray crystal structure of the prokaryotic KcsA ion channel \cite{PDB-KcsA}\cite{LIND2013}. \\
Besides the mathematical conditions and assumptions mentioned in Section \ref{SectionModelAssumptions}, we make the following assumptions on, and limitations to, the geometry and architecture of the computational model of the ion channel. The most relevant are that we  assume:
\begin{enumerate}
\item A mean-field continuous model in space and time, describing the statistical average dynamics at scales above the quantum level. 
The benefits and disadvantages of macroscopic models are discussed in Sections \ref{subsecContModels}, and its validity is justified in Section \ref{ContParams}. The required ergodicity is discussed in Section \ref{ErgodBreaking}.
\item Cylindrical symmetry along the central axis of the ion channel, and therefore apply cylindrical coordinates $z$ and $r$, reducing the computational complexity to 2D by ignoring $\theta$-dependence. 
\item  The linear charge density and amphiphlic coding on the $\alpha$-subunits to be constant, independent of the ligand position. Thus they make that $\phi_A$, and the fraction of $\phi_E$ caused by the contribution of linear/surface charge density $\lambda_E$, are time-independent and only depend on the orthogonal distance $d$ in equations \ref{eq:EHANS1}-\ref{eq:PNPEquation}. 
\item  We assume a constant relative permitivity, and following \cite{REF1SUG1} take $\varepsilon_r = 80$ inside the boundary water, and 2 outside, which is representative
of proteins forming ion channels. 
\item We assume that the ion channel walls are non-solid, and that colliding ions bounce off fully elastically.
\item We generally not assume rotating ion channels. Nor do we take into account the intrinsic ion spin, and we average the ion-ion interaction terms as a mean field, not including the Lennard-Jones potential.
\end{enumerate}

%%%%%%%%%%%% SubSec Implementation
\subsection{Implementation and experimental design}

The EAH equations are implemented in MATLAB, using the open-source Linear Algebra-based NSE algorithms of B. Seibold \cite{Seibold-2008}. 
This efficiently iterates over time the spatially discretized and then vectorized representations of the field variables $\bf{u}$, $\eta$, $\phi_E$, and $P$, all in column vector shape. In the same way, the operators that act on these field quantities are discretized, with appropriate boundary conditions, in corresponding system matrices. \\
The pressure is only given implicitly. It is obtained by solving a linear system using sparse Cholesky decomposition to compute the pressure correction. In the same way, the Poisson equations for the pressure $P$ and electric field $\phi_E$ are solved as linear systems, together with their prescribed homogeneous Neumann boundary conditions and with no-slip boundary conditions assumed for the velocity field $\bf{u}$. 
The incompressibility condition $\nabla\cdot{\bf u}=0$ is treated, not as a time evolution equation, but as an algebraic condition by using a projection approach by G. Strang \cite{GStrang-2008}. By iterating the momentum equation \ref{eq:EHANS1} while ignoring the pressure, it is projected onto the subspace of divergence-free velocity fields. \\
The viscosity terms are treated implicitly and expressed in terms of the dimensionless Reynolds-number Re $\propto$ $1/\nu$. \\
The results are visualized as graphs of {\em streamlines}: the trajectories of the $\textbf{u}(\textbf{r})$-field, or {\em flowlines}: the trajectories of the unit vector  $\hat{\textbf{u}}(\textbf{r})$-field. Flowlines exhibit vortices more clearly, see Fig \ref{fig: GLIC1Misobar}(a). Gray scales (online color) are employed to indicate the hydration field $\eta(\textbf{r})$ or pressure field $P(\textbf{r})$.

%%%%%%%%%%%%  SUBSECTION on 1D channels notably IKs 
\subsection{Observed phenomena in 1D-EAH dynamics }
\label{sec:SEC1DEAHdyn}

\begin{figure}
\begin{center}
\includegraphics[width=0.46\columnwidth]{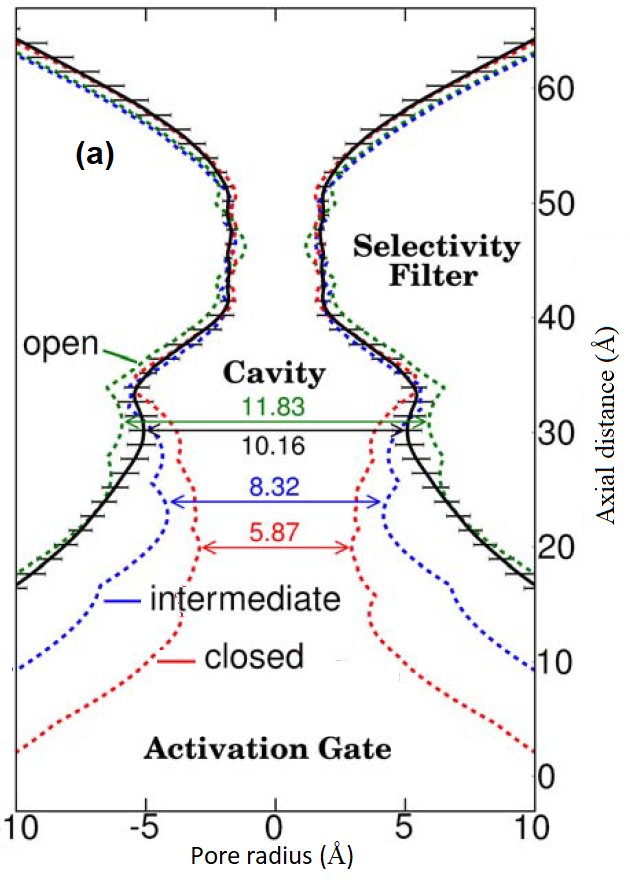}
\includegraphics[width=0.49\columnwidth]{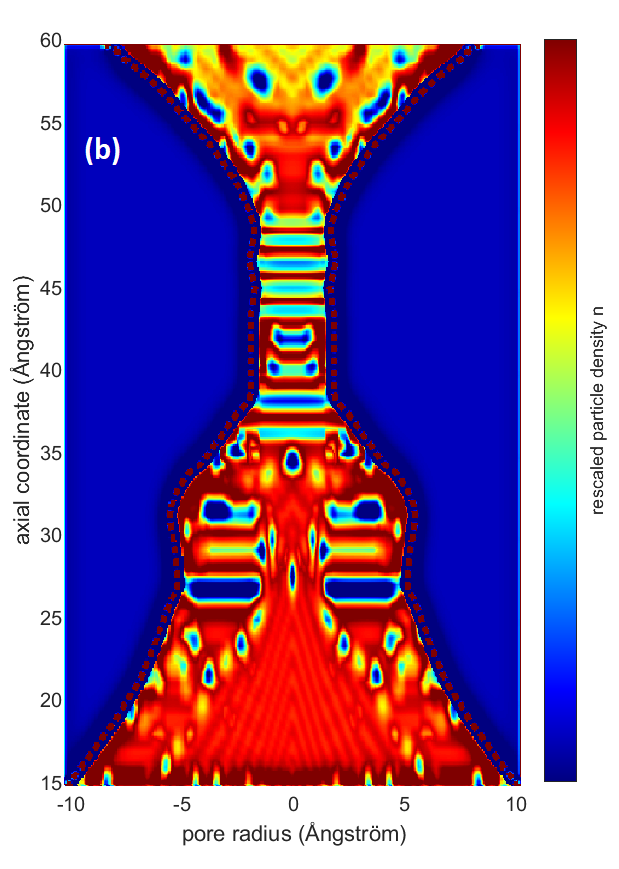}
\caption{Pore radius profile of the KcsA ion channel, derived from backbone atoms of channel states. (a): Comparison of the profiles (dashed lines) formed by the {\em closed}, {\em intermediate}, and {\em open} crystal structures. From Fig 2B in: T. Linder, B.L. de Groot, and A. Stary-Weinzinger \cite{LIND2013}, reproduced by courtesy of the authors.
(b): Particle density in KcsA after 150 msec reveals the formation of a train of oscillating density waves in the Selectivity Filter with wavelength $\lambda\approx$ 3.1 {\AA}. }
\label{fig:KcsAporeradius}
\end{center}
\end{figure}

First, we study the dynamics generated by the EAH-model in a 1D-system in order to understand its implications for the narrow tube-like SFs found in various ion channels, like the bacterial KcsA depicted in Fig \ref{fig:KcsAporeradius}.
Our starting point are the EAH equations restricted to D=1 spatial dimension.
As described in Section \ref{EAH1Dmodeling}, this means the axial coordinate $z$ of an open cylinder of dimensions $L$ and $R$, with entry at $z=0$ and exit at $z=L$. To fit this with the SF of KcsA we choose $L=12$\AA, and $R=2$\AA - as estimated from Fig \ref{fig:KcsAporeradius}(a). 
We observe that, under the right conditions, this gives rise to oscillating pressure waves, fitting to the the length of the tube, that can act as driving force for a 1D-DDHO mechanism.

%%%%%%%%%%%%%%%%%%%%%%%% FIG Bacterial Channel KcsA
%%%%%%%%%%%%%%%%%%%%%%%%%%%%%%%%%%%%%%

%%%%%%%%%%%%%%%%%  SubSec: How a 1D-EAH acts driver of Selection
\subsubsection{1D-EAH model}
The equations for the 1D-EAH follow directly from equations  \ref{eq:EHANS1}-\ref{eq:PNPEquation} by only considering the axial coordinate $z$ of a cylinder with small diameter. 
As the equations for the hydration $\eta$ and particle density $n$ of the $N_S$ ion species are not coupled, we can for simplicity just consider {\em one} single species with electric charges $\{q_A,q_B\}$ and amphiphilic charge $a$. 
Moreover, we assume a constant amphiphilic force along the channel, represented by its potential gradient $g^A$, and constant particle density $n$.
With these simplifications, the 1D-spatial EAH equations represent a 3D-dynamical system $\{u,\eta,\phi_E\}$:
\begin{align}
& \partial_t u+u\partial_{z}u=\nu\partial^2_{zz}u - g^{PE} - g^{EA}\eta  \label{eq:EAH1Du} \\ 
&\partial_t\eta+u\partial_z\eta = D_s\partial^2_{zz}\eta+\sigma_s^+-\sigma_s\eta_s \label{eq:EAH1Deta} \\
&\varepsilon\partial^2_{zz}\phi_{E} = -(\lambda_E + \eta nq_A+(1-\eta)nq_B) \label{eq:EAH1DphiE} 
\end{align}
with the combined electrostatic-pressure gradient body force $g^{PE}=g^P +q_{B}g^E$ and ditto electrostatic-amphiphilic body force $g^{EA}=(q_{A}-q_{B})g^E + ag^A$, with electrostatic gradient $g^E=-\partial_z\phi_E$.  
Note that the sink/source $\sigma$-functions still vary over $z$. Therefore, the equilibrium value $\eta_{EQ1}=\sigma^+/\sigma$ also varies over $z$.
On the other hand, a constant speed $u$ is only feasible if $\eta_{EQ2}=-g^{PE}/g^{EA}$. 
This is only the case if $\eta_{EQ1}=\eta_{EQ2}$, and the pressure gradient settles to a value  $g^{P}=-q_{B}g^E+ ((q_{A}-q_{B})g^E + ag^A)(\sigma^+/\sigma)$.

%%%%%%%%%%%%%%%%  FIG Feigenbaum bifurcation diagram
\begin{figure}
\begin{center}
\includegraphics[width=0.47\columnwidth]{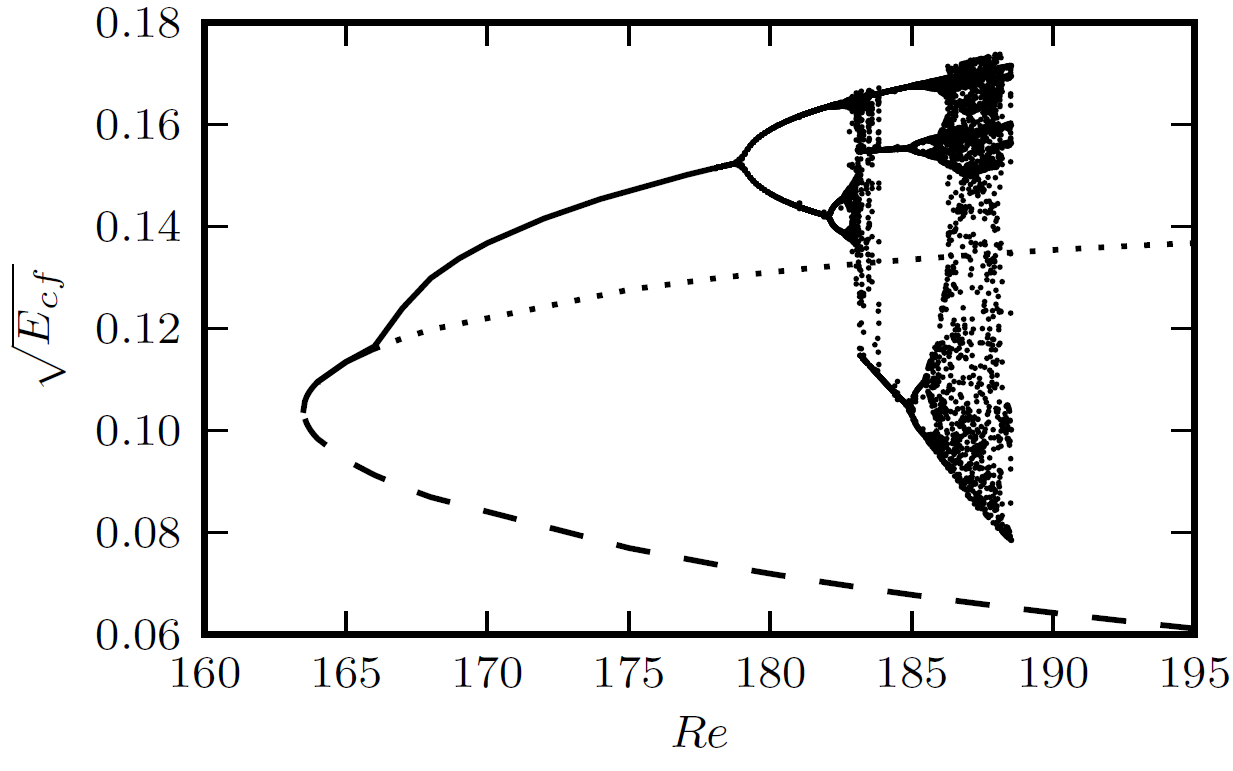}
\caption{Bifurcation diagram for the flow energy in a two-sided open pipe for increasing Re that exhibits period doubling cascade between a Hopf bifurcation at Re = 166 till a boundary crisis at Re =
188, from: Tobias Kreilos and Bruno Eckardt, {\em Periodic orbits near onset of chaos in plane flow}, Chaos \textbf{22}:047505 (2012) \cite{KREIL2012}, reproduced by courtesy of the authors with the permission of AIP Publishing.}
\label{fig:FeigenbaumPipe}
\end{center}
\end{figure}
%%%%%%%%%%%%%%%%%%%%%%%%%%%%%%%%%%%%%

\subsubsection{Oscillating particle-density waves caused by period-doubling bifurcations}
\label{sec:standing1Ddensitywaves}

\begin{figure*} 
 \begin{center}
\includegraphics[width=0.48\textwidth]{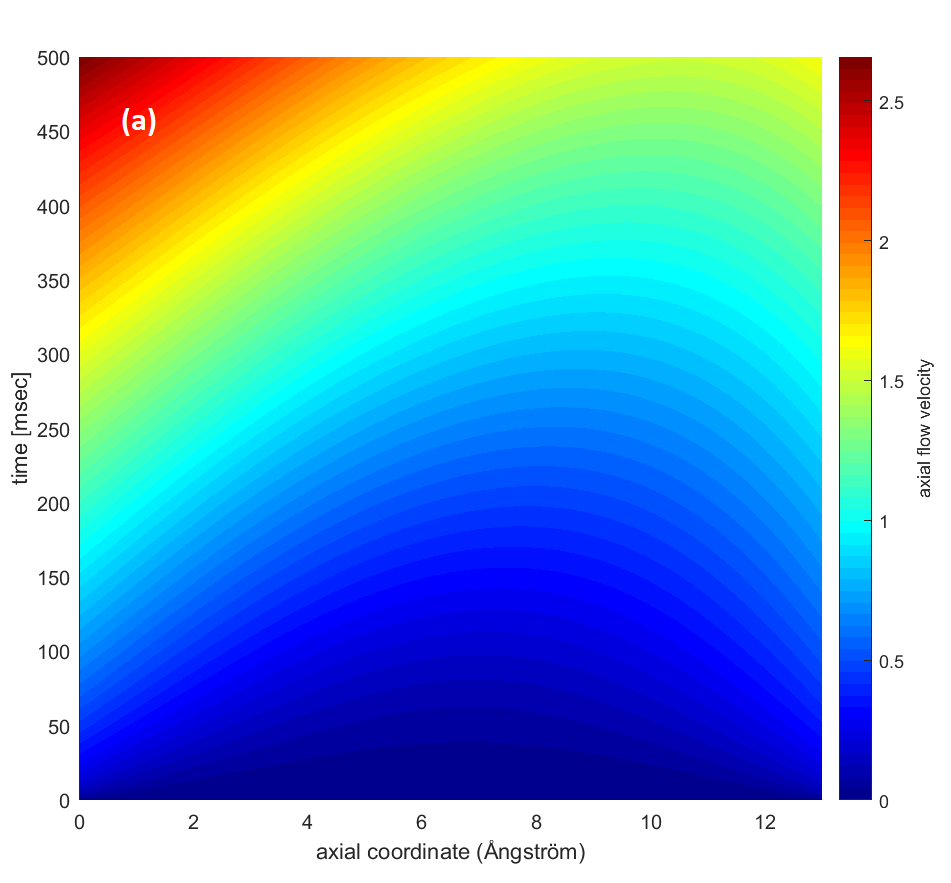}
\includegraphics[width=0.48\textwidth]{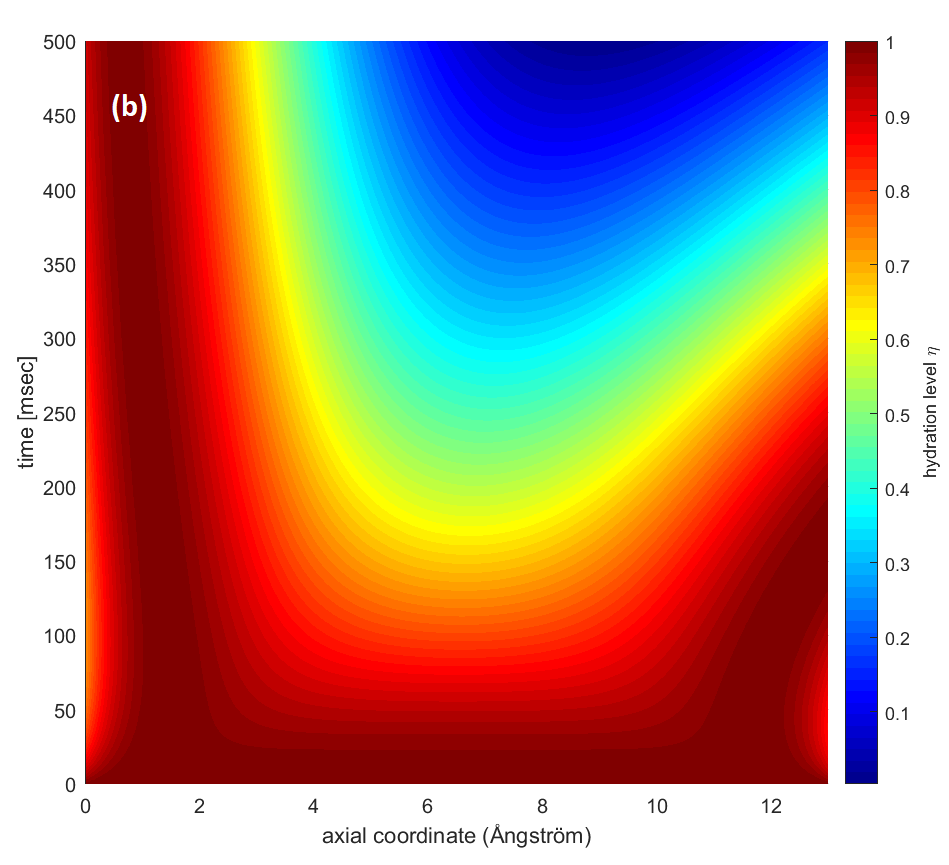}
\caption{Solutions of the 1D-EAH-model inside the SF of KcsA ion channel with low Re = 0.44, (a): Solution for axial velocity $u(z,t)$. (b): Solution for hydration $\eta(z,t)$. Coordinate $z$ denotes the axial coordinate along the SF of KCsA, from its entrance set at $z=0$ to its exit at $z\approx 12$\AA. }
\label{fig: EAH1DlowRe}
\end{center}
\end{figure*}

\begin{figure*} 
 \begin{center}
\includegraphics[width=0.48\textwidth]{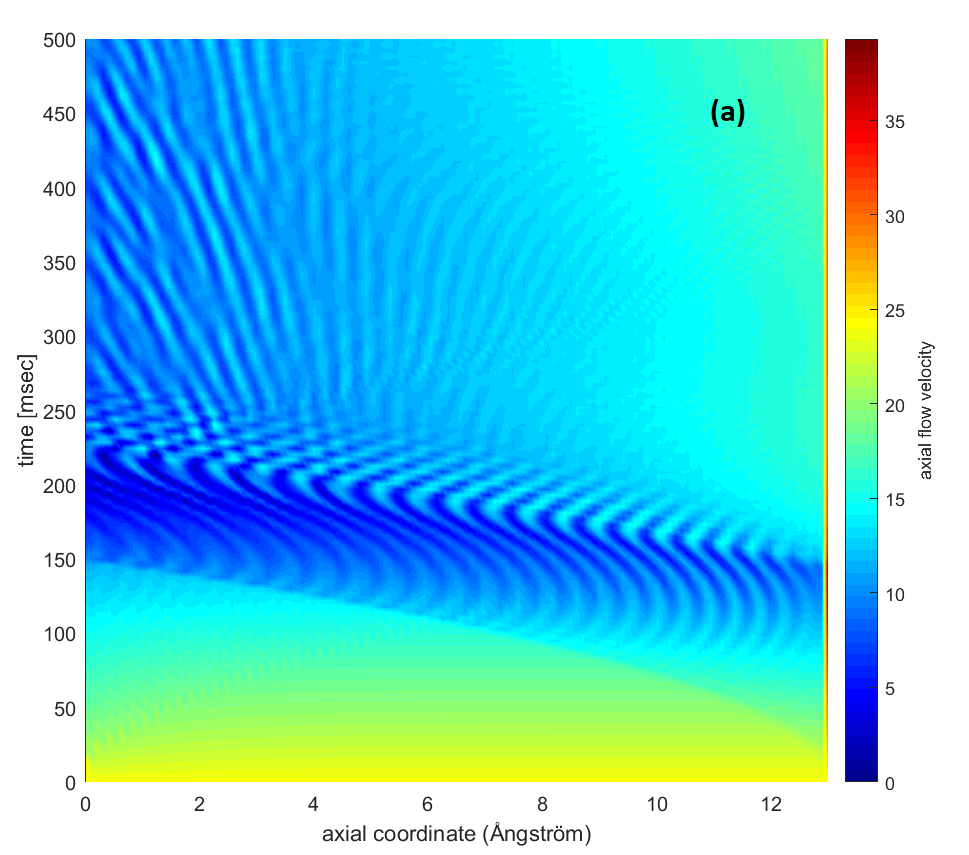}
\includegraphics[width=0.48\textwidth]{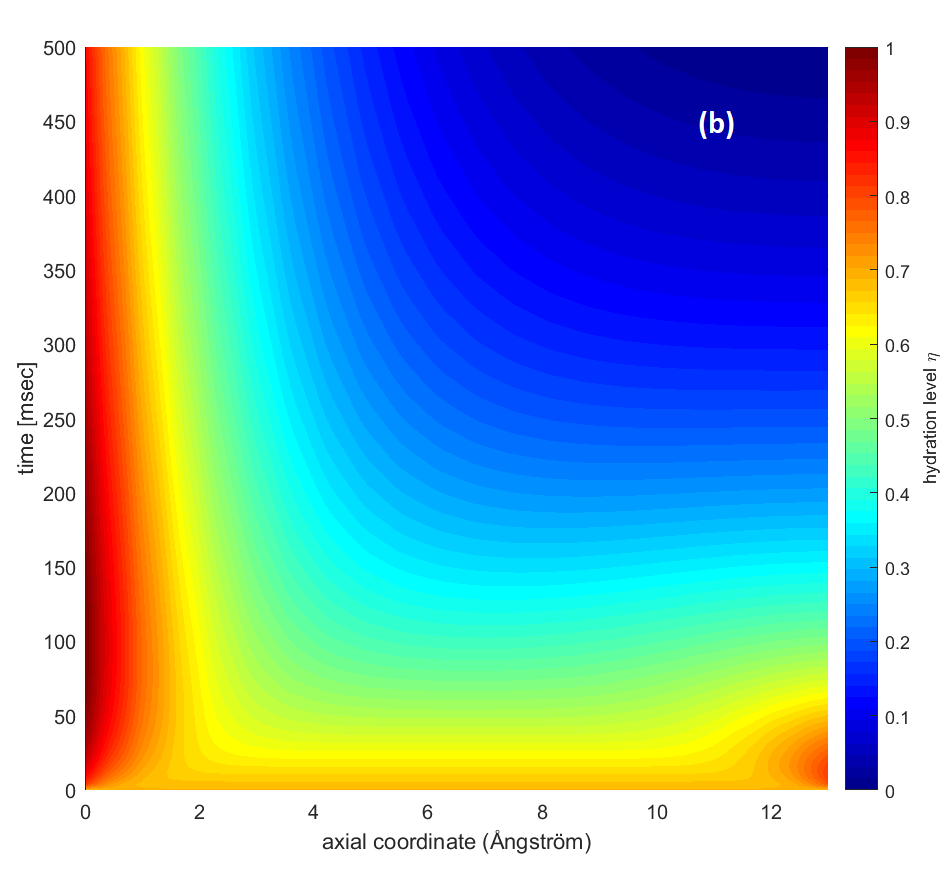}
\caption{Solutions of the 1D-EAH-model inside the SF of KcsA ion channel with high Re = 1E4 reveals the formation of oscillating density waves with short wavelength (a): Solution for axial velocity $u(z,t)$. (b): Solution for hydration $\eta(z,t)$. }
\label{fig: EAH1DhighRe}
\end{center}
\end{figure*}

\begin{figure*} 
 \begin{center}
\includegraphics[width=0.48\textwidth]{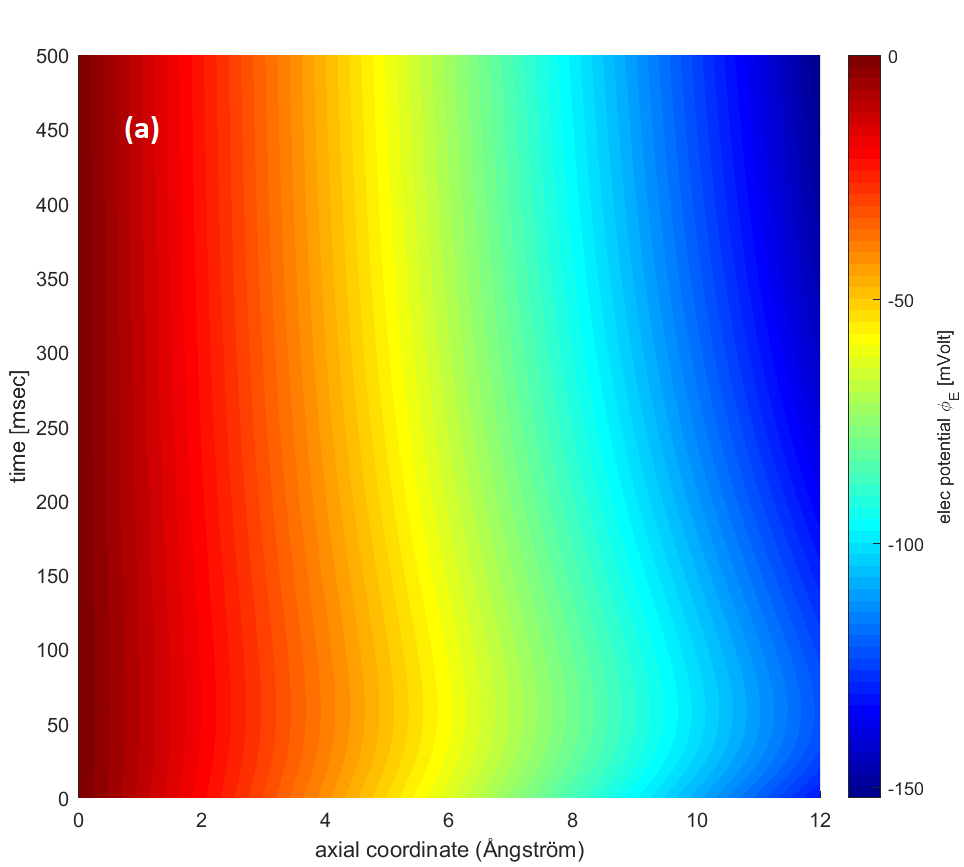}
\includegraphics[width=0.48\textwidth]{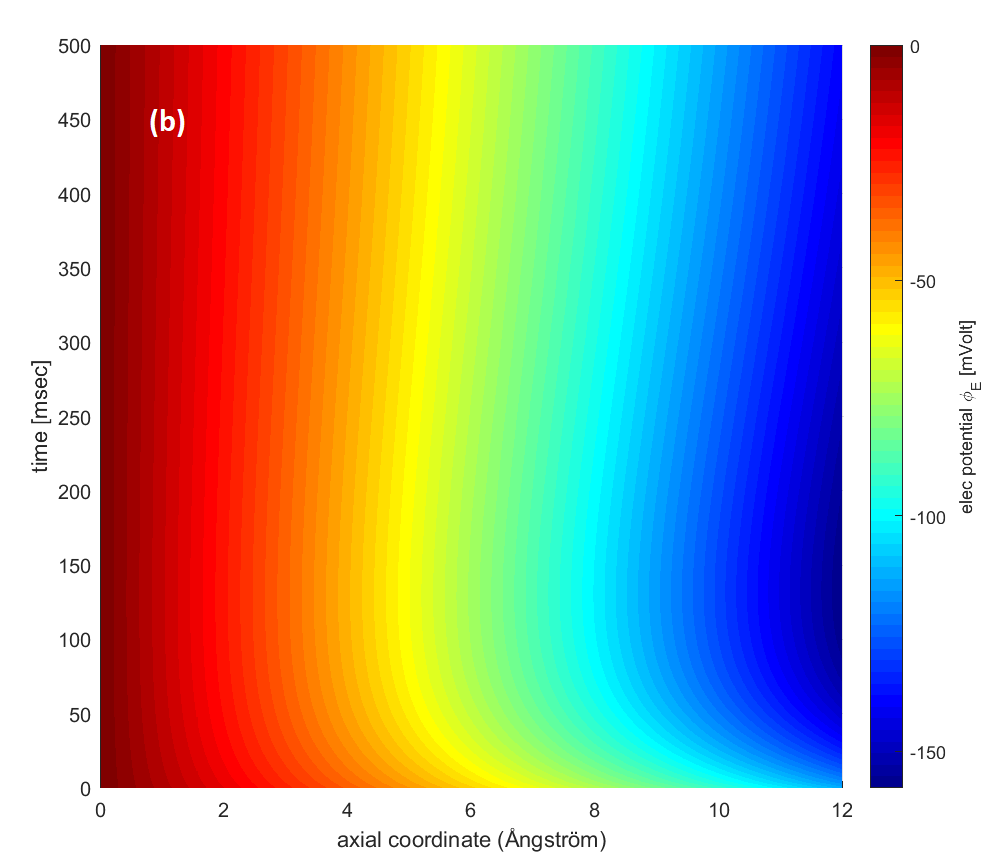}
\caption{Solutions of the electric potential $\phi_E(z,t)$ of the 1D-EAH-model inside the SF of the KcsA ion channel, starting from: $\phi_E(0)=0$ and: $\partial_z\phi_E(0)=-5.6453$ [V/m]. (a): Low $Re$=0.44. (b): High $Re$=1E04.}
\label{fig: EAH1DphiE}
\end{center}
\end{figure*}

For suitable high values of $Re=\nu^{-1}$, the particle density of the 1D-EAH system exhibits persistent oscillating waves that strongly enhance ion dehydration.
The origin of these oscillations lies in complexity theory and follows Landau's description of the transition to turbulence \cite{Eckhardt2005}. 
It is well-known in fluid mechanics that pressure-driven flow in a pipe becomes turbulent in a subsequent cascade of instabilities that results in ever more complicated dynamics, see Fig \ref{fig:FeigenbaumPipe}.
This bifurcation-diagram of the flow energy versus $Re$ in a two-sided open pipe exhibits a periodic doubling $2^k$ cascade that ends with a crisis bifurcation \cite{KREIL2012}.
In our case, this periodic doubling cascade is the cause for the oscillations of the particle density and flow velocity. 
Numerical analysis of the 1D-EAH equations \ref{eq:EAH1Du}-\ref{eq:EAH1DphiE} with realistic initial conditions for different values of $Re$, shows the set of bifurcations and the transition to chaos. The most interesting for us are the stable oscillatory $2^k$ cycles {\em en route} to the chaotic regime.   
Examples are depicted in Figures \ref{fig: EAH1DlowRe} and \ref{fig: EAH1DhighRe} that show the concurrent evolution of flow velocity and ion dehydration in a 'spacetime' diagram $(z,t)$ for resp. low and high $Re$. 
The associated electric potentials follow from equation \ref{eq:EAH1DphiE} with initial conditions: $\phi_E(0)=0$ [V] and: $\partial_z\phi_E(0)=-5.65$ [V/m] 
\footnote{The co-axial electric field strength $E_z(d)$ of an open cylinder($R,L$) with linear charge density $\lambda_E$ at distance $d$ from its entrance is: $E_z(d)=\tfrac{\lambda_E}{2\pi\epsilon R^2}\tfrac{L}{\sqrt{R^2+L^2}}$. For the SF of KcsA we take $L\approx 12$\AA,  $R\approx 2$\AA. 
The charge density $\lambda_E$ inside the SF of KcsA follows from Lincong Wang's data \cite{LiWa2016} for surface charges of various protein-solvents, and estimates $\lambda_E =$\ -1.27E-09 [C/m] .
Combined, for $d=0$, we thus find: $E_z(0) =- 5.6453$ [V/m]. 
The particle density $n$ for $N$ ions inside a cylinder($R,L$) is: $n=N/\pi R^2L$. Following Kopec et. al. \cite{Kopec2018}, we assume N=3 ions inside.  This gives $n$= 1.9894E+28 $m^{-3}$.} and are shown in Fig \ref{fig: EAH1DphiE}. \\
For low $Re$, depicted in Fig \ref{fig: EAH1DlowRe}, the system is in the laminar regime, and the flow builds up speed gradually from the entrance onward, and the fluid slowly dehydrates starting from the middle only after 400 msec. \\
For high $Re$, depicted in Fig \ref{fig: EAH1DhighRe}, the system is in the period-doubling regime, and the flow becomes unstable after $\pm$ 150 msec, and a series of undulations of small wavelengths emerge that represent density-pressure waves in the motion of the ions.
After 250 msec, these undulations slowly disperse. 
Note that the concurrent dehydration is much faster and efficient, resulting in the complete dehydration of the right part of the channel after $\pm$ 150 msec. 
This is caused by the longer exposure to the dehydration term $-\sigma\eta$, due to the undulation that impedes the motion.
Figure \ref{fig:KcsAporeradius}(b) shows how also the 3D EAH equations exhibit the emergence of undulations inside the
narrow tubular SF of the KcsA channel, for Re = 1E4 after 150 msec.\\
In Section \ref{sec:DDHO-model} we will show that this efficient dehydration can be explained by the 1D-DDHO-mechanism, where the oscillating density waves act as the driving force.
\\
%%%%% FIGURE  EAH1DlowRe: 1D-EAH: low Re

 %%%% FIGURE EAH1DhighRe :  1D-EAH: High Re

 %%%% FIGURE EAH1DphiE : 1D-EAH: High Re
%%%%%%%%%%%%%%%%%%%%%%%%%%%%%%%%%%%%%%%%%%%%%%%%%

%%%%%%%%% SUBSECTION 2D and 3D OBSERVED PHENOMENA 
\subsection{Observed phenomena in 2D- and 3D-EAH dynamics} 
\label{sec:observ2D3D}

In 2D and 3D dynamics, for suitable $Re$, vortices and eddies emerge, either isolated (2D) or connected (3D). These vortices efficiently drive ionic dehydration and transport.

\subsubsection{Vortex dynamic enhance efficient dehydration }
The model leads to efficient dehydration in the outward flux, varying with the setting of the model parameters. 
The initially entirely hydrated and motionless fluid naturally dehydrates near the walls of the ion channel due to the sink-term $\sigma^-(d)$. The amphiphilic force attracts new hydrated ions but does not act on the freshly dehydrated ions. This causes a flow towards the ion channel walls. 
%%%% ISOBAR pressure Figure from ELIC source 
\begin{figure*}
\begin{center}
\includegraphics[width=0.48\textwidth]{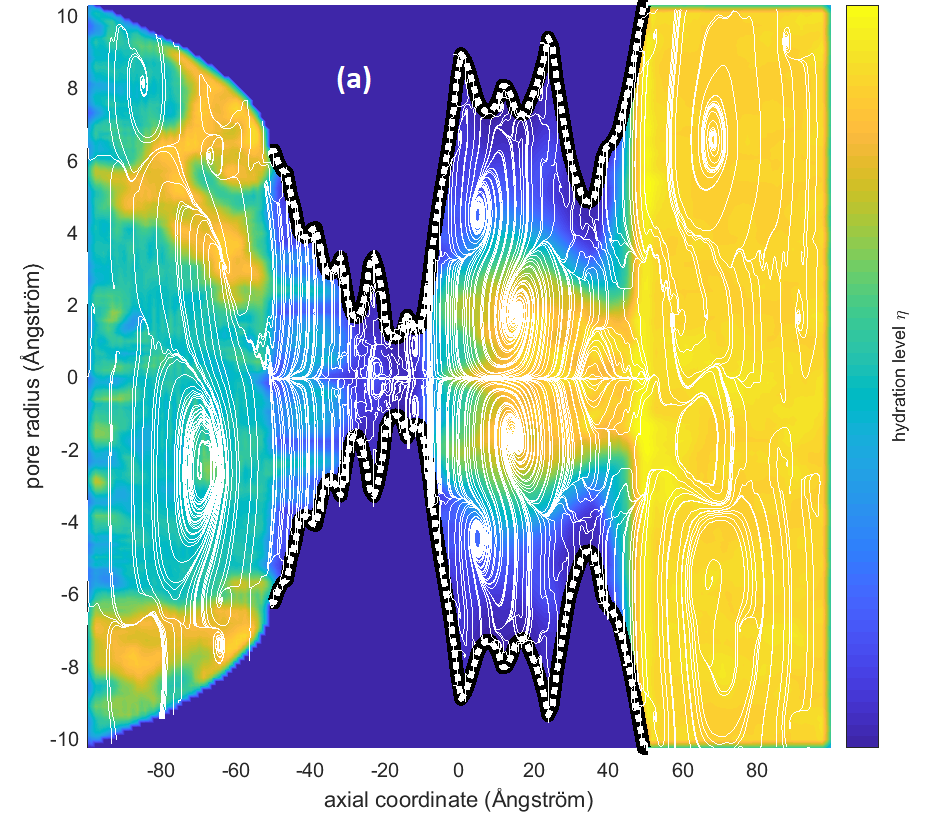}
\includegraphics[width=0.48\textwidth]{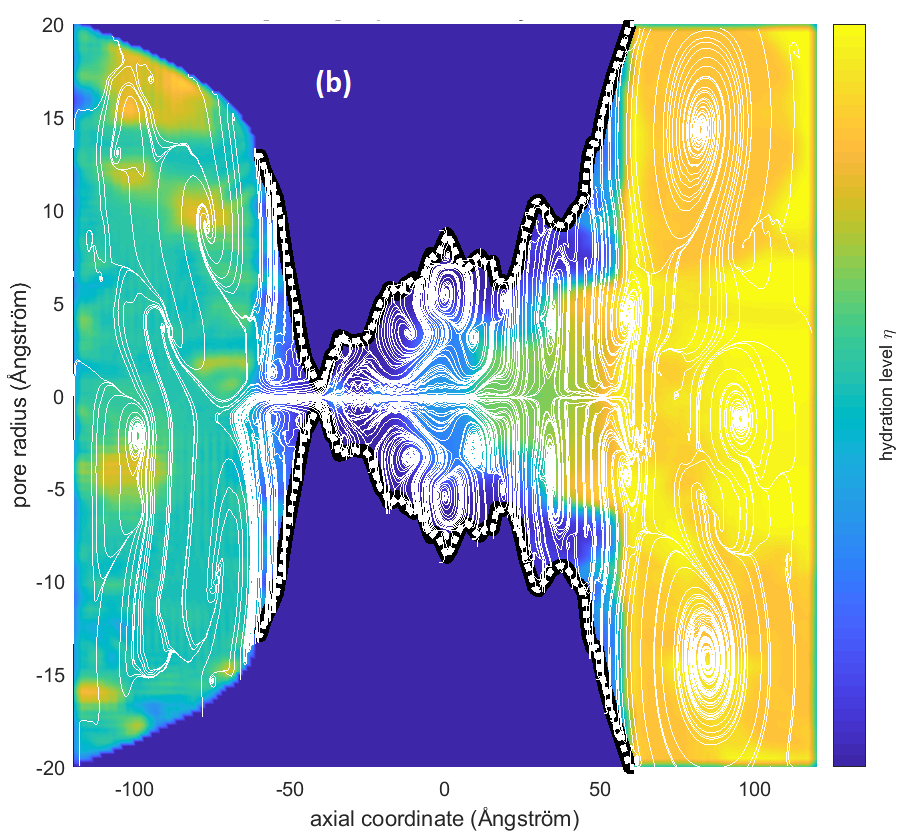}
\caption{Toroidal vortices and mixing visible through streamlines and hydration. (a): strong central toroidal vortices in the ELIC ion channel, Re = 700, after 250K iterations. (b):  toroidal vortices and mixing near the walls of the GLIC1 ion channel, Re = 400, after 450K iterations. Time-step: d$t$=1E-6. Gray scale (online color) depicts hydration $\eta$. Different horizontal and vertical scales.}
\label{fig:ELICisobars}
\end{center}
\end{figure*}
%%%%%%%%%%%%%%%%%%%%%%%%%%%%%%%%%%%%%%%%%%%%%%%
Moreover, the recently dehydrated bare ions have a larger effective electric charge $q_B$ than the aqua-ions $q_A$, due to electrostatic screening. Therefore, the electric charge density $\chi$ is larger in these regions: $\chi = \eta q_A+(1-\eta)q_B$ (per species), as described above. This effect affects the higher electrostatic potential in the dehydrated regions, so, near the ion channel walls, as: $\nabla^2\phi_{\text{E}} = \overline{\chi}/\varepsilon_0$. The electrostatic (body) force component in the NSE equation \ref{eq:EHANS1} therefore is doubly affected, namely as: $\textbf{f}_{\text{E}}= -\overline{\chi}\nabla\phi_{\text{E}}$. The resulting higher electrostatic pressure drives dehydrated ions out of the dehydrated regions, so, to the exit of the ion channel. \\
In a similar way, in case of rotation, centrifugal separation, described in Section \ref{sec:Rot3DEAH}, naturally supports dehydration by segregating bare from aqua-ions, as it forces aqua-ions towards the channel wall and so pushes bare ions away from it, ultimately towards the channel exit.

%%%% SS Vortex dyn
\subsubsection{Vortex dynamics and chaos }
The complexity of the flow pattern increases with the Reynolds number (Re) - i.e. with decreasing viscosity $\nu$. This is a well-known and intensely studied characteristic of the NSE \cite{Landau_1971}\cite{RuelleTakens1971}\cite{Flier2009}\cite{VVEEN2006}\cite{YAS2019}, and therefore naturally holds, and is observed, for the EAH-system. Examples of evolving flow patterns can be found in Figs \ref{fig:ELICisobars}, \ref{fig: GLIC12ANIMS}, \ref{fig: GLIC1Misobar}, and the supplemental material \footnote{See Supplemental Material at [{\em URL to be inserted by publisher}] for three sample time series of the evolution of the ionic velocity vector field $\textbf{u}$ and hydration scalar field $\eta$ under different conditions, and snapshots of kinetic energy and vorticity.}.
At low Re, the flow in the EAH system is laminar. With increasing Re, the non-linearities in the EAH equations start to dominate and cause complex cascades of toroidal vortices and eddies to emerge, including vortex rings. This phenomenon is strongly enhanced in case of external rotation. 
The formation and evolution of vortices strongly influences the progress of dehydration, and vice versa. \\
In the turbulent regime, the system does not attain a steady state but exhibits ever-changing complex dynamics.
The flow field \textbf{u} and hydration field $\eta$ strongly interact and influence each other in non-periodic oscillations. Consequently, the vortices and dehydrated regions, are constantly changing over time. This favours the hydration of that type of ions which is most suited and fitting to the time schedule of the vortex evolution. Control of this scheme allows for optimizing the yield of the ion channel. \\
Note that the 1D oscillating pressure waves in the SF of KcsA, mentioned in  Section \ref{sec:standing1Ddensitywaves}, can also be caused by the back-effect of rolling vortices in the central cavity.

%%%% SubSec OPEN <-> CLOSED STATE
\subsubsection{Open versus closed state ion channels}

\begin{figure*} 
\begin{center}
\includegraphics[width=0.48\textwidth]{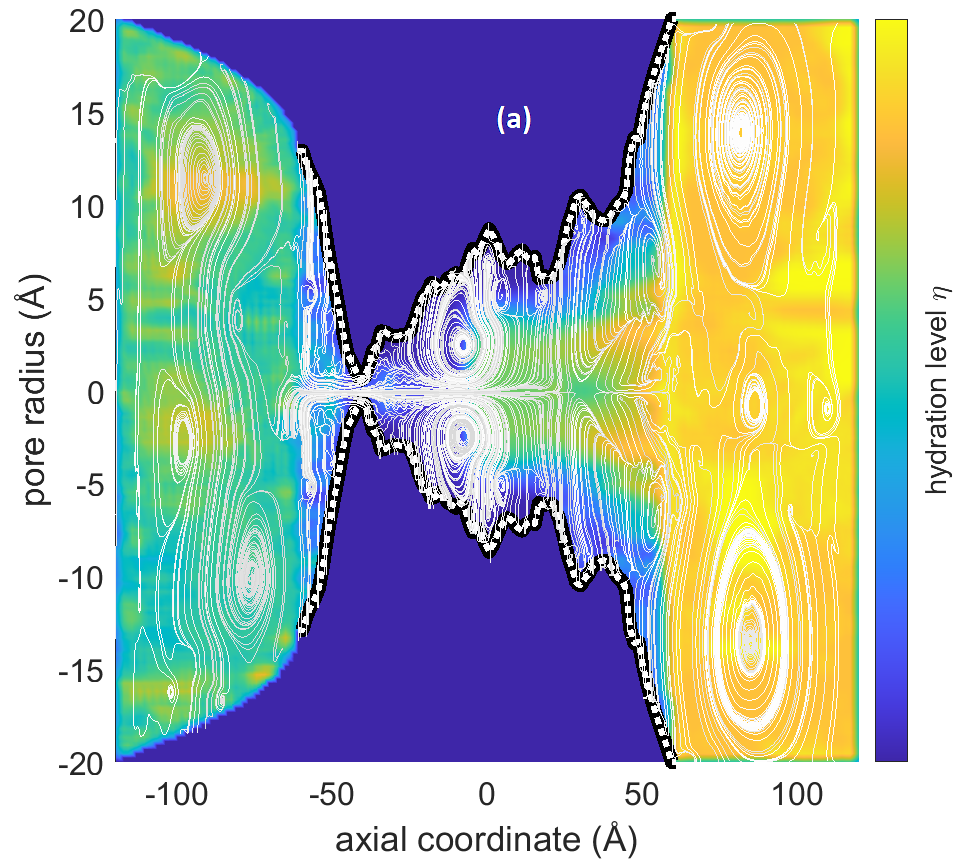}
\includegraphics[width=0.48 \textwidth]{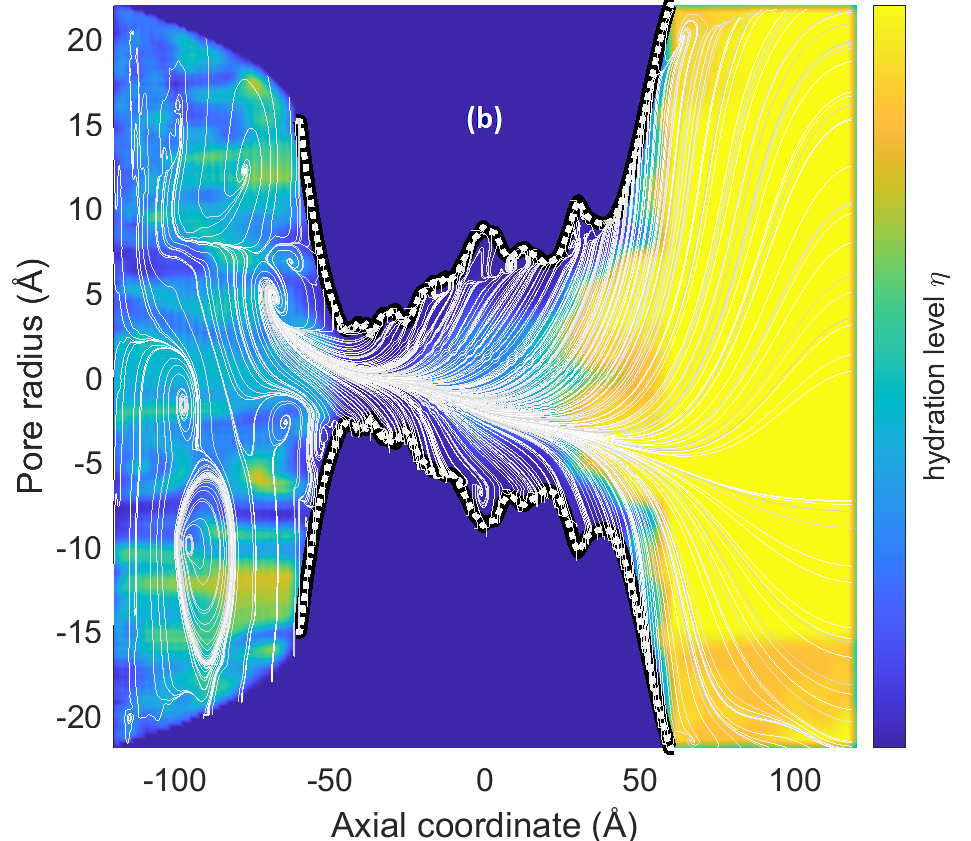}
\caption{Solutions to the EAH equations \ref{eq:EHANS1}-\ref{eq:PNPEquation} for the GLIC ion channel as depicted  in Fig \ref{fig: GLIC12}. (a): Closed state GLIC1 with toroidal vortices dehydrating the ions. (b): Open state GLIC2 flushing the dehydrated ions into the intracellular space. Depicted are the stream lines of the vector field {\bf u} and the degree of hydration $\eta$: light (color: yellow): fully hydrated $\eta$=1, dark (color: blue): fully dehydrated $\eta$=0. Note the different scales on horizontal and vertical axes.}
\label{fig: GLIC12ANIMS}
\end{center}
\end{figure*}

The ion channels GLIC1 and GLIC2 offer an opportunity to compare the open and closed state of the same channel. The latter is one of the few crystal structures that are confirmed as being in the open state.  Fig \ref{fig: GLIC12ANIMS}(a) depicts the closed state GLIC1 after 400K iterations with timestep of d$t$=1E-6. It exhibits strong vortices that drive hydration inside the channel, but hardly affect the intracellular hydration depicted left of the channel. Though some dehydrated ions leak through the outlet. The extracellular environment, depicted right of the channel, is nearly fully hydrated - besides some retrograde flow of dehydrated ions from the channel. 
Fig \ref{fig: GLIC12ANIMS}(b) depicts the open state GLIC2, starting directly from the situation - i.e. from the values of the $\bf{u}$ and $\eta$ fields - of Fig \ref{fig: GLIC12ANIMS}(a), after another 100K steps of 1E-6. It shows that the vortices have dissolved, and a mostly laminar flow has established that now largely dehydrates the intracellur space left of the channel.  
Note that still then miniature vortices persist in the grooves of the walls.

%%%% MY Figure GLIC-ionchannels GLIC1-CLIC2

%%%% SSS: Mutations
\subsubsection{Why mutations in ion channels are detrimental}
\begin{figure*}
\begin{center}
\includegraphics[width=0.48\textwidth]{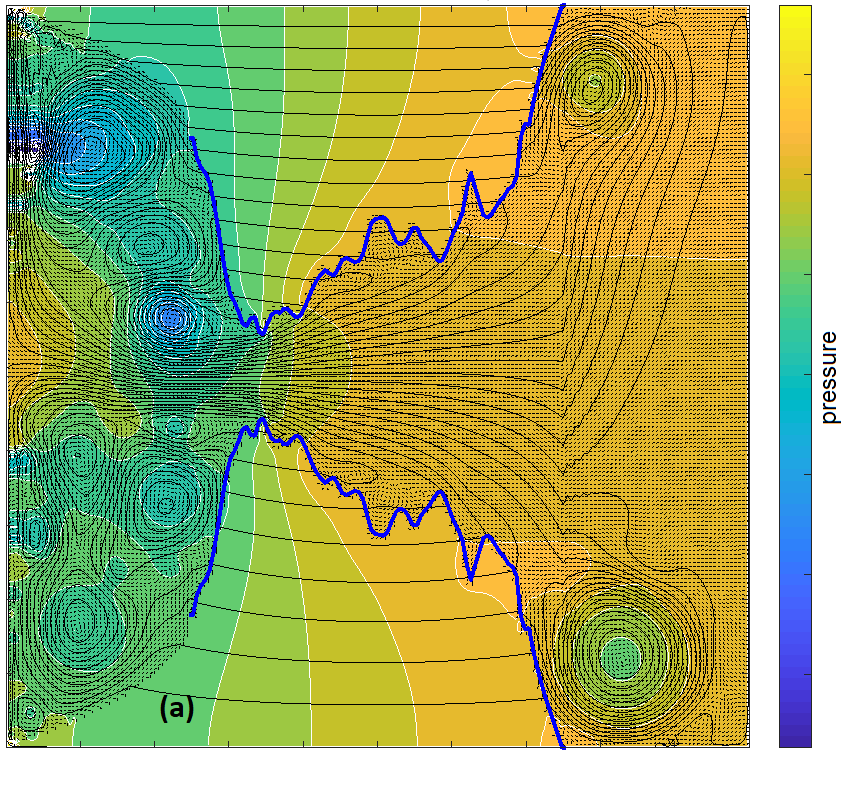}
\includegraphics[width=0.48\textwidth]{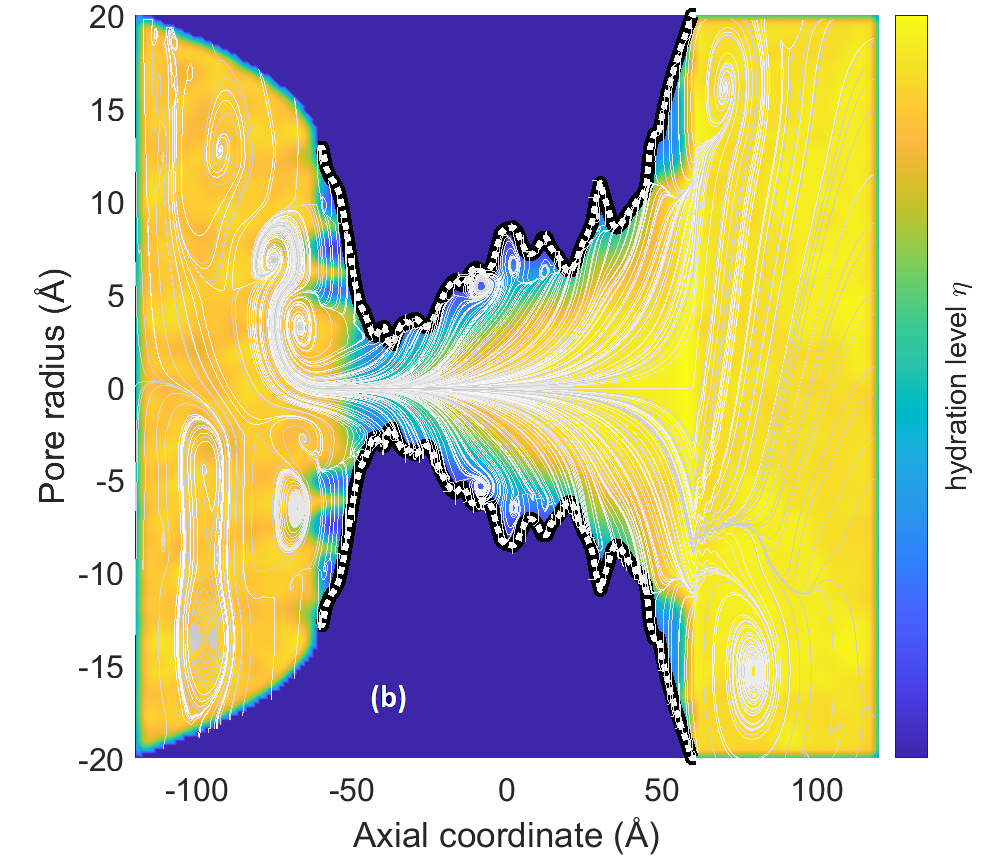}
\caption{Mutant GLIC1M in the open state, showing turbulent-free passage of ionic flow. (a): pressure (gray and color scales) and flow lines. Channel wall depicted as solid line, (b): Hydration (gray and color scales) and streamlines. Both: Re = 500, 400K iterations  (different scales on horizontal and vertical axes)}
\label{fig: GLIC1Misobar}
\end{center}
\end{figure*}

Various observations indicate that mutations that affect the geometric shape of the ion channel are detrimental to ion selectivity in particular \cite{Kapetis-2017}. 
For this reason, we studied the E221A mutation that changes GLIC1 into GLIC1M and increases the minimum pore radius from less than 0.5 {\AA} to 2.3 {\AA} \cite{Hilf-2009}. This showed that GLIC1M cannot sustain a stable closed state and continuously 'leaks' hydrated ions into the intracellular environment, depicted in Fig \ref{fig: GLIC1Misobar}. 
This shows a flow pattern reminiscent of the open state GLIC2, depicted in Fig \ref{fig: GLIC12ANIMS}(b).

%%%% ISOBAR pressure Figure from GLIC1M source 
%%%%%%%%%%%%%%%%%%%%%%%%%%%%%%%%%%%%%%%%%%%%%%%%%%%%%%%%%%%%%%%%%%

%%% SSS: Implicit interaction
\subsubsection{Implicit interaction and competition between the species}
The different species in the mixture do not interact directly in the (de)hydration reaction \ref{eq:EHANS2}. Rather, they interact indirectly in equation \ref{eq:EHANS1}, via the flow field {\bf u} that contains the ensemble averages $(\overline{\eta}, \overline{\nu},\overline{\chi})$ , defined in equations \ref{eq:etamean} - \ref{eq:ensemblechi}. This causes a competition among species which one is most efficient in optimizing its outward flux of dehydrated ions. 
All this results in different outflux performances for different ions, because of their different dehydration, mass, and charge characteristics. See Fig \ref{fig:ZIMM2011FIGS4B}.
%%%%%%%%%%%%  END of SUBSECTION OBSERVED PHEN

%%%%%%%%%%%%  SUBSECTION on COmparison 
\subsection{Comparison between experimental data and theoretical EAH predictions of ion transport in the ELIC channel }
\label{sec:SecCompTheoExp}

In order to assess the consistency of the EAH model, its theoretical predictions are compared to empirical data. For this purpose we draw on the experimental data by I. Zimmermann and R. Dutzler \cite{ZIM2011}, who studied ligand activation and electric conductance of the ELIC ion channel from {\em Erwinia chrysanthemi}, our prototype model ion channel \footnote{Other suitable experimental data is available for the GLIC channel from Tillman et al. \cite{TILM2013}-Fig 2A, and from L Yue et al. \cite{YUE2002} for various bacterial ion channels.}. 
First, we compare measured patch clamp data with predicted outward electric current. Second, we compare observed and predicted conductance and reversal potential of a single ELIC channel. Next, we compare the current-voltage (I-V) relationships of single ELIC channels at different cation concentrations, and finally, the I-V relationships of binary mixtures of monovalent and divalent cations. 
Various experiments described below were measured from single ELIC channels embedded in planar lipid bilayers and not in the cell membrane. In these cases 'i' and 'o' refer to the corresponding 'intracellular' and 'extracellular' side.

%%%% SS Outflux NEW in version 3
\subsubsection{Particle outflux and electrical current: comparison of EAH predictions with empirical patch clamp data}

\begin{figure*}
\begin{center}
\includegraphics[width=0.32\textwidth]{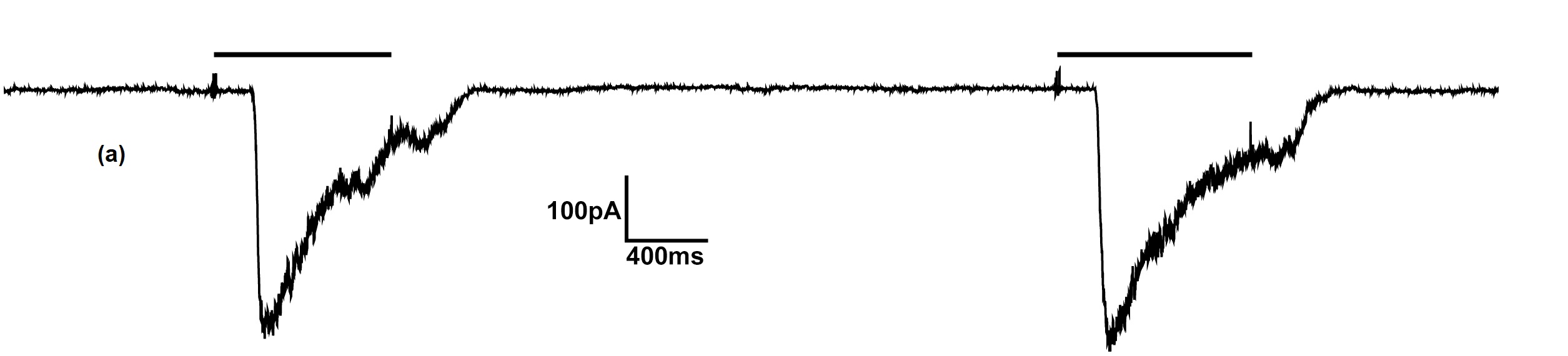}
\includegraphics[width=0.32\textwidth]{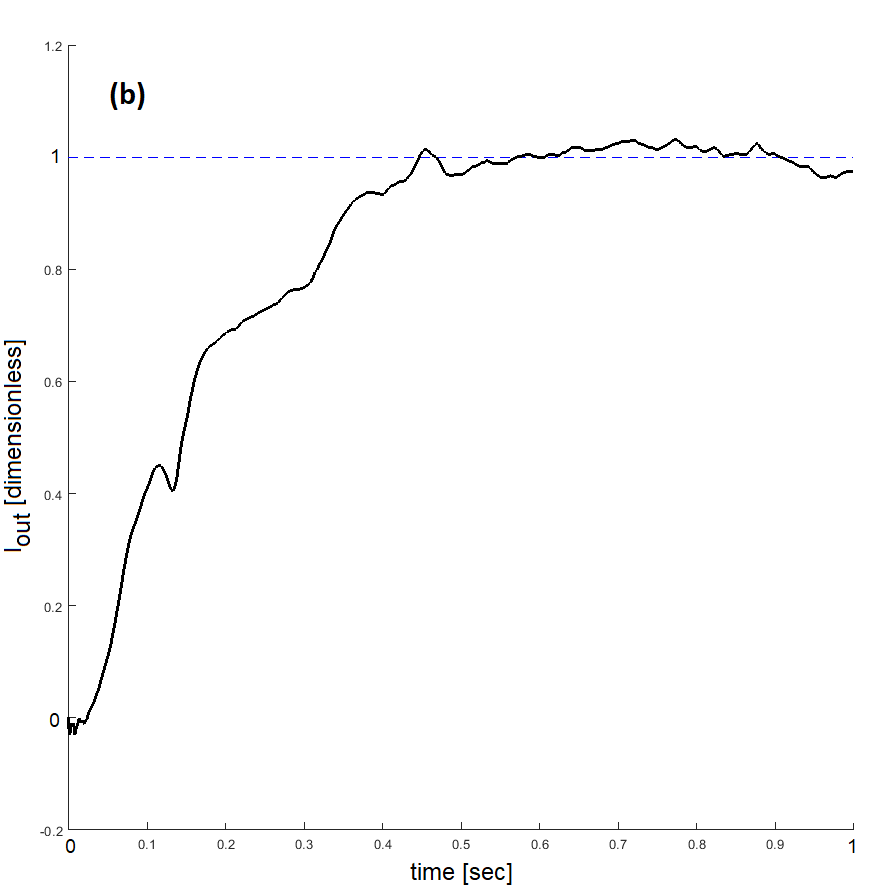}
\includegraphics[width=0.32\textwidth]{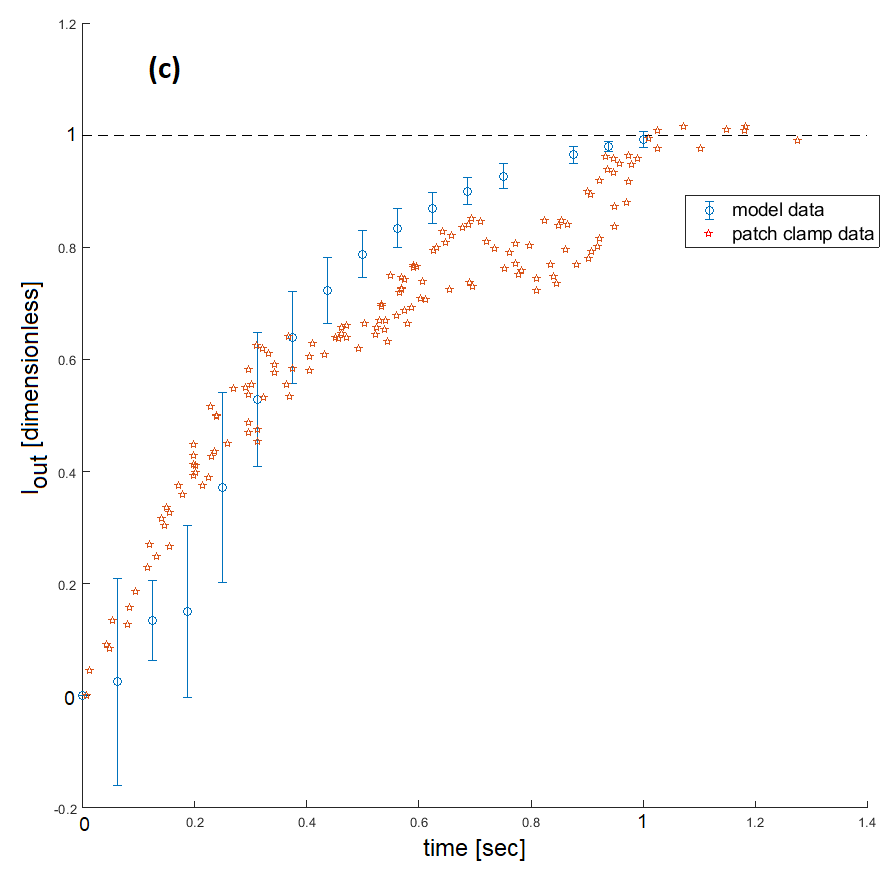}
\caption{(a): Patch clamp data of single ELIC ion channel by Zimmermann and Dutzler \cite{ZIM2011}-FigS4B, reproduced by courtesy of the authors. (b): One typical run of the EAH model outflux data for the ELIC ion channel. (c): EAH model outflux data of ELIC, averaged over 5 runs (disk, error bars represent two standard errors, n=5), compared with the averaged patch clamp data of (a) (star), n=2 (all currents normalized to maximum value).  }
\label{fig:ZIMM2011FIGS4B}
\end{center}
\end{figure*}

Let us focus on the transport of monovalent cations through ELIC ion channel. Consider a mixture of monovalent bare and aqua ions of only one  species, Na$^+$, and predict the outward electrical flux and compare this to patch clamp data.\\
Let us assume a rotation-symmetric model of the ELIC channel, and apply cylindrical coordinates $(z,r,\theta)$. The electrical current at any axial position $z$ is determined by the fluxes of the charged particles. We focus on the symmetry axial coordinate $z$, running from the extracellular entry $z=L$ to the intracellular exit $z=-L$.
Using our definitions of hydration field $\eta(z,t)$ and the effective charges $q_A$ and $q_B$, we can write the current $I(z,t)$ as:
\begin{equation}
 I(z,t) = \int_0^{R(z)}(\eta q_A +(1-\eta)q_B)nu2\pi rdr
\label{eq: currentinside}    
\end{equation}
where $n(z,r,\theta)$ is the particle density and $u(z,r,\theta)$ the horizontal component of the velocity vector field. $R(z)$ is the pore radius at position $z$, as e.g. in Fig \ref{fig: GLIC12}. 
The theoretical value of $I$ can be directly computed from equation \ref{eq: currentinside}, assuming a realistic estimate of the particle density $n$. 
The values for $I$ are evaluated for $z$-values at $z=L$ (inward flux from environment) and $z=-L$ (outward flux into cell) \footnote{The fluxes are actually computed at axial positions $z$=-12{\AA} and $z$=+35{\AA}, as depicted in Fig \ref{fig:ELICflow}, i.e. respectively at the exit and entry of the central cavity. This is to avoid numerical artifacts at the boundary of the channel itself. } of the ion channel and logged during a run from initial state ($t=0$) to final state ($t=1$ sec) with steps of d$t$ = 5E-6 sec. \\
With this approach we obtain results for the outward and inward fluxes that vary strongly for different parameter settings, notably contrasting fixed versus rotating ion channels. 
Examples of inward and outward ion flux through the channel for a variety of system parameter settings can be found in Fig \ref{fig:ZIMM2011FIGS4B} and the supplemental material.
A typical result (of a non-spinning channel), depicted in Fig \ref{fig:ZIMM2011FIGS4B}(b), shows that the (normalized) outflowing current $I$ initially rises steeply and then converges almost exponentially to a value $I^*$. It is observed that the oscillations in this curve are principally caused by the rolling motion and amplification of the vortices in the ionic flow.
Fig \ref{fig:ZIMM2011FIGS4B}(c) shows the average value of the outflow current $\langle I \rangle$ for five different runs of the EAH model for the ELIC ion channel with the same parameter setting as Fig \ref{fig:ZIMM2011FIGS4B}(b).
This graph allows for direct calculation of peak-value of $I_{out}$. 
We estimate the particle density of the throughput as $n=$1.9E-3 ions/{\AA}$^3$ (see Appendix \ref{Appendix1}), and the screened charges $q_A=0.1$e and $q_B=0.9$e \cite{SLAT1930}.
This parameter setting provides a theoretical peak value $I^*_{theor} = (431 \pm 109)$ pA per ELIC ion channel.
The exponential decay time of $I(t)$ to its asymptotic value $I^*$ is estimated using an exponential fit as: $\tau_{theor} = (329 \pm 55)$ msec.
These predicted values can be compared to the empirical results obtained by I. Zimmermann and R. Dutzler \cite{ZIM2011}, shown in Fig \ref{fig:ZIMM2011FIGS4B}(a). 
The patch clamp data in Fig \ref{fig:ZIMM2011FIGS4B}(a) shows a good qualitative correspondence to the theoretical EAH model data in Fig \ref{fig:ZIMM2011FIGS4B}(b), and both are jointly represented in Fig \ref{fig:ZIMM2011FIGS4B}(c). \\
Analysis of the patch clamp data gives a value for the (negative) peak current of $I_{emp}=(423 \pm 19)$ pA, and an exponential decay time $\tau_{emp} = (337 \pm 80)$ msec. 
The theoretical and empirical data show a strong qualitative similarity.
This comparison shows that there is a good qualitative and quantitative correspondence between the theoretical predictions and empirical data of the outward flux of the ELIC ion channel. 

%%%% FIGURE average resonance curve
%%%%%%%%%%%%%%%%%%%%%%

%%% SSS Conductance and reversal potential
\subsubsection{Conductance and reversal potential of a single ELIC channel}
It is also interesting to study the predicted conductance of a single ELIC channel by EAH with empirical data. Zimmermann and Dutzler experimentally examined the conductance of single ELIC channels, including the current voltage relationships in asymmetric concentrations of NaCl, that exhibit a current reversal at the Nernst potential of Na$^+$; see \cite{ZIM2011}-Fig5D.  
%%%%%%%%%%%%%%%%   Figure ZIMM2011FIG5D
\begin{figure*} 
\begin{center}
\includegraphics[width=0.45\textwidth]{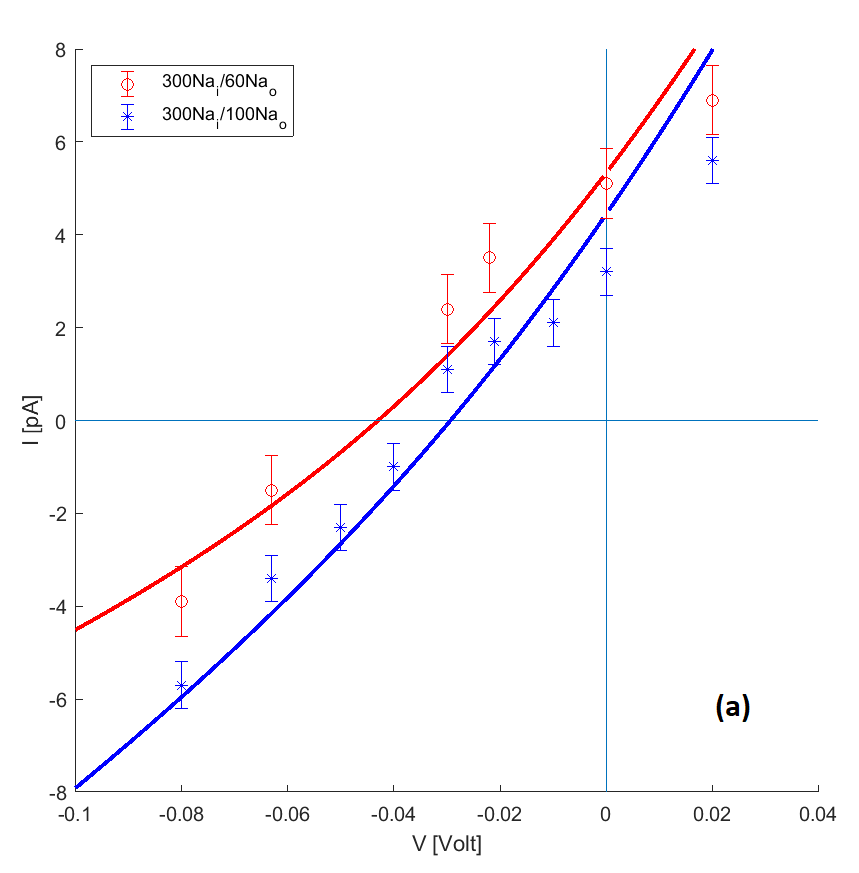}
 \includegraphics[width=0.45\textwidth]{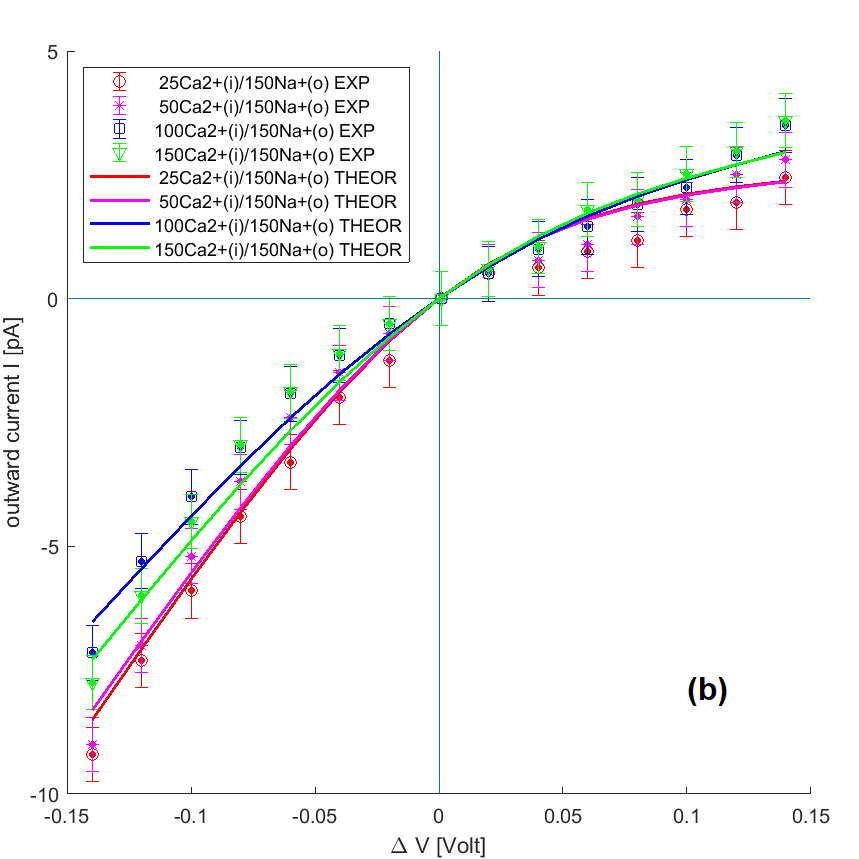}
\caption{(a): Assymetric IV-plot of ELIC for two different intracellular and extracellular concentration gaps of Na$^+$. Lines: theoretical data, markers plus errorbars: experimental data by Zimmermann and Dutzler \cite{ZIM2011}-Fig5D. Upper line and disk markers (online red): 300 mM Na$^+$ intracellular with 60 mM Na$^+$ extracellular, lower line and asterisk markers (online blue): similar for 300 mM and 100 mM. (b): Assymetric IV-plot of ELIC for a binary mixture of 150 mM Na$^+$ extracellular and 25-50-100 or 150 mM Ca$^{2+}$ intracellular by Zimmermann and Dutzler \cite{ZIM2011}-Fig5G.  }
\label{fig: ZIMM2011FIG5D}
\end{center}
\end{figure*}
%%%%%%%%%%%%%%%%%%%%%%%%%%%%%
These settings were simulated with the EAH model; a steady state \footnote{Defined as the (near) vanishing of all time-derivatives.} (after $\pm$ 400 msec) was found with the required intracellular and extracellular concentrations of Na$^+$, and, using this, the theoretical augmented Goldman curve from equation \ref{eq:AugGoldmanEq} was computed.   Fig \ref{fig: ZIMM2011FIG5D}(a) compares the theoretical IV-curve with the experimental data. Here, and in all following comparisons, the empirical single channel main conductance level $g$=96 pS found by Zimmermann and  Dutzler was used to gauge the theoretical channel output current.
The Nernst-potential values are shown in Table \ref{table:NernstPotentialsELIC} and Fig \ref{fig: ZIMM2011FIG5D}(a).
%%%% TABLE rel perm
\begin{table}[h!]
    \centering
    \begin{tabular}{ | l | l | l | }
    \hline
    Source	& 300[Na$_i$]/60[Na$_o$] ($n$)	& 300[Na$_i$]/100[Na$_o$] ($n$)\\
    \hline
    Emperical \cite{ZIM2011}	    &-48.6 $\pm$ 3.3 mV (6) &-31.8 $\pm$ 2.3 mV	 (9)\\
    EAH model	    &-42.5 $\pm$ 2.1 mV    &-29.0 $\pm$ 1.4 mV   \\    \hline
    \end{tabular}
    \caption{Experimental and theoretical values for the Nernst potential for current reversal in ELIC ion channel. Experimental value determined from least squares fit, theoretical value by the augmented GHK-equation. Sample size $n$ in brackets.}
\label{table:NernstPotentialsELIC}
\end{table}
The comparison shows a good qualitative correspondence.

%%%%%%%%%%%%% SSS Current-voltage relations
\subsubsection{Current-voltage relations of single ELIC channels at different concentrations of Na$^+$ and its permeability for Na$^+$}

\begin{figure*} 
 \begin{center}
\includegraphics[width=0.45\textwidth]{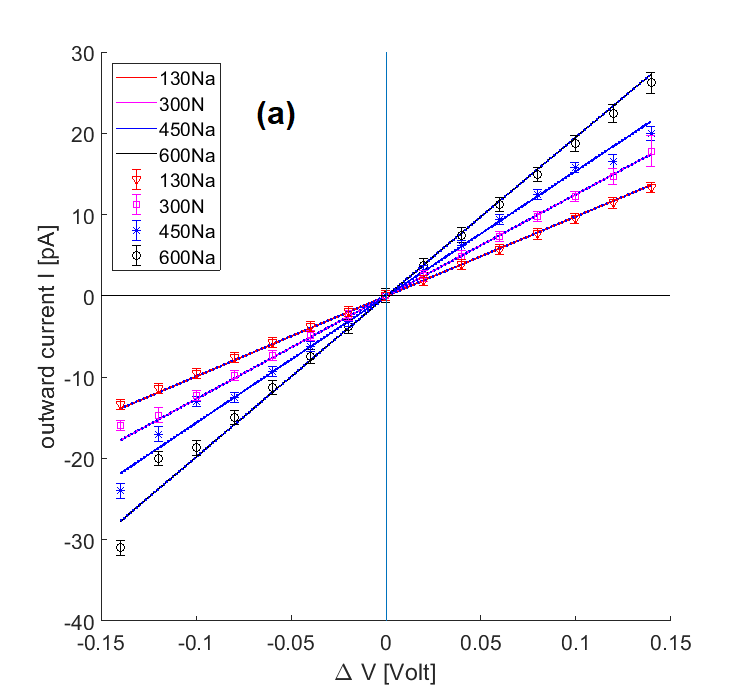}
%%%  ZIM2011FIG5FcompEAH.png}
\includegraphics[width=0.45\textwidth]{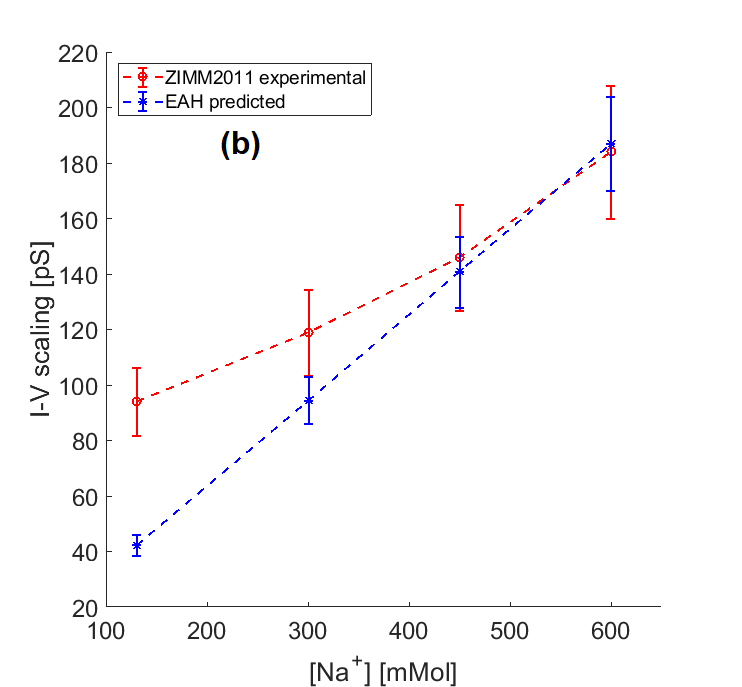}
\caption{(a): Current-voltage relations of single ELIC channels at different concentrations of Na$^+$: 130 mM (online red), 300 mM (online magenta), 450 mM (online blue), 600 mM (black). Lines: theoretical values of EAH-model. Markers and error-bars: experimental data of Zimmermann and Dutzler \cite{ZIM2011}-Fig5F. (b): Slopes of the I-V curves in (a) for various concentrations of Na$^+$, showing that the current I increases monotonically, but not saturates. Disk-markers (online red): experimental data of \cite{ZIM2011}, asterisk: EAH-model.}
\label{fig: ZIMM2011FIG5F}
\end{center}
\end{figure*}

Zimmermann and Dutzler also studied current voltage relationships of single ELIC channel currents at different concentrations of NaCl, with [Na$^+$]$_{in}$ = [Na$^+$]$_{out}$. They found that the single channel conductance linearly increases with the ion concentration and does not show any saturation at high concentrations (Figure 5F of \cite{ZIM2011}).  Note that in the classical Goldman equation \ref{ClassGoldmanEq}, for $n_{i}=n_{o}\equiv n$, the term $(1-e^{\alpha V})$ cancels exactly, so that $I$ depends linearly on $V_E$ \footnote{With an isolated singularity for $V_E=0$ Volt.}. Similar for the augmented Goldman equation \ref{eq:AugGoldmanEq}, if $f_i=f_o$.
In this case, therefore, for $V_E\neq 0$, $I$ becomes linear in $n$ and $V_E$ as: $I \approx \alpha npqV_E $. 
This relation allows for estimating the absolute permeability $p_{Na}$ of Na, when given experimental values for $I$ and $V$. In Fig \ref{fig: ZIMM2011FIG5F}(a) the empirical I-V values (of Figure 5F of \cite{ZIM2011}) are compared with the predicted EAH-values. The latter were obtained by changing the boundary conditions of the EAH-system such, that the extracellular concentration was fixed at $n_{out}$ and equal to the intracellular value $n_{in}$. Under these restrictions, the EAH system is propagated to the steady state (after $\approx$ 0.4 sec), and the augmented Goldman equation \ref{eq:AugGoldmanEq} is applied to determine the $I-V$ relation, shown in the figure. 
This gives a good qualitative correspondence, in that both curves exhibit a monotonic increase of the I-V line with higher concentrations $n$. Fig \ref{fig: ZIMM2011FIG5F}(b) shows the slope of the I-V curve for various concentrations [Na$^+$]. 
The experimental data clearly shows that the current does not saturate, in line with the prediction of the Goldman equation. \\
Note from Fig \ref{fig: ZIMM2011FIG5F}B that the experimental data in fact does {\em not} increase linearly with [Na$^+$], while the predicted EAH relation (dotted line) following above approximation, strictly does. 
It appears that the EAH-Goldman prediction acts as a limit curve for high concentrations [Na$^+$]. 
Alternatively, this relation can also be explained by assuming that the permeability $p_{Na}$ depends on the concentration $n_o$=$n_i$=[Na$^+$].
Using above linear approximation for $I(V)$ and a linear least squares fit of the experimental and theoretical I-V data, we can estimate the permeability $p_{Na}$. The results are shown in Table \ref{table:abspermeabs}. 
\begin{table}[h!]
    \centering
    \begin{tabular}{ | l | l | l | }
    \hline
    [Na$^+$] [mM]	& EXP $p[Na^+]$ 	& THEOR $p[Na^+]$ \\
    \hline
    130	    &0.63 $\pm$ 0.27  &0.74 $\pm$ 0.72 \\
    300     &0.24 $\pm$ 0.10  &0.13 $\pm$ 0.13 \\
    450     &0.23 $\pm$ 0.09  &0.06 $\pm$ 0.06 \\
    600     &0.12 $\pm$ 0.05  &0.03 $\pm$ 0.03\\
    \hline
    \end{tabular}
    \caption{Experimental and theoretical values for absolute permeabilities $p[Na^+]$ in ELIC ion channel for various concentrations of Na$^+$. Experimental data from \cite{ZIM2011}-Fig5F, sample size $n$=15 in all cases.
    Theoretical values from EAH model. 
    Both values determined from linear least squares fit from Goldman curves Fig \ref{fig: ZIMM2011FIG5F}(a).
    %: $I = C_0+C_1V$, so: perm = $(C_1-\alpha C_0/2)/(\alpha nq)$.
    %, $\delta n = \alpha C_0n/(C_1-\alpha*C_0/2)$.
    }
\label{table:abspermeabs}
\end{table}
This Table exhibits a strong correlation between the theoretical and experimental values, both showing a decrease of $p_{Na}$ with increasing [Na].
%%% , for which a randomization test statistics gives a p-value of $0.73\pm 0.32$.  
% %%%% FROM Zimm2011 Figure 5F relation conductance - concentration

%%%%%% SSS: IV relationships
\subsubsection{IV relationships of binary mixture of monovalent and divalent cations}

A similar approach as above is possible for determining the {\em relative} permeability of a binary mixture of a monovalent and a divalent cation. 
Consider a mixture of Na$^+$ and Ca$^{2+}$ with given extra- and intracellular densities [Ca$^{2+}$]$_{in}$ and  [Na$^+$]$_{out}$. This experimental setting for the ELIC channel was studied by Zimmermann and Dutzler in \cite{ZIM2011}-Fig5G. 
They determined the current-voltage relationships in asymmetric salt conditions, where the extracellular side contains 150 mM NaCl, and the intracellular side contains different concentrations of CaCl$_2$ (25-150 mM).  
The theoretical expected I-V relation of this mixture is described by the augmented Goldman formula \ref{eq:AugGoldmanEq}. A comparison of experimental values and the predicted I-V curve is presented in Fig \ref{fig: ZIMM2011FIG5D}(b). \\
 Theoretical values for I-V are obtained from a run of EAH with boundary conditions fixing [Na$^+$] and [Ca$^{2+}]$ to the experimental values; namely: $[Na^+]_{outside}$ = 150 mM and $[Na^+]_{inside}$ = 0 mM  fixed during the run, and $[Ca^{2+}]_{outside}$ = 0 mM, while $[Ca^{2+}]_{inside}$ varies as a series: \{25 mM, 50 mM, 100 mM, 150 mM\}. After reaching the steady state and the time-derivatives vanish, typically after 0.4 sec,  we find the hydration levels $\eta_{Na}^{out}$ and $\eta_{Ca}^{out}$, and flow velocities $u^{out}$ and $u^{in}$. These values now can be entertained in the augmented Goldman equation \ref{eq:AugGoldmanEq}, giving the total predicted current of the mixture. 
The predicted total current of the mixture is: $I_{theor} = p_{Na}I_{Na} + p_{Ca}I_{Ca}$, where $I_{Na}$ and $I_{Ca}$ are the predicted currents with the given settings of the known particle densities in-and-out, and $p_{Na}$ and $p_{Ca}$ are the unknown absolute permeabilities. \\
Comparison with the observed current $I_{exp}$ allows for the estimation of the permeabilities  $p_{Na}$ and $p_{Ca}$. 
 With minimum norm least-squares estimation and data randomization we find values for means and errors of the absolute and relative permeabilities as presented in Table \ref{table:FigZ5Gtable}. The relative permeability of Na to Ca is: $p^{\text{EAH}}_{rel} \equiv p_{Na}/p_{Ca}$. \\
The theoretical and experimental I-V curves in Fig \ref{fig: ZIMM2011FIG5D}(b) exhibit a strong qualitative correspondence. 
%%%%%%%%%%%%%%%%%%%%%%%%%%%%%%%%%%%%%%%%%%%%%%
\begin{table}[h!]
    \centering
    \begin{tabular}{ | l | l | l | l | l | l | }
        \hline
        $[Ca^{2+}]_i$ (mM) & $p[Na^+]$ & $p[Ca^{2+}]$ & $p^{\text{EAH}}_{rel}$ & $p^{\text{GHK}}_{rel}$ & $p^{\text{AUG}}_{rel}$ \\
        \hline
        25  & $0.86 \pm 0.20$ & $1.81 \pm 0.21$ & 0.48 & 0.33 & 0.45 \\
        50  & $1.12 \pm 0.21$ & $1.50 \pm 0.23$ & 0.75 & 0.67 & 0.69 \\
        100 & $1.49 \pm 0.19$ & $0.98 \pm 0.22$ & 1.50 & 1.33 & 1.45 \\
        150 & $1.79 \pm 0.21$ & $0.86 \pm 0.20$ & 2.09 & 2.00 & 2.21 \\
        \hline
    \end{tabular}
    \caption{Theoretical values for absolute and relative permeabilities in ELIC ion channel for various concentrations of Ca$^{2+}$ 'inside' (mM), and fixed 150 mM Na$^+$ 'outside'. Experimental data from \cite{ZIM2011}-Fig5G, sample size $n$=15 in all cases.}
    \label{table:FigZ5Gtable}
\end{table}
  %%%%%%%%%%%%%%%%%%%%%%%%%%%%%%%%%%%%%%%%%%%%%%%
 The thus obtained EAH-predicted values $p^{\text{EAH}}_{rel}$ can be compared with the expected relative permeabilities by applying the GHK-equation on the experimental setting.
 For the 'classical' GHK-equation this gives \footnote{Obtained by solving $I_{\text{tot}}(V_{\text{Nernst}})=0$, and observing in Fig \ref{fig: ZIMM2011FIG5D}(b) that this occurs at $V_{\text{Nernst}}=0$.}:
$$
 p^{\text{GHK}}_{rel} = 2\frac{[\text{Ca}^{2+}]^{in}}{[\text{Na}^{+}]^{out}}
 $$
Values for the roots of the augmented GHK-equation, called $p^{\text{AUG}}_{rel}$, are determined numerically.
The obtained expected values $p^{\text{GHK}}_{rel}$ and $p^{\text{AUG}}_{rel}$ are listed in the last two columns of Table \ref{table:FigZ5Gtable}, showing good qualitative correspondences.

\begin{figure*}
\begin{center}
\includegraphics[width=0.45\textwidth]{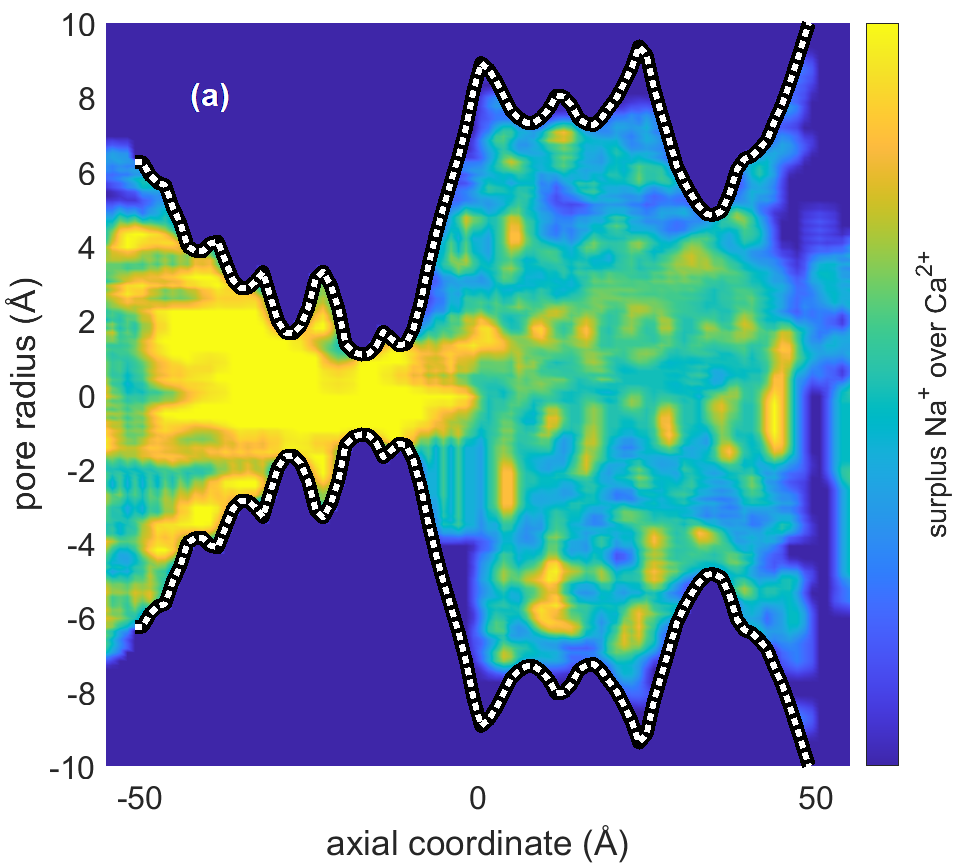}
\includegraphics[width=0.45\textwidth]{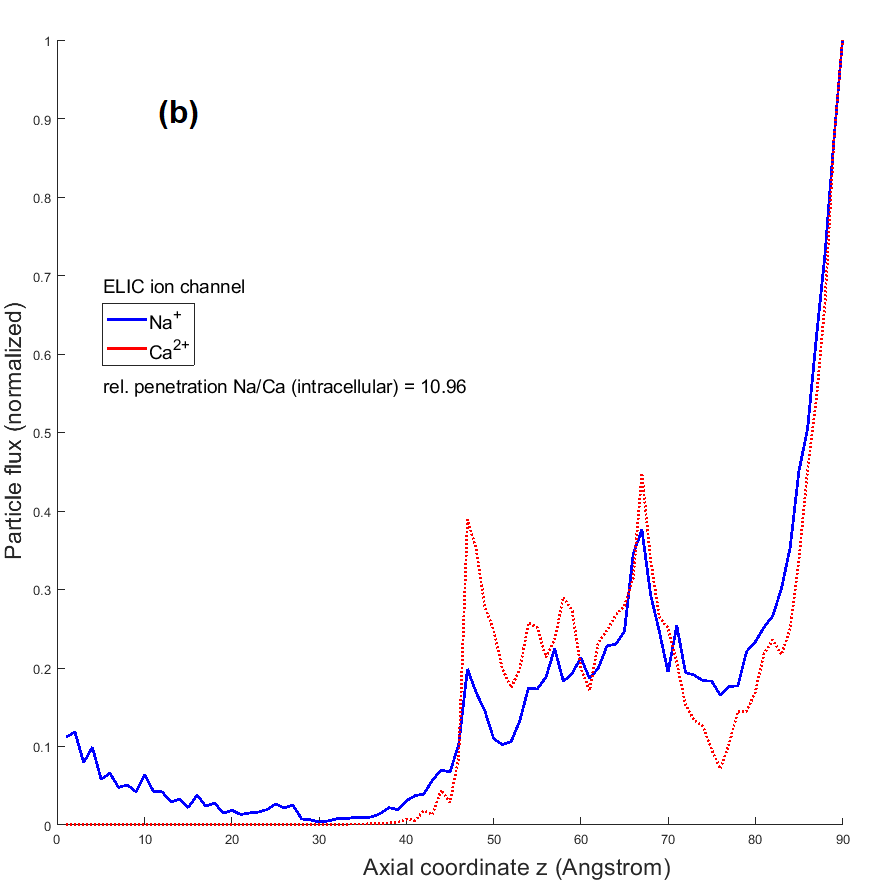}
\caption{Binary mixture of Na$^+$ and Ca$^{2+}$ in the ELIC channel after 400 msec. (a): Surplus of Na$^+$ over Ca$^{2+}$ ions. (b): Penetration of Na$^+$ and Ca$^{2+}$ ions alongside the central ionic axis in the central cavity. Dotted (online red): Ca$^{2+}$, solid line (online blue): Na$^+$. In Figure (b) the intracellular side is at z=0, the extracellular at z=90 \AA.  }
\label{fig:penetrationELIC}
\end{center}
\end{figure*}

We conclude this Section with the remark that the (relative) {\em permeability} differs substantially from the (relative) particle {\em penetration} in the ion channel \cite{MCLES1999}. The latter is defined as the fraction of the particle concentrations: $n_{rel} = n_{Na}/n_{Ca}$ at the intracellular side of the ion channel. Figure \ref{fig:penetrationELIC} shows the surplus of Na$^+$ over Ca$^{2+}$ at $t$=0.5 sec simulation of EAH for the ELIC channel and the associated ion penetration. 

%%%% FIGURE average resonance curve
%%%%%%%%%% END of SECTION comparison ...

%%%%%%%%%%%% SS: EAH and mechanism
\subsection{EAH and the mechanism behind ion channel selectivity }
\label{EAHvsKOPEC}
Do the EAH equations directly expose the mechanism behind ion channel selectivity? 
The key to this is the specific amphiphilic charge $a_s$ of an ion species $s$, which depends on the extent $\eta$ of its water-shell.
In addition, the energy required to strip or gain a water molecule depends sensitively on the detailed amphiphilic coding of the proteins that constitute the wall. 
Thus, the amphiphilic potential step near the channel wall: $\Delta\phi_A = \nabla \phi_A\cdot \textbf{u}$ in the EAH-main equation \ref{eq:EHANS1}, poses an energy threshold to de/re-hydration in the flow: $\Delta E_{s} = a_{s}(\eta)\Delta\phi_A$, which differs for different species, and thus 'selects' the ion type.
Note that this selection is independent of the size of the ion, e.g. a larger ion with a higher amphiphilic charge may slip easier through a channel than a smaller ion with less amphiphilic charge.
This is similar to the explanation of W. Kopec et. al. \cite{Kopec2018}, mentioned in Section \ref{sec:whatIOSis}, proposing a higher energy threshold for dehydrating Na$^+$ then to K$^+$ at the entrance of the channel, causing it to be more permeant to K$^+$ than to Na$^+$, despite its larger size. 
However, the higher amphiphilic charge of a species would hold for {\em all} ion channels, so would {\em not} contribute to the ion selectivity of that channel.  
In Kopec et al.'s explanation, the energy threshold is channel-specific, and therefore selectivity is also channel-specific. \\
This inadequacy of EAH to explain ion selectivity can be easily mended by adding higher-order amphiphilic tensorial terms to its equations.
Such terms are legitimatized by the hydrophobic moments proposed by Eisenberg and Silverman \cite{Eisenberg-1982}\cite{Silverman-2001}.
Near the channel walls, these new terms would cause tidal forces on the ion, subject to dipole moments induced by the specific distribution of amphiphilic and electric charges on the protein-subunits - and to its proper amphiphilic and electrical charges $a_s$ and $q_s$. This is reminiscent to the explanation of B. Hille \cite{Hille2001} of the dehydration of an aqua-ion, namely that while it rolls along the channel wall, the hydrophilic moments successively 'peel off' its water molecules. \\
Though this shows that an augmented EAH model alone can - in principle - explain ion selectivity, here we pursue a different route, one that is purely based on the DDHO framework presented in the next Section, where EAH merely contributes the harmonic driving force.
%%%%%%%%%%%%%%%%%%%%%%% END of SECTION III %%%%%%%%%%%%%%%%%%%%%%%%%%%%

%%%%%%%%%%%%%%%%%%%%%%%%%%%%%%%%%%%%%%%%%%%%%%%%%%%%
%%%%%%%%%%%%%%%%%%% SECTION Ion channel selectivity by 
%%%%%%%%%%%%%%%%%%% the Driven Damped Harmonic Oscillator mechanism
%%%%%%%%%%%%%%%%%%%%%%%%%%%%%%%%%%%%%%   

\section{Ion channel selectivity by the Driven Damped Harmonic Oscillator mechanism}
\label{sec:DDHO-model}

\subsection{The Resonance-driven damped harmonic oscillator}

The EAH continuum model for the transport and (de/re)hydration of ions through an ion channel demonstrates the emergence of strong and stable turbulent structures in the dehydrating flow of ions, each with its specific natural frequency. These structures are manifest as undulating pressure waves in 1D and toroidal vortices in 2D and 3D.
As an overarching term, we denote all these 1D, 2D, and 3D turbulent structures as 'turbulences'. 
These turbulences emerge and disappear, depending on the local degree of hydration indicated by the hydration field $\eta$, which vice versa strongly influences the fluid velocity field $\textbf{u}$ through the channel. 
In 3D, this phenomenon is even enhanced in the case of high-frequency rotating ion channels \cite{Shaw-2013}. \\
However, turbulence-engendered dehydration alone is {\em not} sufficient to explain ion selectivity. In a mixture of various species, all ions experience exactly the same oscillatory motion. So, {\em what} determines the winning ion type that can dehydrate fastest and dominate the out-flux of dehydrated ions, and thus determine the ion channel type?  
We propose that the dominant frequencies associated with these turbulences act as driving force in a harmonic oscillator system, and cause resonances depending on ion-specific parameters, especially their mass and electric charge. \\ 
For these reasons, let us consider the dehydration process inside a stable attractive turbulent structure in a dehydration zone of an ion channel; the selection filter or the central cavity. We assume that the efficiency  of dehydration depends on the duration $\tau$ that an average aqua-ion spends inside this structure: the longer it spends inside - the higher the probability it becomes dehydrated. Now consider undulating perturbations around the steady-state solutions, as for 3D depicted in Fig \ref{fig:Vortexmodes}. If $x(t)$ is a measure for the 1D quasi-periodic spatial perturbation to the steady state attractor, then $\tau$ is in first order proportional to the length of the perturbation, hence in first order proportional to $x(t)$. Let us assume that the motion inside this trajectory is governed by Newton’s laws and the dynamics near the equilibrium is attractive, ergodic, and dissipative. 
The derivation below is independent of the spatial dimension D, so holds for both the 1D undulations mentioned in Section \ref{sec:standing1Ddensitywaves} and stable vortices of Section \ref{sec:observ2D3D} alike.
We propose the following contributions to the dynamics of $x$:

\begin{enumerate}

\item 
	An elastic mean field force $F_{MF}$ that drives the ion back to the stable attractor, caused by the mean field of all combined hydrophilic and electric interactions. Padhi \cite{Padhi_2015} showed that the hydrophilic force is directly proportional to the charge $q$ of the ion. This also holds for the Coulomb force. Thus, the elastic force is in first order proportional to the linear deviation $x$ to the attractor: $F_{elast} \propto -\kappa q x $, where $\kappa$ is some constant of proportionality. We assume that this force is isotropic, and so also $\kappa$.

\item
    A damping force $F_{V}$ resulting from the average interactions and collisions between the ions. These cause viscous friction that impedes general ion motion. In first order, this is proportional to the ion velocity: $F_{V} \propto -\gamma\dot{x}$. The constant $\gamma$ of this proportionality acts as an implicit viscosity of the ion fluid, and is assumed independent of the mass $m$.

\item
	The motion inside the turbulent structure is driven by its set of natural frequencies $\Omega_n$. This gives rise to periodic forces on the ion: $F_{DF} = \sum_n F_n\text{e}^{i(\Omega_n t + \Phi_n)}$, where $F_n$ is a measure for the strength of the $n$-th mode of the turbulent structure, and $\Phi_n$ its phase. For instance, the two dominant modes in a toroidal vortex are the {\em transverse} mode $\Omega_T$ around the $z$-axis and the {\em longitudinal} mode $\Omega_L$ of the helical orbit wrapped around the toroidal hull of the vortex, as in Fig \ref{fig:Vortexmodes}.

\end{enumerate}

Combining all these contributions into Newton’s second law we obtain the well-known DDHO equation with one (in 1D and 2D) or two (in 3D) driving frequencies.
Specifically, in 3D the driving frequencies are: $\Omega_T$ (transverse) and $\Omega_L$ (longitudinal), with phase difference $\Phi$, and we obtain: 
\begin{equation}\label{DDHO}
m(\eta)\ddot{x} + \kappa q(\eta) x + \gamma(\eta)\dot{x} = F_T\text{e}^{i\Omega_T t} + F_L\text{e}^{i(\Omega_Lt+\Phi)}	 
\end{equation}
where $m$ and $q$ represent the mass and effective electric charge of the ion, and $\gamma$ the damping/viscosity of the environment. \\
Note that equation \ref{DDHO} is derived from perturbations of ion trajectories from the standard streamlines, dictated by velocity field $\textbf{u}$. Therefore, they are manifest as local particle density fluctuations and will propagate with the local speed of sound $c = \sqrt{\partial P/\partial \rho}$, which is higher than the local flow speed $u$.
These sound waves are driven by the driving force and transfer energy into those ions that resonate with the driving frequency.
\\
Most of the parameters in the DDHO equation depend on the degree of hydration $\eta$ of the ion. 
For an individual ion the mass depends on the aqua-shell number $n_H$, i.e. the number of bound water molecules: $m = m_{0} + n_Hm_{H_2O}$, with $m_{0}$ is the mass of the bare ion and $m_{H_2O}$ the mass of one water molecule. Tables \ref{table:TableResPeaks} and \ref{table:TableAnionPeaks} lists some values of average aqua-shell numbers $\overline{n}_{H}$. So, for an average hydrated ion the mass we take: $m(\eta) = m_{0} + \eta \overline{n}_{H}m_{H_2O}$. \\
The effective electric charge $q$ of an aqua-ion decreases with hydration level $\eta$, due to the effect of electrostatic screening, as the envelope dipole water molecules impede short range electrostatic interactions \cite{WOHLERT2004} \cite{YANG2013}. Thus, the effective charge becomes:  $q(\eta) = \eta q_A + (1-\eta)q_B$, with $q_A$ and $q_B$ the electrical charges of an aqua-ion and a bare ion respectively.
Also the damping $\gamma$ increases with higher hydration $\eta$. \\
We are looking for the longest possible duration of aqua-ions in the vortex and hence, as we argued above, to the largest deviations from the stable cycle. These occur when the cycle $x(t)$ resonates with a driving frequency $\Omega$ ($T$ or $L$). 
As the the DDHO equation \ref{DDHO} is linear in $x(t)$ we can study the solutions to the two modes separately: $x(t)=x_T(t)+x_L(t)$. 
In 1D and 2D there is but one driving frequency.
Thus, for each mode separately we obtain the familiar DDHO steady-state solution: $x(t) = X(\omega^*)F^*\exp(i\omega^*t + \Phi(\omega^*))$, with $F^*$ and $\omega^*$ referring to either $T$ or $L$ mode. The variables $X(\omega)$ and $\Phi(\omega)$ apply to both modes:
\begin{equation} \label{eq:DDHOamplitude}
  X(\omega) = \frac{1}{m\sqrt{(\omega_N^2-\omega^2)^2+(\omega\gamma/m)^2}} 
\end{equation}
\noindent and $\tan \Phi = \gamma\omega/m(\omega_N^2 - \omega^2)$. 
Here, $\omega_\text{N} \equiv \sqrt{\kappa q/m}$ is the natural frequency of the non-driven, free oscillator.  $\omega_\text{R}$ is the actual resonance frequency of the system.
The resonance peak occurs at a frequency $\omega_R$ given by:  
\begin{equation}\label{DDHO_resomg}
\omega_R = \sqrt{(\omega_\text{N}^2-\frac{\gamma^2}{2m^2})} 
\end{equation} 
Note that in 3D the frequencies $\omega_\text{N}$ and $\omega_\text{R}$ are valid for both the $T$ and $L$ mode, as this is entirely defined by the DDHO parameters $\{m,q,\kappa,\gamma\}$. 
The resonance frequency $\omega_R$ clearly depends on the physical properties of the ion: its charge $q$, its mass $m$, and the friction parameter $\gamma$ that depends on the size and shape of the toroidal vortex. The maximum value of amplitude $X$ occurs at the resonance peak $\omega_R$: 
\begin{equation}\label{DDHO_Xmax}
X^{*} = X(\omega_\text{R}) = \frac{2Q^2}{m\omega_N^2\sqrt{4Q^2-1}}
\end{equation}
where we use the dimensionless parameter $Q$ for 'oscillator quality': $Q \equiv m\omega_\text{N}/\gamma$. $Q$ counts the decay time for the oscillator energy by the number of oscillations. In atomic systems it is not unusual to observe high $Q$-values in the range of $10^8$. Similarly, a dimensionless frequency is defined as $\xi \equiv \omega/\omega_\text{N}$.  \\
\noindent From equation \ref{DDHO_resomg} it appears that resonance can only occur when $\kappa q \geq \gamma^2/2m$, i.e. the damping is not too large. 
This means that there is a maximum value $\gamma^* = \sqrt{2\kappa qm}$ for which resonance can occur, so a minimum value for the oscillator quality $Q^* = 1/\sqrt{2}$. This value $Q^*$ also minimizes $X^*$.  Note that equation \ref{DDHO_Xmax} requires $Q > 1/2$, but this is automatically satisfied by the minimum requirement $Q > Q^* \geq 1/\sqrt{2}$.  \\
The width $\Delta\omega$ of the peak is defined as the width at half of the height at resonance:
\begin{equation}
\Delta\omega = \frac{\omega_\text{N}}{Q}
\label{DDHO_delta}
\end{equation}
Or, similarly, $\Delta\xi = 1/Q$.\\
The elastic constant $\kappa q$ determines the energy stored in the oscillator. For instance in 3D, the total stored energy $E$ combines the transverse mode $\xi_T$ and lateral mode $\xi_L$:
\begin{equation} 
E = E_\text{pot,max} = \frac{1}{2}\kappa q\{F_T^2 X(\xi_T)^2 + F_L^2 X(\xi_L)^2\}
\label{eq:TotEnergy}
\end{equation} 
Optimal resonance curves $E(\xi)$ have well-separated high and sharp peaks with high quality factors $Q$ \-- hence small widths $\Delta\xi$. Plots of stored energy $E$ versus frequency $\xi$ in Figs \ref{fig:resonancepeaks} and \ref{fig:resonanceseries} show exactly these desired properties. \\
Note, as argued above, that these results depend on the degree of hydration $\eta$, and so will change during the dehydration of the ionic flow in the channel. \\
The DDHO equation \ref{DDHO} can also be derived in the context of the Brownian Dynamics Framework, see Appendix \ref{AppendixBDddho}. 
However, BD is not able to account for the required external harmonic driving forces.  

%%%% SS: DDHO consequences in 1-2-3D physical systems
\subsection{Consequences of the DDHO Mechanism for Physical Systems of $N$ Spatial Dimensions}

DDHO is a powerful and ubiquitous mathematical principle that is not bound to the spatial dimension of the acting physical system. 
For this reason it applies equally well to the 1D narrow linear selection filters found in various ion channel architectures, e.g. KcsA, as to the 3D-spacious helically twisted ELIC and GLIC channels.
All the resonance characteristics presented in Figures \ref{fig:MahalVVQeff}-\ref{fig:resonanceseries} and Tables \ref{table:TableResPeaks}-\ref{table:TableResAllMahalDists} therefore apply equally to these systems, as they are independent of the dimension and any physical characteristic - besides the DDHO-parameters - of the underlying system. 
More dimensions of the system only offer more options for resonance, {\em not} more resonance peaks.  
In this Section we explore how the DDHO mechanism manifests itself in physical channel structures of different spatial dimension.

\subsubsection{Resonance-driven selection in 1D-Ion Channels}
\begin{figure}
\includegraphics[width=0.4\textwidth]{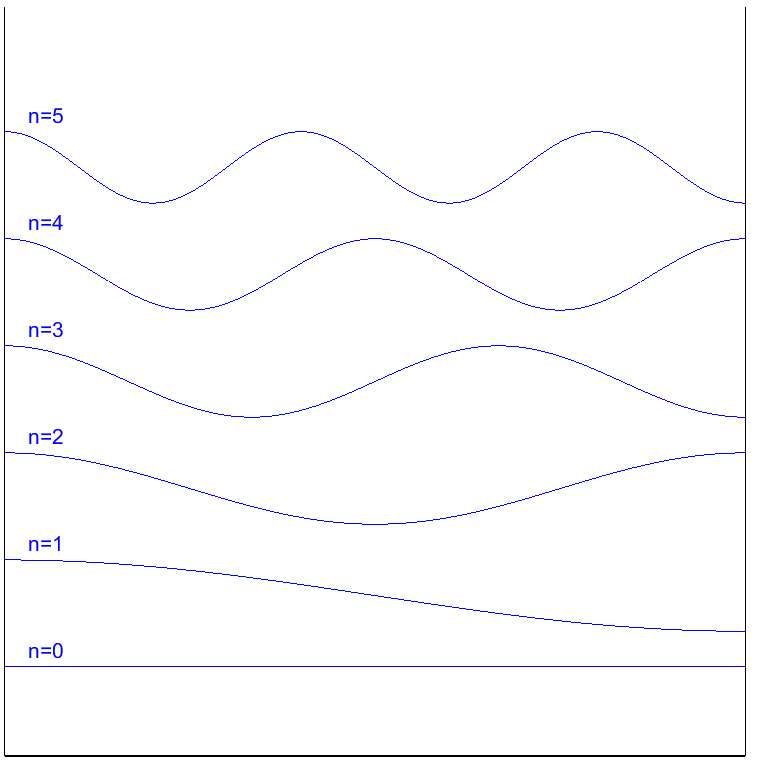}
\caption{Stacked series of resonances $\lambda_n=2L/n$ in a two side open-ended pipe of length $L$ for mode-number $n=0\cdots 5$.}
\label{fig:stackresonance1D2sideOpipe}
\end{figure}

\begin{figure*}
\begin{center}
\includegraphics[width=0.9\textwidth]{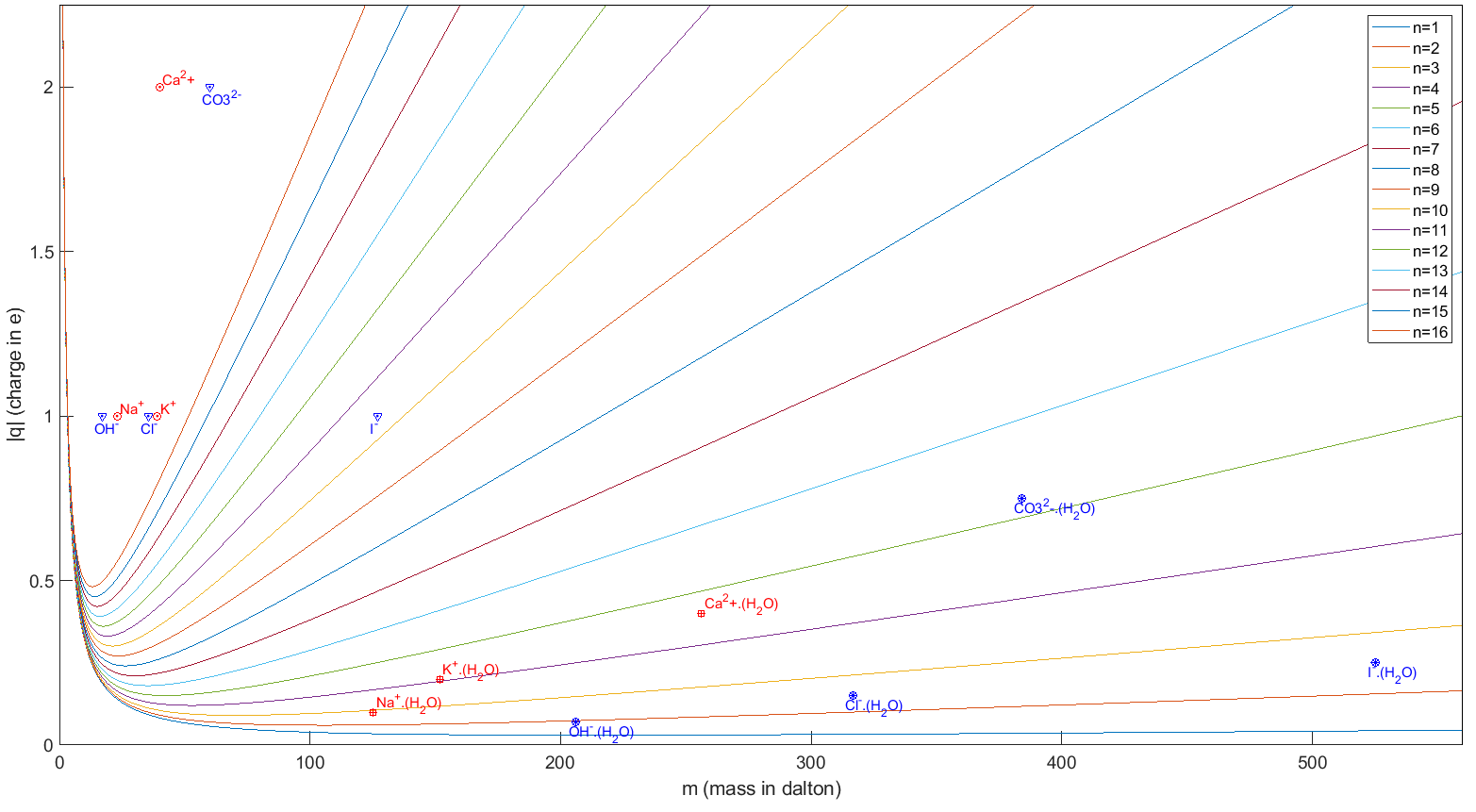}
\caption{Spectrum of mass-charge resonance curves for standing particle density waves in 1D two-side open-end tube, $n$ is the mode-number. Markers indicate various bare or aqua-ions. DDHO-parameters of equation \ref{eq:DDHO1Dspectrum}: $\kappa$ = 1.60 pN/e\AA, $\gamma$ = 1.3E-14 kg/s, $L$ = 12.0 \AA, $c$ = 10.00 m/s, $q$=e=1.6E-19 C. Mode number of the curves increase upward from $n=1$ to $n=16$}
\label{fig:1D_resonances}
\end{center}
\end{figure*}

First, let us study how the DDHO mechanism operates in 1D selection filters, e.g. the tube-like SF of KcsA in Fig. \ref{fig:KcsAporeradius}. For this purpose, we compare these narrow tubes to a two-sided open-ended cylinder, and approximate its dynamics with the 1D-EAH equations \ref{eq:EAH1Du}-\ref{eq:EAH1DphiE}, defined in Section \ref{sec:SEC1DEAHdyn}. 
Applying the DDHO-framework requires a stable 1D driving force mechanism. 
Otherwise, energy dissipation perturbs the system and the modes decay in time, known as quasinormal modes \cite{KON2011}.
We propose that this stability is provided by the oscillating particle density waves, observed in Section \ref{sec:standing1Ddensitywaves}.
We note that a tube-like SF acts as the ideal setting for the classical case of standing waves in a two-sided open-ended cylinder. The medium for these waves can be either particle density waves in the metaphor of microscopic perspective, or longitudinal pressure waves in the macroscopic picture of the ionic fluid. 
In the classical two-side open-ended cylinder, the wavelength $\lambda$ for a standing wave must relate to the tube length $L$ such that: $n\lambda/2=L$, for some integer $n=1,2,\dots$. This gives rise to a spectrum of possible resonance frequencies: $\Omega_n = n\pi c/L$, where $c$ is the wave velocity (not equal to the particle speed $u$). The first five eigen-resonances are depicted in Fig \ref{fig:stackresonance1D2sideOpipe}.  
%%%%% FIGURE resonance series
%%%%%%%%%%%%%%%%%%%%%%%%%%
These pressure waves act as driving force for a DDHO-mechanism, so, resonance occurs if for some $n$ the resonance frequency $\omega_R$ of equation \ref{DDHO_resomg} equals the driving frequency $\Omega_n$; so: $\omega_R = \Omega_n$, and therefore: 
\begin{equation}
    \frac{\kappa q}{m} - \frac{\gamma^2}{2m^2}=\frac{\pi^2c^2}{L^2}n^2
    \label{eq:DDHO1Dspectrum}
\end{equation}
This defines a set of spectral-resonance curves in the mass-charge plane $(m,q)$, indexed by the spectral mode-number $n=1,2,\dots$. Each spectral-resonance curve represents a case where a standing wave of frequency $\Omega_n$ fits with the resonance frequency $\omega_R$ for a particle of mass $m$ and charge $q$.
Fig \ref{fig:1D_resonances} depicts some spectral-resonance curves, namely for $n=1,2, \dots$, 16. In the same Figure, three cations and four anions are shown, both free and hydrated. The plot clearly shows that the ions are well separated, specifically, the hydrated and the free ions form distinct well-separated sub-clusters. Moreover, all the hydrated ions are located on different spectral curves, i.e. with distinct spectral mode number. This implies that a 1D-DDHO would be able to distinguish, hence uniquely select, each ion species. 
Thus, by selecting a specific mode number $n$, a specific hydrated ion can be targeted and selected.
Also note that this affects especially the hydrated ions and not the bare not-hydrated ions, which favours the dehydration process.
%%%% END of: SubSec: 1D-systems

\subsubsection{Vortex-driven Selection in 2D-Systems}
In two spatial dimensions, planar vortices, strictly isolated in an azimuthal plane, act as engine for the DDHO mechanism, see e.g. Fig \ref{fig:ELICflow}. 
The rotational frequency $\Omega$ of the vortex dictates the driving frequency in the DDHO. 
Thus, only those aqua ions with the right combination of mass and charge will resonate with - and adhere to - the vortex, until they dissolve their water shell in the dehydration zone, are released, and following the flow exit the ion channel.

\subsubsection{Toroidal vortices as driving engines in 3D-Systems}
\begin{figure}
\begin{center}
  \includegraphics[width=0.45\columnwidth]{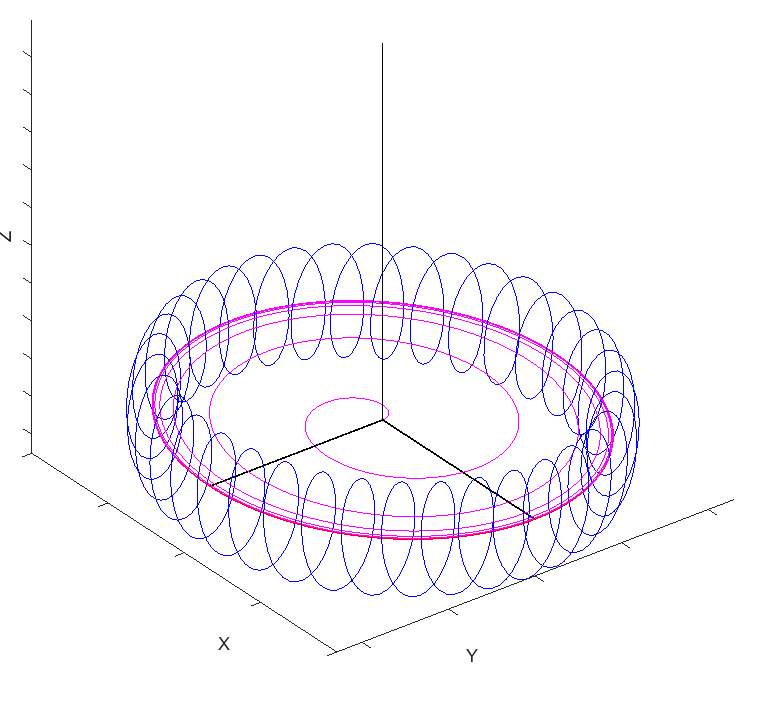}
  \caption{3D-motion of ions in a toroidal vortex: T: $\textit{transverse}$ mode i.e. converging to uniform rotation about the $z$-axis, and L: $\textit{longitudinal}$ mode as converging to helical-winding motion about the torus of the vortex}
\label{fig:Vortexmodes}
\end{center}
\end{figure}

In three spatial dimensions, stable 3D-connected toroidal vortices may appear, as depicted in various figures, e.g. Figs \ref{fig:ELICisobars} and \ref {fig: GLIC12ANIMS}(a), and in the supplemental material. These vortices act as the driving engine for the 3D-DDHO mechanism. 
This means that there are now {\em two} independent driving frequencies; the $\textit{transverse}$ T-mode, caused by uniform rotation about the $z$-axis, and the $\textit{longitudinal}$ L-mode, caused by the helically-winding motion about the torus of the vortex, see Fig \ref{fig:Vortexmodes}.  
As noted before, the two independent driving frequencies $\Omega_T$ and $\Omega_L$ do not offer more  resonance peaks, but more options for resonance to occur. Thus, they cover a broader spectrum simultaneously, comprising high and low frequencies.

%%%%%%%%%%%%% FIGURE vortex motion
%%%%%%%%%%%%%%%%%%%%%%%%%%%%

%%%%%%%%%%%%%%%%%%%%%%%%%%%%%%%%%%%%%%%%%%%%%%%%%%%%%%%%%%
%%%% SUBSECTION: APPLICATION of DDHO ON REAL DATA
%%%%%%%%%%%%%%%%%%%%%%%%%%%%%%%%%%%%%%%%%%%%%%%%%%%%%%%%%%
\subsection{Application of the Resonance-driven resonance model to real ion data}
\label{Appl-DDHO}
Let us now apply the DDHO selectivity model to real data of cations and anions. 
These results hold for any spatial dimension of the physical system.

%%% SubSec Resonance peak separation based on average hydration numbers
\subsubsection{Resonance peak separation based on average hydration numbers}

We selected published data that matches the environmental setting inside an ion channel as close as possible. Table \ref{table:TableResPeaks} lists information \cite{Maehler-2012}-\cite{CRC_ChPhys} for three species of cations: \{$\text{Na}^+$, $\text{K}^+$, $\text{Ca}^{2+}$\}, namely: the \textit{average} hydration number $\overline{n}_H$ with error d$ \overline{n}_H$, the atomic weight $M_0$ and error d$M_0$, and the average mass of a hydrated ion $\overline{M}_A = M_0+\overline{n}_HM_{H_2O}$, and its error d$\overline{M}_A$.  
Information for four anions: \{$\text{Cl}^-$, $\text{I}^-$, $\text{CO}_3^{2-}$, $\text{OH}^-$\} is listed in Table \ref{table:TableAnionPeaks} \cite{CRC_ChPhys}-\cite{Yadav2018}. We will study the ion selectivity for mixtures of these anions and cations. For $\text{H}_2{\text{O}}$ we used an atomic weight of $\text{M}_{H_2O}$ = (18.02 $\pm$ 1E-8) Da. All experiments were performed in the dimensionless DDHO variables $\xi$ and $Q$, with constant parameters $\kappa$=2, $\gamma$=1, amplitude $F$=5, and phase $\Phi$=0.\\
We ignore complexities involved with real aqua-shell topologies, such as multiple hydration layers, see \cite{Yadav2018}.\\
Note also that in the nano-scale domain the values of the physical parameters, including diffusion coefficient, viscosity, and conductivity, may differ from the bulk values we used here.  
%%%% TABLE I
\begin{table}[h!]
    \centering
    \begin{tabular}{ | l | l | l | l | l | l | l | }
    \hline
    cation	& $\overline{n}_H$ 	& d$\overline{n}_H$ 	&$M_0$ [Da] &d$M_0$ [Da]	& $\overline{M}_A$  [Da]	& $d\overline{M}_A$  [Da] \\
    \hline
    $\text{Na}^+$	&5.68	&1.21	&22.99 	&2E-8	  &125.32  	&21.80  \\
    $\text{K}^+$	&6.25	&0.43	&39.10 	&1E-4	  &151.69  	&7.75  \\
    $\text{Ca}^{2+}$	&12.0	&0.8	&40.08 	&1E-3	  &256.26   	&14.41  \\
    \hline
    \end{tabular}
    \caption{Properties of the species of cations used in the numerical simulations. Mass in daltons [Da].}
    \label{table:TableResPeaks}
\end{table}
%%%% TABLE Ib
\begin{table}[h!]
    \centering
    \begin{tabular}{ | l | l | l | l | l | l | l | }
    \hline
    anion	& $\overline{n}_H$ 	& d$\overline{n}_H$ 	&$M_0$ [Da] &d$M_0$ [Da]	& $\overline{M}_A$  [Da]	& d$\overline{M}_A$  [Da] \\
    \hline
    $\text{Cl}^-$	    &15.6  &1.65	 &35.45	 &1E-2 	& 316.56  &  29.73\\
    $\text{I}^{-}$		&22.1  &2.10	&126.9  &3E-5 & 525.15  &  37.84   \\
    $\text{CO}_3^{2-}$	    &78  &29	&60.00  &8E-3	&2484.0  &  522.0 \\
    $\text{OH}^-$	    &10.5  &2.78	&17.01  &4E-4	&206.22  &  50.10 \\    \hline
    \end{tabular}
    \caption{Properties of four species of anions used in the numerical simulations. Mass in daltons [Da].}
    \label{table:TableAnionPeaks}
\end{table}

\begin{figure*}
\begin{center}
\includegraphics[width=0.9\textwidth]{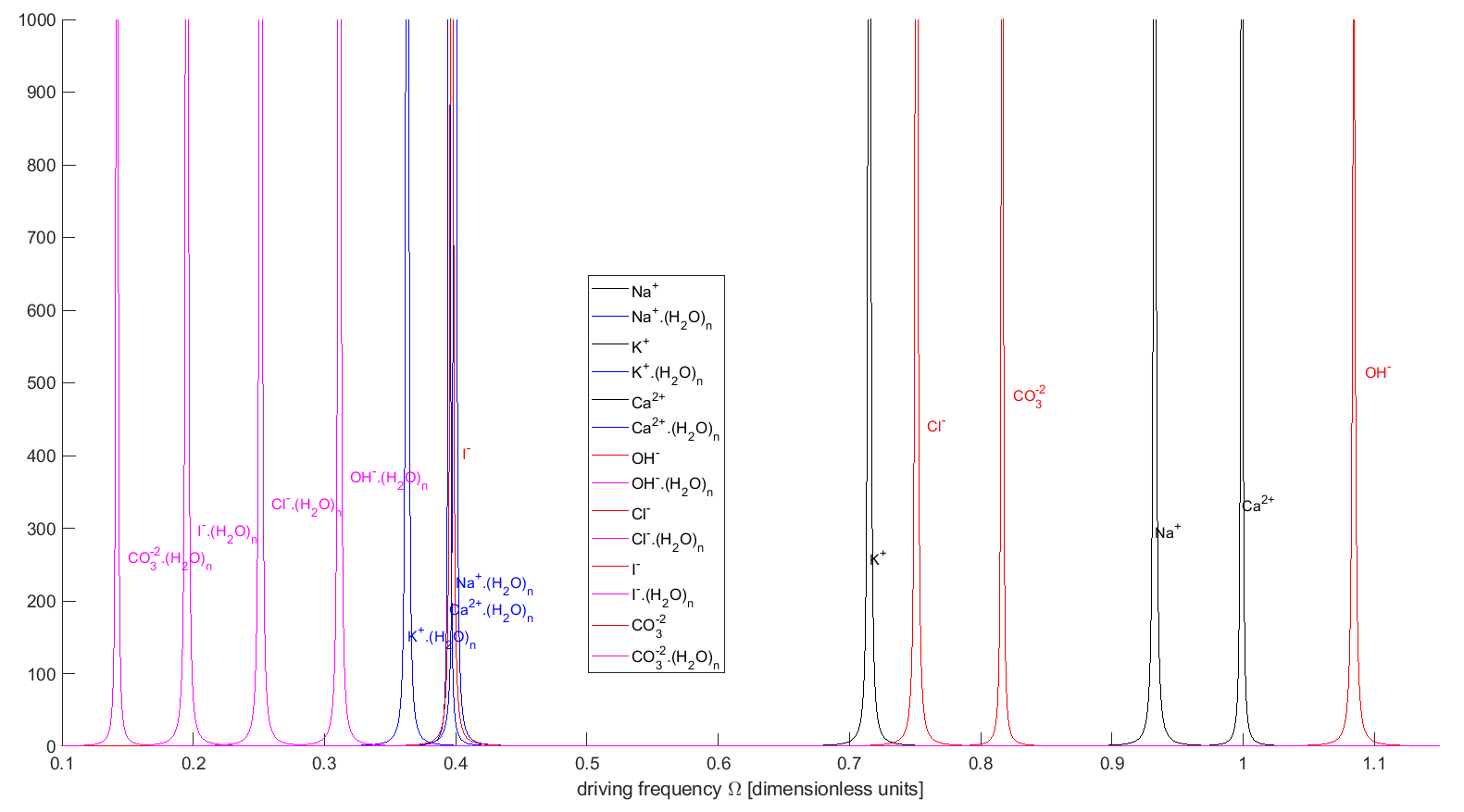}
\caption{Resonance peaks of various bare ($n_H$=0) and averagely hydrated ($n_H$=$\overline{n}_H$) cations and anions, without electrostatic screening: $q_A$=$q_B$. At higher resolution the seemingly overlapping peaks of $Ca^{2+}.(H_2O)_n$, $Na^{+}.(H_2O)_n$, and  $I^-$ become separated.}
\label{fig:resonancepeaks}
\end{center}
\end{figure*}

\begin{figure}
\includegraphics[width=0.4\columnwidth]{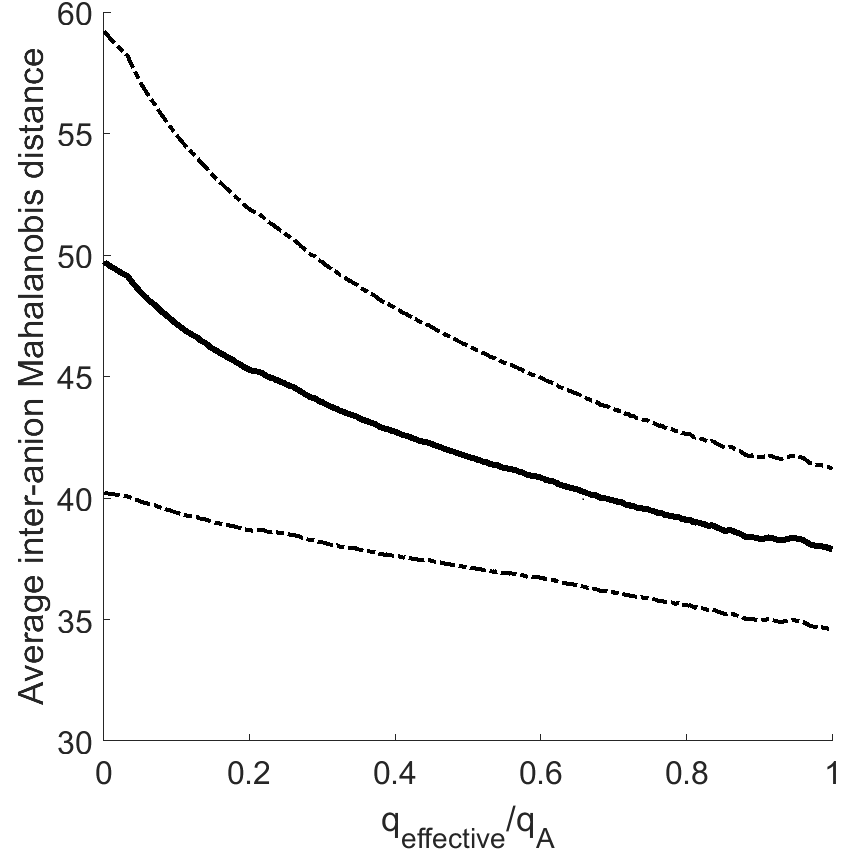}
\caption{Mahalanobis separation as function of the effective charge $q_A/q_B$ (solid line). Dotted lines give upper and lower 1$\sigma$ error boundaries.}
\label{fig:MahalVVQeff}
\end{figure}

Using this data we plot the ion energy $E_{\max}$ from equation \ref{eq:TotEnergy} versus the driving frequency $\Omega$. Note that this holds for both the $T$ and the $L$ mode. The resulting graph, for the bare and aqua anions and cations with their corresponding average shell number $\overline{n}_H$, is plotted in Fig \ref{fig:resonancepeaks}.  
%%%% FIGURE average resonance curve
%%%%%%%%%%%%%%%%%%%%%%%%%%%%%%%%%%%%%%
We observe that the resonance peaks for different ions are clearly separated. We define the separation between two peaks with centers $\omega_1$ and $\omega_2$ and widths $\Delta\omega_1$ and $\Delta\omega_2$ respectively as the dimensionless normalized Mahalanobis distance \cite{Mahal-1936}: $\Delta\omega_{12} = |\omega_2 - \omega_1|/\sqrt{(\Delta\omega_1^2 + \Delta\omega_2^2)}$. A peak separation $\Delta\omega_{12}$ larger than 1 means that the peaks are well separated, for values less than 1 they partly overlap, for $\Delta\omega_{12} = 0$ they coincide. Table \ref{table:TableResAllMahalDists} lists the Mahalanobis separations between resonance peaks of various aqua- and bare cations and anions for the DDHO model as depicted in Fig \ref{fig:resonancepeaks}. This table shows that all these separations (for our data) are larger than 1, most being significantly larger. The average Mahalanobis separation is: for cations only: $\overline{\Delta} =  28.5 \pm  20.1$, for anions only: $\overline{\Delta} =  48.7 \pm  25.8$, for a mixture of all ions in our dataset  combined: $\overline{\Delta} =  42.1 \pm  29.1$. \\
Electrostatic screening is not included in any of these calculations, i.e. $q_A = q_B$, as no reliable data on the effective charge of an aqua-ion in our context was found. The inter-ion Mahalanobis separation increases with decreasing effective charge $q_A/q_B$ as shown in Fig \ref{fig:MahalVVQeff}. Thus, any amount of screening of the dipole water molecule envelope will only strengthen the ion type selectivity.

%%%%%% TABLE IIa
\begin{table}[h!]
    \centering
    \begin{tabular}{ | l | l | }
    \hline
    Species-to-species & Mahalanobis distance  \\ 
    \hline
%CATION DATA: \\ 
$\Delta[Na^+,Na^+.(H_2O)_n] $ & 30.54 \\
$\Delta[Na^+,K^+] $ & 10.90 \\
$\Delta[Na^+,K^+.(H_2O)_n] $ & 32.74 \\
$\Delta[Na^+,Ca^{2+}] $ & 3.35 \\
$\Delta[Na^+,Ca^{2+}.(H_2O)_n] $ & 31.14 \\
$\Delta[Na^+.(H_2O)_n,K^+] $ & 29.96 \\
$\Delta[Na^+.(H_2O)_n,K^+.(H_2O)_n] $ & 9.17 \\
$\Delta[Na^+.(H_2O)_n,Ca^{2+}] $ & 58.49 \\
$\Delta[Na^+.(H_2O)_n,Ca^{2+}.(H_2O)_n] $ & 1.31 \\
$\Delta[K^+,K^+.(H_2O)_n] $ & 33.76 \\
$\Delta[K^+,Ca^{2+}] $ & 20.17 \\
$\Delta[K^+,Ca^{2+}.(H_2O)_n] $ & 31.35 \\
$\Delta[K^+.(H_2O)_n,Ca^{2+}] $ & 62.72 \\
$\Delta[K^+.(H_2O)_n,Ca^{2+}.(H_2O)_n] $ & 10.66 \\
$\Delta[Ca^{2+},Ca^{2+}.(H_2O)_n] $ & 60.91 \\
\hline
%ANION DATA: \\ 
$\Delta[OH^-,OH^-.(H_2O)_n] $ & 33.07 \\
$\Delta[OH^-,Cl^-] $ & 12.89 \\
$\Delta[OH^-,Cl^-.(H_2O)_n] $ & 35.68 \\
$\Delta[OH^-,I^{-}] $ & 29.26 \\
$\Delta[OH^-,I^{-}.(H_2O)_n] $ & 38.12 \\
$\Delta[OH^-,CO^{-2}_3] $ & 11.06 \\
$\Delta[OH^-,CO^{-2}_3.(H_2O)_n] $ & 40.45 \\
$\Delta[OH^-.(H_2O)_n,Cl^-] $ & 38.71 \\
$\Delta[OH^-.(H_2O)_n,Cl^-.(H_2O)_n] $ & 24.79 \\
$\Delta[OH^-.(H_2O)_n,I^{-}] $ & 24.11 \\
$\Delta[OH^-.(H_2O)_n,I^{-}.(H_2O)_n] $ & 52.93 \\
$\Delta[OH^-.(H_2O)_n,CO^{-2}_3] $ & 73.54 \\
$\Delta[OH^-.(H_2O)_n,CO^{-2}_3.(H_2O)_n] $ & 88.12 \\
$\Delta[Cl^-,Cl^-.(H_2O)_n] $ & 44.22 \\
$\Delta[Cl^-,I^{-}] $ & 30.54 \\
$\Delta[Cl^-,I^{-}.(H_2O)_n] $ & 49.40 \\
$\Delta[Cl^-,CO^{-2}_3] $ & 5.03 \\
$\Delta[Cl^-,CO^{-2}_3.(H_2O)_n] $ & 54.37 \\
$\Delta[Cl^-.(H_2O)_n,I^{-}] $ & 43.41 \\
$\Delta[Cl^-.(H_2O)_n,I^{-}.(H_2O)_n] $ & 30.21 \\
$\Delta[Cl^-.(H_2O)_n,CO^{-2}_3] $ & 83.49 \\
$\Delta[Cl^-.(H_2O)_n,CO^{-2}_3.(H_2O)_n] $ & 71.58 \\
$\Delta[I^{-},I^{-}.(H_2O)_n] $ & 63.16 \\
$\Delta[I^{-},CO^{-2}_3] $ & 57.86 \\
$\Delta[I^{-},CO^{-2}_3.(H_2O)_n] $ & 84.61 \\
$\Delta[I^{-}.(H_2O)_n,CO^{-2}_3] $ & 92.86 \\
$\Delta[I^{-}.(H_2O)_n,CO^{-2}_3.(H_2O)_n] $ & 46.75 \\
$\Delta[CO^{-2}_3,CO^{-2}_3.(H_2O)_n] $ & 102.11 \\
\hline
\end{tabular}
\caption{Mahalanobis distances of inter-cation (upper) and inter-anion (lower) resonance peaks for $q_A/q_B=1$}
\label{table:TableResAllMahalDists}
\end{table}

%%% SubSec Resonance peaks of the full hydration series
\subsubsection{Resonance peaks of the full hydration series}

\begin{figure}
\begin{center}
\includegraphics[width=0.4\columnwidth]{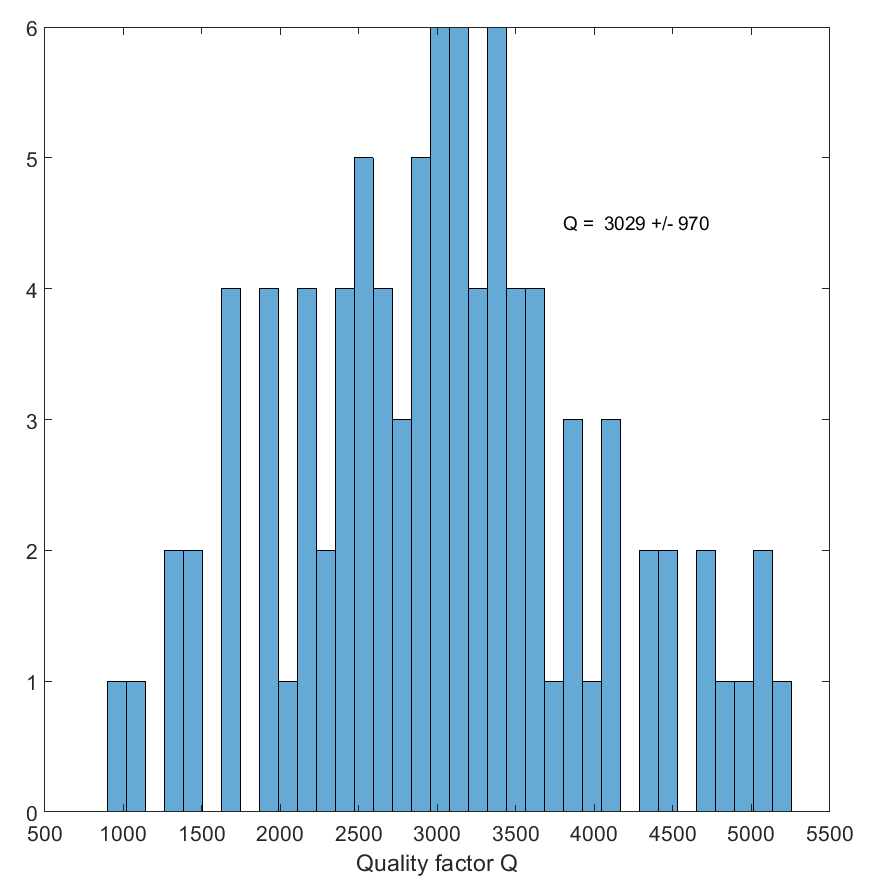}
\caption{Histogram of oscillator quality $Q$ for the hydration series: $n_H$=0,$\ldots$,12 of the 6 ions of the dataset, with $q_A=q_B$.}
\label{fig:QHist}
\end{center}
\end{figure}

\begin{figure*}
\begin{center}
\includegraphics[width=0.95\textwidth]{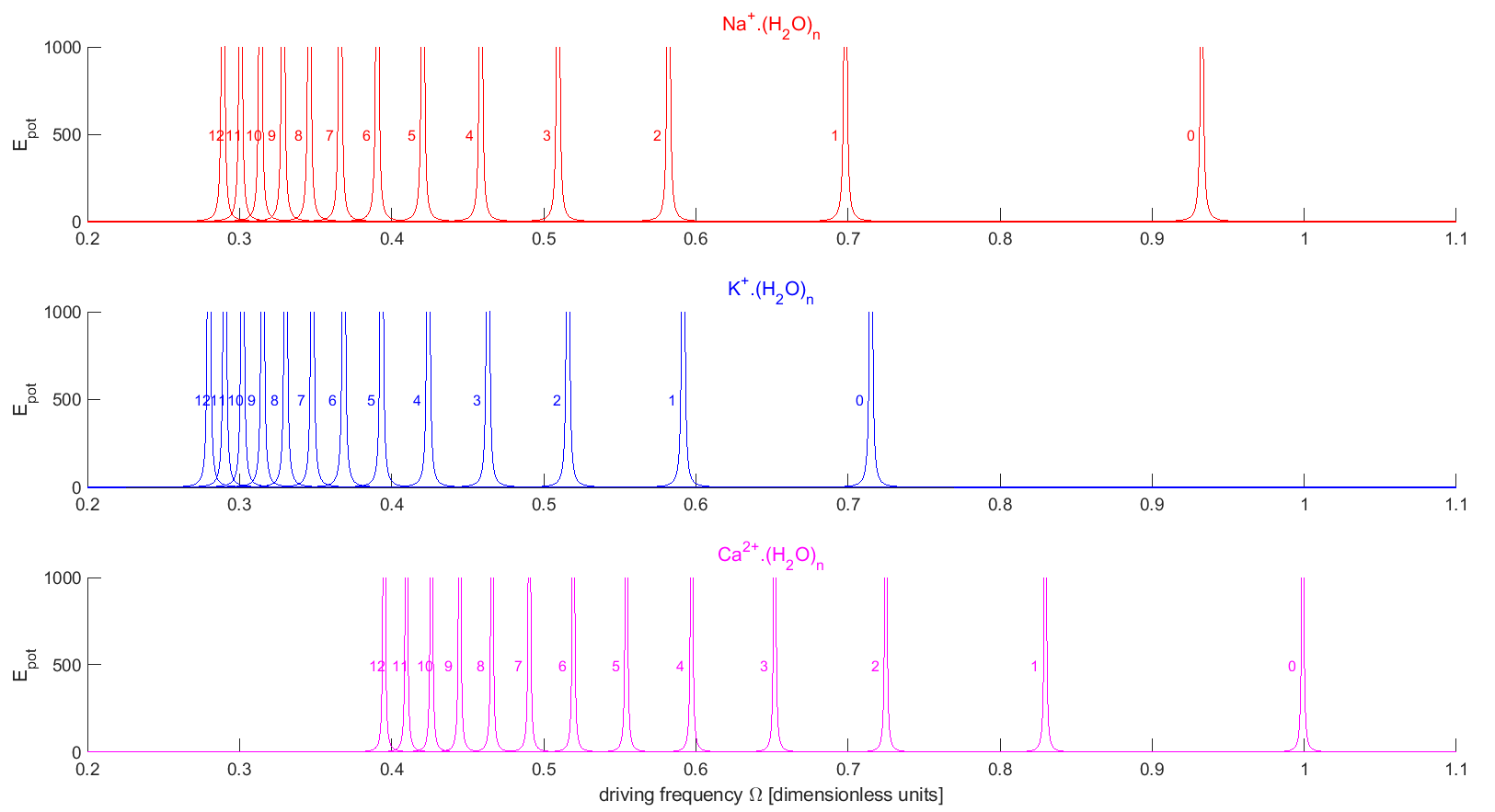}
\caption{Resonance peaks of the dehydration stages for the cations Na, K, and Ca respectively. $n$ is the hydration number, assuming an effective charge $q_A/q_B = 1$. In these dimensionless units the required driving frequency $\Omega$ of the series of Na and K converges to 0.0142 for $n\rightarrow\infty$, and for Ca it converges to  0.0201.}
\label{fig:resonanceseries}
\end{center}
\end{figure*}

Rather than considering the resonance peaks for the \textit{average} hydration number $\overline{n}_H$, we can study the resonance peaks for the \textit{entire} series of hydration of an ion, i.e. the series of peaks for hydration numbers $n_H=0$ (i.e. fully dehydrated, $\eta = 0$) to $n_H=12$ (fully hydrated, $\eta=1$). This is depicted for the three cations $\text{Na}^+$, $\text{K}^+$, and $\text{Ca}^{2+}$ in Fig \ref{fig:resonanceseries}, again assuming $q_A = q_B$. This figure clearly shows that the hydration series of one species are clearly separated, though occasionally they can overlap with a peak of another species. 
%%%%% FIGURE resonance series
%%%%%%%%%%%%%%%%%%%%%%%%%
Due to this cluttering of the peaks, the average Mahalanobis distance of the entire resonance series of the seven ions in our dataset drops considerably. For a series of 13 hydrations ($n_H=0,\ldots,12$) for the seven ions we obtain 91 partly overlapping peaks, giving 4095 pairs of peaks with average $\overline{\Delta} = 1.11 \pm 0.53$, i.e. barely separated. However, this result should be corrected for the average hydration number $\overline{n}_H  \pm \delta n_H$ per ion species. Weighing the Mahalanobis distance with a normal distribution $N(\overline{n}_H ,\sigma )$, favors the resonance peaks near the average hydration. This leads to $\overline{\Delta} = 5.1 \pm 6.1$, for $\sigma = 3\delta n_H$. This means that a mixture of various anions and cations still exhibits well-separated resonance peaks \footnote{We constantly used $q_A=q_B$ in all these calculations}. Note that this decrease in $\overline{\Delta}$ stems mostly from the {\em intra-species} cluttering, while the {\em inter-species} separation is {\em not} affected, and remains high:  the unweighted average  species-to-species resonance series separation is:  $\overline{\Delta} = 29.7 \pm 16.8$. \\
The 91 narrow and high peaks of the full resonance series display a high oscillator quality $Q$, defined above, namely: $Q = 3029 \pm 970$, see Fig \ref{fig:QHist}.
%%%%% FIGURE Q-hist of resonance series
%%%%%%%%%%%%%%%%%%%%%%%%%%%%%%%%%%%%%%%%%%%%
The dynamical evolution of the resonances - their location, magnitude, and angular frequency, observed in the EAH-model, can drive the fully hydrated ions through these cascade of resonance peaks until the last one is reached, and the ion is fully dehydrated. 
Figs \ref{fig:resonancepeaks} and \ref{fig:resonanceseries}, and Table \ref{table:TableResAllMahalDists} show that the proposed resonance-driven resonance mechanism for ion specificity indeed allows for targeted selection of ions, reminiscent of the LC-circuit as basis for the channel selectivity of classical analog radio receivers.
%%%%%%%%%%%%%%  END SECTION DDHO

%%%%% FIGURE resonance series

%%%%%%%%%%%%%% %%%%%%%%%%%%%% 
%%%%%%%%%%%%%% SECTION Discussion
%%%%%%%%%%%%%% %%%%%%%%%%%%%% 
\section{Discussion}
\label{SEC-Discusion}

The classical physical mechanism of the damped harmonic oscillator, driven by the natural frequencies inside a stable turbulent structure in the dehydrating ionic flow through an ion-channel, proves to be extremely efficient in uniquely selecting one specific type of ion.
This mechanism requires a strong harmonic driving force which current frameworks for ion channel transport cannot offer. Therefore, we have proposed the new continuum EAH model for the transport - and specifically the (de/re)hydration - of ions through an ion channel. 
The EAH model provides numerical support that ion transport is accompanied by strong and dynamic turbulences, each with its specific natural frequency. 
These turbulent structures appear and disappear depending on the local degree of hydration, and vice versa strongly influence the distribution of hydration throughout the channel. 
New in our framework is that this also depends sensitively on the geometry of the channel; this defines the location of the standing pressure wave or the position of the toroidal vortices. \\
Analysis of experimental patch-clamp data and current-voltage relations in ELIC with the theoretical predictions of the EAH framework in Section \ref{sec:SecCompTheoExp} revealed statistically significant correlations. 
EAH can also successfully predict current-voltage relationships of single ELIC channels at different input and output cation concentrations, and of binary mixtures of monovalent and divalent cations. 
Similarly, it can be employed to compute absolute and relative cation permeabilities and the Nernst reversal potential.
The wide variety of patch clamp morphologies produced by EAH for different parameter settings offers perspective on elucidating microscopic mechanisms, viz. spinning versus fixed ion channels. \\ 
The EAH formalism contributes to understanding ion selectivity, as discussed in Section \ref{EAHvsKOPEC}, by assuming different amphiphilic charges to different cations, and so different energies for releasing water molecules.
This holds even more when augmented by higher-order tensorial terms, representing hydrophilic moments.\\
Cyclic motion inside a turbulence alone cannot distinguish between ion species as it affects all ions captured in its current equally.  
However, in a mixture of various species, those ions that can dehydrate fastest dominate the outflux of dehydrated ions and thus determine the ion channel type. 
This gives leverage for the DDHO mechanism.
The EAH model provides all the required ingredients of the mechanism of the DDHO: an elastic restoring force, a damping and a harmonic driving force and driving frequency.
The turbulence itself only provides the harmonic driving force and frequency, the other ingredients depend on ion mass, charge, and degree of hydration.
These fundamental parameters make that the resonance peaks are substantially different for different ions. \\
The DDHO equation \ref{DDHO} describes perturbations on the regular flow pattern, resulting in fluctuations in the local particle density, hence pressure, that will propagate with the local speed of sound - which is higher than the local flow speed.
These sound waves are driven by the driving force and efficiently transfer their energy into those ions that resonate with the driving frequency. 
Consequently, these energized aqua-ions instantly shed off their aqua shell and, as bare ions, subject to the electric forces and dominant currents, quickly exit the ion channel into the cell interior.\\
The response curves, the relation between energy transfer and driving frequency as in Figs \ref{fig:resonancepeaks} and \ref{fig:resonanceseries}, prove to have small half-width $\Delta\omega$ and a high oscillator quality $Q$. 
Numerical experiments with seven different ions, cations and anions, yielded response curves with sharp resonance peaks that were widely separated with average Mahalanobis separation $\overline{\Delta}$ = 42.1 $\pm$ 29.1. The response curves also exhibited high oscillator quality $\overline{Q}$ = 3029 $\pm$ 970, indicative of a strongly persistent oscillator. \\
In narrow tube-like selection filters, the 1D-EAH-equations predict a Hopf-bifurcation and a cascade of period-doubling bifurcations {\em en route} to a chaotic regime. Modeled as a two-sided open pipe, the energy inside will show resonances associated with eigen-frequencies related to the length of the pipe. These can act as driving forces in a 1D-DDHO-mechanism, provided they match the resonance frequency of an aqua-ion. Consequently, this aqua-ion type has a higher probability to hydrate. \\ 
In more spacious segments of the ion channel, e.g. the central cavity of ELIC and GLIC, vortices in the ionic flow can act as engines of the 2D/3D-DDHO and promote dehydration. \\
The magnitude of the dehydrated ionic outward flux varies with the setting of the model parameters, and the experiments show that the parameter space consists of various local maxima. 
The EAH-DDHO model provides insight in the effect of genetic-induced damage, as these easily destroy the subtle balance between geometric shape, force magnitudes, and flow parameters - and thus destroy or impede the formation and evolution of the self-organizing oscillations, and so the driving engine at the basis of the selection mechanism. 
This matches the recent observation in the Propane study \cite{Kapetis-2017} that point mutations (SNPs) that affect the architecture, especially the grooves of the interior walls, can potentially be severe. Also observed is that these SNPs can be remarkably isolated on the DNA; some ligand positions on the protein subunits may cause severe phenotypic effects while neighboring ligands have no measurable effect. 
In our simulations we find that indeed changing the geometry of the wall, including removing existing or introducing  novel grooves in the wall, gives entirely different flow patterns and resulting vortices. 
\\ 
Our framework adds support to the hypothesis of spinning ion channels \cite{Shaw-2013}, as fast rotation induces strong vortices and therefore strong resonances and consequently fast and effective target-ion yield. Similarly, the observation that smoothing the interior walls impedes vortex formation, and thus removes the driving force in the DDHO mechanism, adds more support, as these grooves act as the fulcrum on which the ion flow acts as the lever for the mechanical torque that drives the rotation.
Yet, note that our model does not depend on this hypothesis, because even without rotation it predicts vortices in the ionic flow.
\\
An essential element in our approach is the application of a macroscopic continuum model rather than one based on microscopic and discrete events. 
The main motivation herein is to detect and identify spatiotemporal turbulences that can act as driving force of a DDHO mechanism. 
This triggers two questions. 
First, ergodicity breaking questions the validity of the continuum assumption vital to a mean-field derivation as phase space becomes locked to the subspace of its attractor. However, this concerns the evolution of one specific run, and does not violate the derivation of the EAH equations itself over the grand ensemble of all possible configurations.
Second, the high degree of stochastic fluctuations and low density inside the ion channel challenges the existence of turbulent structures on the nanoscale scale.
Boiteux et al \cite{REF1SUG4} argue that there are typically only 20 - 40 water molecules inside the hydrophobic cavity at a time, and that dehydrated ions move in a concerted single file through the selectivity filter. 
However, the fore-mentioned high rate of transport through an ion channel, exceeding $10^7$ ions per second, fully legitimizes a macroscopic 3D mean field model over realistic time-scales. In this respect, the  Navier-Stokes equations provides an established physical model, mathematically well founded on the underlying microscopic dynamics \cite{Bisi-2014}-\cite{Zhakin-2013}.
The single file transport also does not pose a problem, as this is covered by the 1D-EAH equations, and reveals pressure oscillations in the SF of KcsA.
The turbulences; standing waves and toroidal vortices, in this view are more related to pressure and density waves, and do not manifest themselves in the individual particle trajectories. This is also why they pass unnoticed in MD, PNP, and BD, as these do not include fields to represent pressure $P$ and velocity $\textbf{u}$.
Finally, a macroscopic continuum approach is also consistent with computational physics that shows that consistency with continuum theories emerges quickly, viz. the ideal gas law appears for a few dozen hard core molecules. \\
The most common type of ion channel has a long and narrow tubular selectivity filter, viz. in KcsA. 
In such structures particles move in a single file \cite{Corry2018}\cite{Kopec2018}\cite{REF1SUG5}\cite{REF1SUG6}\cite{REF1SUG8}.
For this type, EAH predicts self-organizing standing pressure waves in the SF. These oscillations act as driving engine for 1D-DDHO selection.
Other types of ion channels exhibit spacious selectivity filters, such as in the ELIC and GLIC. Here, the EAH-equations predict stable vortices, not in the SF, but in the central cavity, as depicted in Fig \ref{fig:ELICflow}. These vortices act as harmonic driving forces of a 2D/3D- DDHO selection mechanism.\\
The role of water molecules has been shown to be a crucial factor in the de/re-hydration process \cite{REF1SUG7}. Some authors explain molecular selectivity with adsorption properties in nanotubes \cite{REF1SUG8}\cite{REF1SUG9}.
In our study, water act as a background continuum, implicit in the EAH-equations in the conservation of the mass density, momentum and energy.
The energy step required for (de/re)hydration is represented by the term $a\nabla\phi_A$ in EAH-equation \ref {eq:EHANS2}, as described in Section \ref {sec:AmphInteractions}. 
This term alone already selects for ion type, so defines the selectivity-type of the channel. This is similar to the explanation of K$^+$--selectivity in certain ion channels by Kopec et al. \cite{Kopec2018} as described in Section \ref{EAHvsKOPEC}.
\\ 
A final question concerns the exact subject of the resonance; is it the bare ion or the aqua-ion? 
'Aqua-ion', however, is a collective term that encompasses a whole sequence of different amounts of water molecules captured in its shell, as depicted in Fig \ref{fig:resonanceseries}. 
Only on average it contains $n_H$ water molecules with a wide standard deviation $\delta n_H$.  
We propose that it is the average aqua-ion that is targeted. 
Firstly, the resonance peaks are very sharp and high, and therefore the aqua-ion is vastly injected with an enormous amount of energy that instantly kicks-off {\em all} of its water shell molecules. 
The then free and more electrically charged bare ion moves promptly, subject to the forces in the fluid - or statistical collective - out of the ion channel and into the cell. 
Secondly, the variation in solvation shell number $n$ is a statistical steady-state distribution $P_{EQ}(n,T)$ with average at $n_H$.
Rather than equation \ref{eq:EHANS0} suggests, only one water molecule at a time is lost or gained.
This means that a loss at $n=n_H$ is quickly compensated from the interval of $n$'s directly adjacent to $n_H$ in the distribution, so, by aqua-ions not affected by the resonance due to the sharpness of its peak. \\
The proposed EAH-DDHO mechanism for ion selectivity has consequences for computational cell models.
The outflow of ion channels is namely not an on- or off-switch, as the EAH-DDHO system has to build up till the moment that it is ready to produce the requested ions, causing a certain dead-time before the maximum output can be reached.
This, however, is very fast and possibly unnoticeable for the typical time-scales of current cell-models. 
Furthermore, our simulations show that the pressure gradient $\nabla P$ in equation \ref{eq:EHANS1} strongly determines the onset of turbulence and thus the resulting ion selectivity. 
This is relevant especially for cardiac myocyte models as these are substantially contracted during the cardiac cycle, causing high pressure gradients over the cell wall. \\
In future work we will incorporate the model in existing myocyte and neuron cell models \cite{Heijman-2013}\cite{Kapetis-2017} to explore its effects on multiscale dynamical systems from cell to whole organ.  
Furthermore, the EAH and DDHO models allow experimental tests to compare theoretical predictions with empirical data. The key experimental target here is to confirm that, in bulk transport, ions collectively behave as a physical fluid.   \\
%%%%%%%%%%%%%%% END SECTION discussion

%%%%%%%%%%%%%%%%%%%%%%%%%%%%%
%%%%%%%%%% SECTION Conclusion
%%%%%%%%%%%%%%%%%%%%%%%%%%%%%%
\section{Conclusions}
\label{SecCONC}

We have presented a simple yet adequate model for explaining ion channel selectivity. 
The model explains selectivity in terms of the resonance frequencies generated by a mechanism of the damped harmonic oscillator driven by periodic forces provided by strong persistent turbulent structures in the dehydrating ion currents through the ion channel. 
These turbulences are manifest as standing pressure waves in 1D linear tubular selection filters and toroidal vortices in 2D and 3D spacious cavities.
Their characteristics and dynamics depend sensitively on the geometry and the electrostatic and amphiphilic coding of the ion channel interior walls. 
As these are directly transcribed from the genes involved, it offers natural selection leverage to adapt to the animal's lifestyle - but it also causes susceptibility to harmful SNP mutations.  \\
The turbulences are modeled in a novel macroscopic ionic transport framework that includes spatiotemporal fields for velocity and hydration.
This theory correlates statistically significant with experimental patch clamp data of prokaryotic ELIC ion channels.\\
The resulting resonance frequencies are functions of the ion mass, charge, and hydration level, and therefore allow for uniquely differentiating ion species. 
Numerical analysis with empirical ion data show that the resonances are indeed well-separated and that they persist over many thousands of oscillation cycles.\\ 
In effect, our theory provides an explanation in purely physical terms for the remarkable precise ion selectivity that evolved in ion channels, a feat where all living organisms heavily rely on in their daily struggle for survival and procreation.
%%%%%%%%%%%%%%%%%% END of SECTION conclusion %%%%%%%%%%%%%%%%%%%%%%%%

\begin{acknowledgments}

\noindent
The author wishes to express sincere gratitude to Markos Xenakis, Bert H.J.M. Smeets, and Patrick J. Lindsey of the Faculty of Health, Medicine and Life Sciences, Translational Genomics (Maastricht University/The Netherlands) for their collaborative contributions within the European PROPANE project. Their insights and expertise were invaluable.
The author is also deeply grateful to Thomas Burwick (Frankfurt Institute for Advanced Studies, Germany) and Jordi Heijman (Maastricht University/The Netherlands, CARIM) for their insightful discussions and constructive feedback, which significantly enhanced the quality of this work.

\end{acknowledgments}

\appendix

%%%%% APPENDIX A weg
% \section{COMPUTATION OF OUTWARD CHARGE FLUX }
% \label{Appendix0}
% \noindent

\section{COMPUTATION OF TURBULENCE MARKERS}
\label{Appendix1}
\noindent
Here we compute various markers for the onset of turbulence for the environment inside the ion channels of our study: ELIC and GLIC. We model the ion channel as a cylinder and assume the following values for the physical parameters: \\  
 - ionchannel length:   L = 1E-08 [m]\\
 - ionchannel radius: R = 5E-10 [m]\\
- \# ions inside at any time: N = 30 , namely:\\
 - \# cations: 15 [ions]\\
 - \# anions: 15 [ions]\\
 - \# H$_2$O: 30 [molecules]\\
 - throughput-rate: tr = 1E+08 [particles/sec]\\
% - organic temperature: T = 310 K \\
%  - auto correlation time: 1E-08 [sec]\\
%  - energy dissipation fraction/molecule: 1 [eV]\\
Following C.Boiteux et al. \cite{REF1SUG4}, we assume about 20-40 ions in the cavity at any time. We take $N_{tot}$ = 30$\pm$10 = 15$\pm$5 Na + 15$\pm$5 Cl. In addition we assume 30 H$_2$O molecules.
As model we take the values of our prototype ion channels ELIC and GLIC, and model the central cavity as a cylinder of radius R = 5\AA, and length L = 100\AA, so volume $V\approx\pi R^2 L =$ 7.9E-27m$^3$, so particle density $n[Na] = n[Cl] \approx $ 1.91E+27 m$^{-3}$. For the other variables, we choose: T=310K (organic temperature), q = +1e (cation, Na+), q = -1e (anion, Cl-).
The average ion speed thus becomes: $u =  tr\times L/ N_{tot} \approx$ 0.017 m/sec. \\
This ELIC/CLIC environment setting gives the following values for the turbulence-markers: \\ 
* Debye length = 3.41E-11 m \\
* Kolmogorov scale = 2.51E-12 m \\
* Reynolds number = 6052\\
%* Froude number = 1.26\\ \\
The detailed calculations are provided below.

\subsection{Debye length  inside an ion channel}
\noindent For an electrolyte or plasma at temperature $T$, with - per charge carrier $\#k$ - particle density $n_k$ and charge $q_k$, the Debye length $\lambda_D$ is:
$$\lambda_D = \sqrt{\frac{\epsilon_0\epsilon_rk_BT}{\sum_kn_kq_k^2}}$$ 
We follow the approach from: Corry, Kuyucak, et al. \cite{REF1SUG1} and assume the dielectric constant is $\epsilon_r \approx 80$ inside the boundary water and 2 outside, which is representative of proteins forming ion channels. 
Applying these values in above equation, we find: $\lambda_D \approx$ 3.41E-11 m $\approx$ 0.34\AA, inside the central cavity.

\subsection{Kolmogorov length inside an ion channel}
\noindent The Kolmogorov length scale $\kappa$ is defined as: 
$$\kappa = (\frac{\nu^3}{\varepsilon})^{1/4}$$
with $\nu$ the kinematic viscosity of the ionic fluid and $\varepsilon$ the average dissipation of turbulent kinetic energy per unit mass. 
A microscopic theory for computing the kinematic viscosity is provided by the Einstein-Helfand theory, and is expressed by the Green-Kubo relations \cite{Kirova2015}\cite{Deen2014}:
$$ \nu = \frac{V}{k_BT}\lim_{\tau\to\infty}\int_{0}^{\tau}dt\langle\sigma(t)\sigma(0)\rangle $$
This involves the average auto-correlation of the microscopic shear stress tensor $\sigma^{\alpha\beta}(t)$. This has two contributions due to momentum transport: i. a kinetic term through particle motion and dissipated momentum due to inelastic collisions, and an interaction part caused by electromagnetic and chemical interactions. 
$$\sigma^{\alpha\beta} = \sum_{i=1}^N\frac{p^\alpha_ip^\beta_i}{m_i} - \sum_{i>j}^N(x^\alpha_i - x^\alpha_j)\frac{\partial\Phi_{ij}}{\partial x^\beta_j}  $$
where $p^\alpha_i$ is a momentum component, $\Phi_{ij}$ is the symmetric interaction potential, $m_i$ is a mass, $x^{\alpha}$ is a component of the radius vector, $i$ and $j$ are particle indices.\\
We assume that the auto-correlation $\langle\sigma(t)\sigma(0)\rangle$ decays exponentially with half-time $\tau_C$ from starting value: 
$\sigma_0^2 \equiv \sum_{\alpha\beta}\sigma^{\alpha\beta}(0)^2$. 
This estimates the viscosity as: 
$$\nu \approx \frac{V}{k_BT}\int_{0}^{\infty}3\sigma_0^2\text{exp}(-t/\tau_C)dt = \frac{3\sigma_0^2\tau_CV}{k_BT}$$ 
In order to estimate $\sigma_0^2$, we note that the off-diagonal elements of the stress tensor $\sigma^{\alpha\beta}$ consist of two parts: an ensemble sum of the kinetic term, and an ensemble sum over the distance-weighted gradient of the interaction potential. \\
The kinetic part either cancels if the momentum components are uncorrelated, or, if not, at most sums to $2N <E_{kin}>$. With $30 \pm 10$ ions moving on average 1 $\textsf{cm/s}$ this gives approximately $1 \times 10^{-28} \textsf{J}$, which we can ignore. The interaction part relates to the average particle distance: $\delta \approx (V/N)^{1/3} \approx 1.3 \times 10^{-10} \textsf{m}$. At these small distances the interaction forces are colossal. Yet, on average the interaction part is zero, because the antisymmetric factors $(x^\alpha_i - x^\alpha_j)\frac{\partial\Phi_{ij}}{\partial x^\beta_j}$ in the integrand on average cancel out, but with a huge variance (in numerical simulations stdev/mean $\approx$ 400). Because of this uncertainty we over-estimate $\sigma_0$ to 1. The relaxation time $\tau_C$ of the autocorrelation on atomic scale can be small. In the case of simple point-charge (SPC) water models \cite{Toukan1985} extremely small autocorrelation times are found in the order of 1 $\textsf{ps}$. We over-estimate $\tau_C = 10 \textsf{ns}$. Thus, for our setting using above values and expressions, we obtain: $\nu = \frac{3\tau_C\sigma_0^2V}{k_BT} \approx 2.8 \times 10^{-12} \textsf{m}^2/\textsf{s}$ for the kinematic viscosity. We roughly estimate the cumulative dissipation of turbulence kinetic energy to 1 $\textsf{eV}$ as a typical atomic scale energy value, and take a sodium bare ion with mass 23 $\textsf{Da}$. In this way, we find $\varepsilon = 4.2 \times 10^6 \textsf{J/kg}$. Applying all these values, we find a Kolmogorov length $\kappa = (\frac{\nu^2}{\varepsilon})^{1/4} \approx 2.5 \times 10^{-12} \textsf{m} = 0.025\textsf{\AA}$.

\subsection{Reynolds number inside an ion channel}
\noindent For a flow in a cylindrical pipe of length $L$, with kinematic viscosity $\nu$ and flow velocity $u$, the dimensionless Reynolds number Re is defined as:
$\text{Re} = uL/\nu$.
Using the ELIC/GLIC ion channel parameter setting and the value for $\nu$ obtained above, we find Re = 6052. \\

\section{The DDHO in the Brownian Dynamics Framework}
\label{AppendixBDddho}
%%%% SUBSEC BD resonances
In this study, we focus on 3D hydrodynamics models for ion channel transport that exhibit persistent vortices that act as driving force in the DDHO model. However, the DDHO is a template that fits all physical settings where there is: (i) a quadratic potential energy minimum plus (ii) a dissipative force plus (iii) a harmonic driving force, in any spatial dimension $n$D. In this way, the DDHO model also applies to theoretical frameworks such as deterministic continuum permeation models, many-particle system molecular dynamics, and stochastic Brownian dynamics that contain these ingredients.
For instance, in the context of a single file of co- and counter-ions interspersed with water molecules, moving inside a narrow tube-like selectivity filter, as in the KcsA ion channel \cite{REF1SUG5}\cite{REF1SUG6}\cite{REF1SUG7}, the DDHO template could also apply - provided the trio of DDHO ingredients are present. For ion transport through a uniform cylindrical channel, the solution to the PNP equation displays a sigmoid dielectric self-energy barrier \cite{Kuyucak2003}, but for more realistic, variable cross-sections, the self- and Coulomb energy profiles of an ion can exhibit local quadratic minima. We may safely assume that the elastic restoring force is isotropic and first-order charge-dependent, so: $F_{elastic} \propto-\kappa q x$ for small perturbations $x$ from the potential minimum. A damping force, first-order proportional to $\dot{x}$, is provided by the Langevin term. Thus, {\em if} harmonic driving forces are present, we obtain the familiar stochastic driven damped harmonic oscillator \cite{Gitterman}: 
$$m\frac{d^2x}{dt^2} = -\kappa q x - \gamma\frac{dx}{dt} + \sum_k F_k\exp(\Omega_k t) + \sqrt{2C}\nu(t)$$
\noindent using the notation from above, with $-\gamma\dot{x}$ being the Langevin damping force and parameter $C$ being Einstein's term $\gamma k_BT/m$. The random white noise term $\nu(t)$ covers all unknown interactions between the ion and the rest of the system. This equation is essentially identical to equation \ref{DDHO}, and so leads to the same conclusions on the resonance peaks as selection mechanism for ion species. \\
However, the BD framework is {\em not} able to account for the required external harmonic driving forces. In the Electro-Amphiphilic Hydrodynamic framework of the EAH-model that we build on here, vortices do appear spontaneously, providing adequate harmonic driving, and so offer a natural setting for the DDHO.

\section{CONVERTING DIMENSIONLESS UNITS TO FREQUENCIES IN Hz}
\label{Appendix2}
Exact values for the frequencies of the resonance peaks are difficult to assess, accounting for representing Figures \ref{fig:MahalVVQeff}-\ref{fig:resonanceseries} in the dimensionless units. This is caused by the difficulty to directly obtain experimental data on the required parameters $\gamma$ and $\kappa$ in equation \ref{DDHO_resomg}.
Let us try to estimate these values. \\
Following \cite{SUKH2012}, a typical force $F$ in the molecular mechanics of ion channels is about 4 picoNewtons (pN). 
We estimate a typical amplitude $A$ of ionic oscillation to be a few \AA. From equation \ref{DDHO}, this gives a value $\kappa=F$/e$AK\approx$ 1.6 pN/e\AA $\approx$ 1E17 [N/Cm] for ion valence $K$=1. 
For a hydrated sodium ion Na$^+$.(H$_2$O)$_{n_H}$, with mass $M$ (table \ref{table:TableResPeaks}): $M$=125.32 [Da], this gives a natural resonance frequency $\omega_N = \sqrt{\kappa\text{e}/M}\approx$2.77E+11 [rad/s].
\\
%Data on time scales in ion selection: 
According to \cite{MAFF2012}, in most ion channels, the time scale of ion permeation ranges from tens of nanoseconds (porins) to microseconds and even milliseconds (in active transport). Let us take a timescale $\tau$=1E-07 sec, giving a resonance frequency $\omega_R$ = 2$\pi/\tau \approx$ 6.28E07 [rad/s].
Rewriting equation \ref{DDHO} to: $\gamma = M\sqrt{2(\omega_N^2-\omega_R^2)}$ thus produces a value $\gamma\approx$1.93E-15 [kg/s].\\
With these values we can convert the dimensionless units in Figures \ref{fig:MahalVVQeff} and \ref{fig:resonanceseries} to dimensional units [rad/s] or [Hz]:\\
1 [dimensionless unit] $\simeq$ 6.74E07 [rad/s] $\simeq$ 1.1E07 [Hz]. \\
Note that the thus calculated ionic resonance values are magnitudes smaller than the harmonic frequencies of thermal molecular vibrations. For instance, using above conversion, the resonance peak for dehydrating hydrated sodium Na$^+$.(H$_2$O)$_{n_H}$ occurs at 4.32 MHz, while typical molecular resonance peaks occur in the THz regime, e.g. 46.8 THz for a single water molecule \cite{WANG2019}. Thus, regular thermal vibrations of aqua-ions will not cause spontaneous dehydration and subsequent selection.

\bibliography{arXivReferences.bib}

\end{document}